\documentclass[pra,superscriptaddress,reprint,dvips]{revtex4-1}
\usepackage{bm}
\usepackage{amsmath}
\usepackage{amssymb}
\usepackage{graphicx}
\usepackage[dvipdfmx, unicode=true,pdfusetitle,
 bookmarks=true,bookmarksnumbered=false,bookmarksopen=false,
 breaklinks=false,pdfborder={0 0 1},backref=false,colorlinks=false]
 {hyperref}

\begin{document}

\title{Evaluation of surface roughness of metal films using plasmonic Fano
resonance in attenuated total reflection}
\author{Munehiro Nishida}
\email{mnishida@hiroshima-u.ac.jp}

\affiliation{Graduate School of Advanced Science of Matter, Hiroshima University,
Higashi-Hiroshima, 739-8530, Japan}
\author{Taisei Matsumoto}
\affiliation{Graduate School of Advanced Science of Matter, Hiroshima University,
Higashi-Hiroshima, 739-8530, Japan}
\author{Hiroya Koga}
\affiliation{Graduate School of Advanced Science of Matter, Hiroshima University,
Higashi-Hiroshima, 739-8530, Japan}
\author{Terukazu Kosako}
\affiliation{YAZAKI Research and Technology Center, 1500 Mishuku, Susono-city,
Shizuoka, 410-1194, Japan}
\author{Yutaka Kadoya}
\affiliation{Graduate School of Advanced Science of Matter, Hiroshima University,
Higashi-Hiroshima, 739-8530, Japan}
\begin{abstract}
Attenuated total reflection (ATR) by surface plasmon polariton (SPP)
is a method for evaluating the dispersion relation of SPP from the
position of a dip in the reflection spectrum. However, recent studies
have shown that the dips are displaced from SPP resonance because
they are produced by a type of Fano resonance, i.e., the interference
between the resonant reflection process accompanied by resonant excitation
of SPP and the direct reflection process without resonant excitation\@.
This result suggests that the system properties difficult to be achieved
in the dispersion relation of SPP can be characterized using the ATR
method. In this study, we investigate the effect of surface roughness
due to nanosized dimples created in the initial stage of pitting corrosion
on the ATR spectrum, from the viewpoint of Fano resonance. Using the
temporal coupled-mode method, it is shown that the Fano resonance
in ATR is caused by the phase change of direct reflection because
of the absorption on the metal surface, and the spectral shape is
determined by this phase, along with the ratio of the external (radiative)
decay rate to the total decay rate of the resonant mode. Moreover,
it is clarified that the internal and external decay rates extracted
from the ATR spectrum provide information on corrosion, such as the
effective thickness of the metal film and the randomness in dimple
distribution. 
\end{abstract}
\maketitle

\section{Introduction}

Surface plasmon resonance sensors using resonant absorption by surface
plasmon polariton (SPP) are practically realized as highly sensitive,
refractive index sensors, and used widely in the fields of chemistry
and biotechnology \cite{Liedberg_83,Homola_99,Homola_08}. Kretschmann
configuration is the most popular structure of a surface plasmon resonance
sensor, which uses the attenuated total reflection (ATR) caused by
SPP on a thin metallic film evaporated on a prism \cite{Kretschmann_68}.
ATR is a phenomenon that produces sharp dips in the reflection spectra
at a specific incident angle or incident wavelength due to the resonant
excitation of SPP through the evanescent wave produced by the total
reflection of the prism.

It is considered that the dips in ATR originate from the material
loss of the metal film during the resonant excitation of SPP, and
that the dip position directly corresponds to the dispersion relation
of SPP\@. In 1971, Kretschmann derived a Lorentzian spectral function
typical for resonant phenomena by approximating an exact reflection
coefficient for a three-layer structure near the resonant wavenumber
of SPP \cite{Kretschmann_71,Raether_88}. This result is the basis
of the above view and was used in interpreting the results of subsequent
studies. In this view, the change in the dispersion relation of SPP
(relation between the incident angle and resonant frequency) appears
directly in the position change of the dip, and enables to evaluate
the change in the refractive index of the ambient medium using the
expression of the dispersion relation.

However, recent studies \cite{Vinogradov_18,Nesterenko_18} have shown
that resonant dips are produced by a type of Fano resonance \cite{Fano_61}.
There appears a shift between the complex wavenumbers of the reflection
coefficient pole (complex resonant wavenumber) and the reflection
coefficient zero (complex zero-point wavenumber) due to the metallic
loss. As a result, there appears an asymmetric peak-dip structure
in the spectrum. Such behavior of the reflection coefficient can be
interpreted as a result of the interference between the resonant reflection
process accompanied by the excitation of the resonant mode and the
direct reflection process without resonant excitation.

There is a shift between the dip and resonant positions in the Fano
spectrum, whose size is determined by the interference of the two
reflection processes and changes in a complicated manner with a change
in the amplitude and phase of the reflected waves. Therefore, calibration
using the measured value of the real system is necessary to precisely
evaluate the refractive index of the ambient medium. On the other
hand, using the notion of Fano resonance, the properties that are
difficult to be achieved in the dispersion relation of SPP (the real
part of the complex resonant frequency) can be evaluated. For example,
diffusive scattering caused by the surface roughness affects the interference
between resonantly reflected wave and directly reflected wave, and
may shift the position of the dip (peak) produced by destructive (constructive)
interference.

\begin{figure*}
\includegraphics[width=0.8\textwidth]{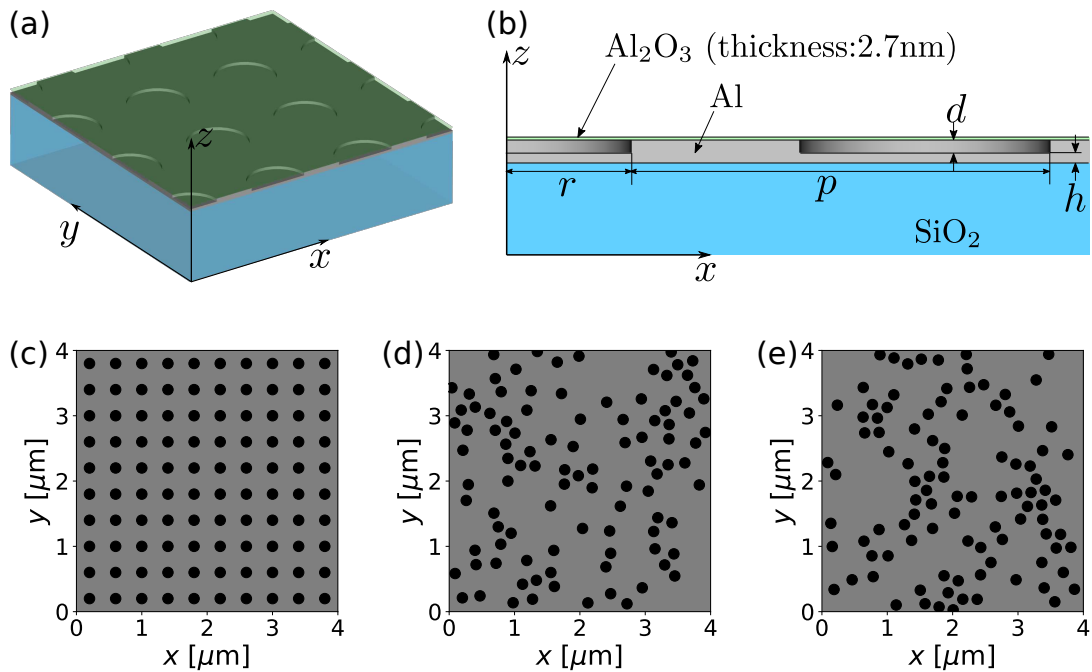}

\caption{\label{fig:system}System configuration. (a) Schematic of a cylindrical
nanodimple array created on an aluminum film. (b) A cross section
of the system. (c) Configuration of a periodic dimple array. (d) and
(e) Random arrangements of dimples that produce the smallest (d) and
largest (e) internal losses in 100 samples.}
\end{figure*}

Indeed, there have long been studies evaluating the roughness of a
metal surface using the ATR method \cite{Braundmeier_74,Raether_88}.
Experimental studies have shown that the effects of surface roughness
are clearly seen as changes in dip position and width \cite{Hornauer_74,Braundmeier_74,Orlowski_79,Sari_80,Chen_10,Agarwal_17}.
In addition, there have been attempts to study the corrosion resistance
of the metal by observing these changes in the ATR method \cite{Bussjager_96,Oliveira_17}.
However, theoretical studies that treated the effect of roughness
as perturbation have shown that second-order approximation is necessary
for describing the influence of roughness on the SPP dispersion relation,
which gives about a factor 10 smaller change in the dip position as
compared to the observed data \cite{Raether_83,Raether_88}. Although
attempts have been made to explain the difference between theory and
experiment in a specific situation \cite{Fontana_88}, as far as we
know, there is no theory that can quantitatively estimate the effect
of roughness of a metal surface on the dip position in general systems.
What is lacking here is the viewpoint that the dip and the dispersion
relation of SPP are not directory related but displaced with each
other due to Fano resonance. From this viewpoint, the results of experiments
and theories obtained so far are consistent.

Thus, reconsideration of the ATR method from the viewpoint of Fano
resonance will provide information on the system property, e.g., the
progress of corrosion, with high accuracy. Especially, at the initial
stage of pitting corrosion \cite{Frankel_98,Kaesche_03,McCafferty_10},
nanosized pits are created on the metal surface and can cause strong
diffusive scattering to SPP\@. This is expected to influence the
shape of the reflection spectrum via Fano resonance, and may enable
the deduction of the initial progress of pitting corrosion from the
spectral shape. Pitting is a type of localized corrosion that occurs
in a metal with a passivation film, such as aluminum alloy or stainless
steel. It is a dangerous process, causing accelerated localized dissociation
of metals that trigger mechanical failures or stress corrosion cracking.
Therefore, it is quite important to detect its initial process. If
the roughness of the metal surface can be read accurately from the
ATR spectrum, it will become possible to detect the initial process
of pitting corrosion on nanoscale.

In this paper, to build a base to quantitatively deduce the roughness
of metal surfaces from a spectral shape, the relation between the
ATR spectrum and metal surface condition is clarified by analyzing
the Fano resonance in the ATR of Kretschmann configuration using the
temporal coupled mode (TCM) method \cite{Haus_83,Fan_03,Nishida_Kadoya_18}
and the spatial coupled mode (SCM) method \cite{Garcia-Vidal_Martin-Moreno_05,Nishida_Kadoya_15,Nishida_Kadoya_18}.
Especially, distribution of cylindrical dimples is used for a model
of surface roughness or pits created in pitting corrosion. The effect
of the periodic and random arrays of dimples on the ATR spectrum is
studied in detail. Specifically, first, we identify the parameters
that determine the shape of the resonant spectrum based on the expression
of the reflection coefficient obtained using the TCM method. Next,
we calculate the reflection spectrum by using the SCM method and extract
parameters for a system composed of a flat metal film with various
thicknesses and material loss and for a system composed of an aluminum
film with a periodic or random array of cylindrical dimples. Finally,
we study the correlation between the parameters and the condition
of the metal film to verify the possibility of evaluating the pitting
corrosion from the ATR spectrum shape.

\subsection*{System}

Figure\ \ref{fig:system} shows a schematic of our concerned system.
Cylindrical dimples are created periodically or randomly in the aluminum
thin film evaporated on a SiO$_{2}$ substrate. To account for the
natural oxide film created on the surface of the aluminum film, a
homogeneous Al$_{2}$O$_{3}$ film with a thickness of 2.7 nm covers
the aluminum film even at the top of the dimple. Although it is plausible
that the oxide film at the top of the dimple is removed, we omit this
effect because it produces only a small change in the reflection spectrum.

In what follows, it is assumed that the radius and depth of the cylindrical
dimple are $r$ and $d$, respectively; the thickness of the aluminum
film left below the bottom of the dimple is $h$; and the period of
the periodic dimple array is $p$, as shown in Fig.\ \ref{fig:system}(b).
The refractive indices of the SiO$_{2}$ substrate and Al$_{2}$O$_{3}$
film are 1.457 and 1.6764, respectively, and the space above the Al$_{2}$O$_{3}$
film and the inner space of the dimples are filled with a NaCl solution
whose refractive index is 1.338. The Drude-Lorentz model, proposed
in \cite{Rakic_98}, is used for the dielectric function of aluminum
$\varepsilon_{\text{Al}}\text{\ensuremath{\left(\omega\right)}}$.

We focus on the resonant structure that appears in the incident-angle
dependence of zeroth-order reflection when an incident light with
wavelength $\lambda_{\text{i}}=650$ nm is irradiated from the substrate
side under the total reflection condition. We study the spectral change
in the following three cases to reveal the influence of the surface
condition on the spectral shape: without dimples ($d=0$ nm, corresponding
to a flat metal film), with a periodic dimple array (Fig.\ \ref{fig:system}(c)),
and with a random dimple array (Fig.\ \ref{fig:system}(d) and (e)).
Here, the random array is composed of randomly placed cylindrical
dimples with a radius of 80 nm. We produce 100 samples of a random
dimple array under the condition that the distance between adjacent
dimples is not less than 60 nm and 100 dimples exist in the 4 $\mu$m
$\times$ 4 $\mu$m area, which is the calculation area imposed by
the periodic boundary condition along the $x$ and $y$ directions.
We select three samples from the 100 samples in which the effect of
diffusive scattering is considered minimum (d), maximum (e), and near
the average, judging from the analysis of the spectral shape factor
described below. In what follows, we will present only the data of
these three samples for random arrays.

\section{Methods}

\subsection{Temporal coupled mode method}

Kretschmann configuration in plasmonic ATR is a system in which a
cavity with a single resonant mode corresponding to SPP is attached
to a single input/output port representing a connection with the incident
and reflected waves. The TCM method describes resonant scattering
phenomena in a unified way by considering the dynamics of cavities
attached to ports \cite{Haus_83,Fan_03,Nishida_Kadoya_18}. Therefore,
our system can be described by the TCM method, which reproduces general
features of the resonant spectrum by setting a few parameters, and
is effective in understanding the origin of the resonance.

In the case where the cavity couples with the port weakly, the amplitude
of the resonant mode with a resonant angular frequency of $\omega_{\text{r}}$
is described by the following coupled mode equation \cite{Haus_83,Fan_03,Joannopoulos_08}:

\begin{align}
\frac{\text{d}a}{\text{d}t} & =-\text{i}\omega_{\text{r}}a-\gamma a+\kappa s_{+},\label{eq:a_motion-1}\\
s_{-} & =r_{\text{d}}\text{e}^{\text{i}\phi_{\text{d}}}s_{+}+da,\nonumber \\
\gamma & =\gamma_{\text{i}}+\gamma_{\text{e}}.\nonumber 
\end{align}
Here, $\gamma_{\text{i}}$ denotes the internal decay rate due to
the loss of the materials composing the cavity and $\gamma_{\text{e}}$
denotes the external decay rate due to the loss caused by the radiation
to the port. The variables $s_{+}$ and $s_{-}$ denote the amplitudes
incoming and outgoing the radiative modes through the port, respectively,
where the mode fields are normalized so that $\left|s_{\pm}\right|^{2}$
denote the powers of the modes. The parameters $r_{\text{d}}$ and
$\phi_{\text{d}}$ denote the magnitude and phase of the direct reflection
coefficient, which determines the reflection process in which the
incoming wave from the port is reflected directly to the port without
the resonant excitation. The parameter $\kappa$ ($d$) denotes the
coupling constant between the incoming (outgoing) mode and cavity
mode through the port.

In the case where the internal loss of the resonant mode, the energy
absorption in the direct reflection process, and the coupling to the
port are all weak, the parameters $\gamma_{\text{e}}$, $d$, and
$\kappa$ can be approximately treated as independent of the internal
decay rate $\gamma_{\text{i}}$ and absorption in the direct reflection
process. Considering the case where $\gamma_{\text{i}}=0$ and the
direct reflection coefficient is expressed as $r_{\text{c}}=\text{e}^{\text{i}\phi_{\text{c}}}$,
the relations $|d|^{2}=2\gamma_{\text{e}}$, $r_{\text{c}}d^{*}=-d$,
and $d=\kappa$ are obtained from the principle of conservation of
energy and the time-reversal symmetry \cite{Fan_03}. For continuous-wave
incidence with the angular frequency of $\omega$, the reflection
coefficient is expressed as 
\begin{align}
r\left(\omega\right) & =r_{\text{d}}\text{e}^{\text{i}\phi_{\text{d}}}\left(1-\frac{2\text{i}\frac{\gamma_{\text{e}}}{r_{\text{d}}}\text{e}^{\text{i}\phi}}{\omega-\omega_{\text{r}}+\text{i}\gamma}\right)\nonumber \\
 & =r_{\text{d}}\text{e}^{\text{i}\phi_{\text{d}}}\left(\frac{\omega-\omega_{\text{0}}+\text{i}\gamma_{0}}{\omega-\omega_{\text{r}}+\text{i}\gamma}\right),\label{eq:r_coupled_mode}\\
\omega_{0} & =\omega_{\text{r}}-2\sin\phi\gamma_{\text{e}}/r_{\text{d}},\ \gamma_{0}=\gamma-2\cos\phi\gamma_{\text{e}}/r_{\text{d}}.\label{eq:omega_0_gamma_0}
\end{align}
Thus, the change in the phase of direct reflection due to absorption
$\phi=\phi_{\text{c}}-\phi_{\text{d}}$ makes the shape of the reflection
spectrum asymmetric and shifts the minimum from the resonance condition.

\subsection{SCM method}

The SCM method describes the electromagnetic field in the metal film
with nanoholes by waveguide modes in the nanohole. It enables a semi-quantitative
calculation for reflectance and transmittance with a small computational
resource and high speed. However, because it is assumed that the electromagnetic
field becomes zero inside the metal film in the SCM method developed
so far, it is not possible to accurately describe a system with a
thin metal film in which the effect of tunneling through the film
is large. To make the SCM method applicable to the system with a thin
metal film, we make the following improvements (see Supplemental Material
for the derivation): 
\begin{itemize}
\item To adopt the boundary condition considering the penetration of electromagnetic
wave inside the metal (transition boundary condition \cite{Eriksson_07}
with in-plane wavenumber dependence). 
\item Not to use mean-field approximation for the inner product of the waveguide
mode and plane wave \cite{Nishida_Kadoya_15}. 
\item To fix reciprocity using the scattering matrix. 
\end{itemize}
As mentioned in the last item, we use a scattering matrix that enables
to calculate spectra for arbitrary multilayer films by using recurrence
formula \cite{Weiss_11}. Therefore, the usual Kretschmann configuration
without the layer of nanohole array can be treated similarly.

\begin{figure}
\begin{minipage}[t]{0.5\columnwidth}%
 
\begin{flushleft}
(a)\\
 \includegraphics[width=0.95\columnwidth]{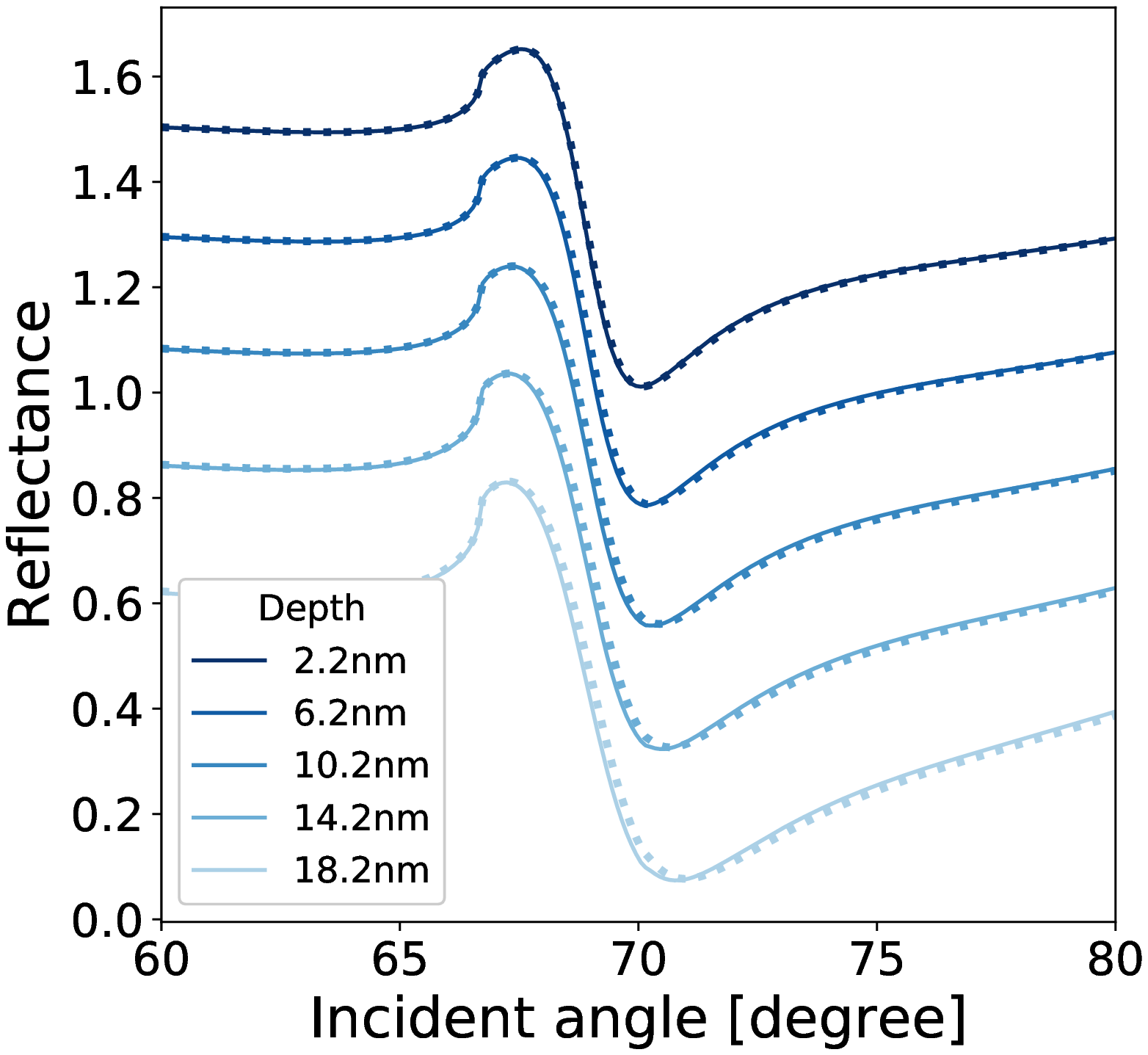} 
\par\end{flushleft}%
\end{minipage}%
\begin{minipage}[t]{0.5\columnwidth}%
 
\begin{flushleft}
(b)\\
 \includegraphics[width=0.95\columnwidth]{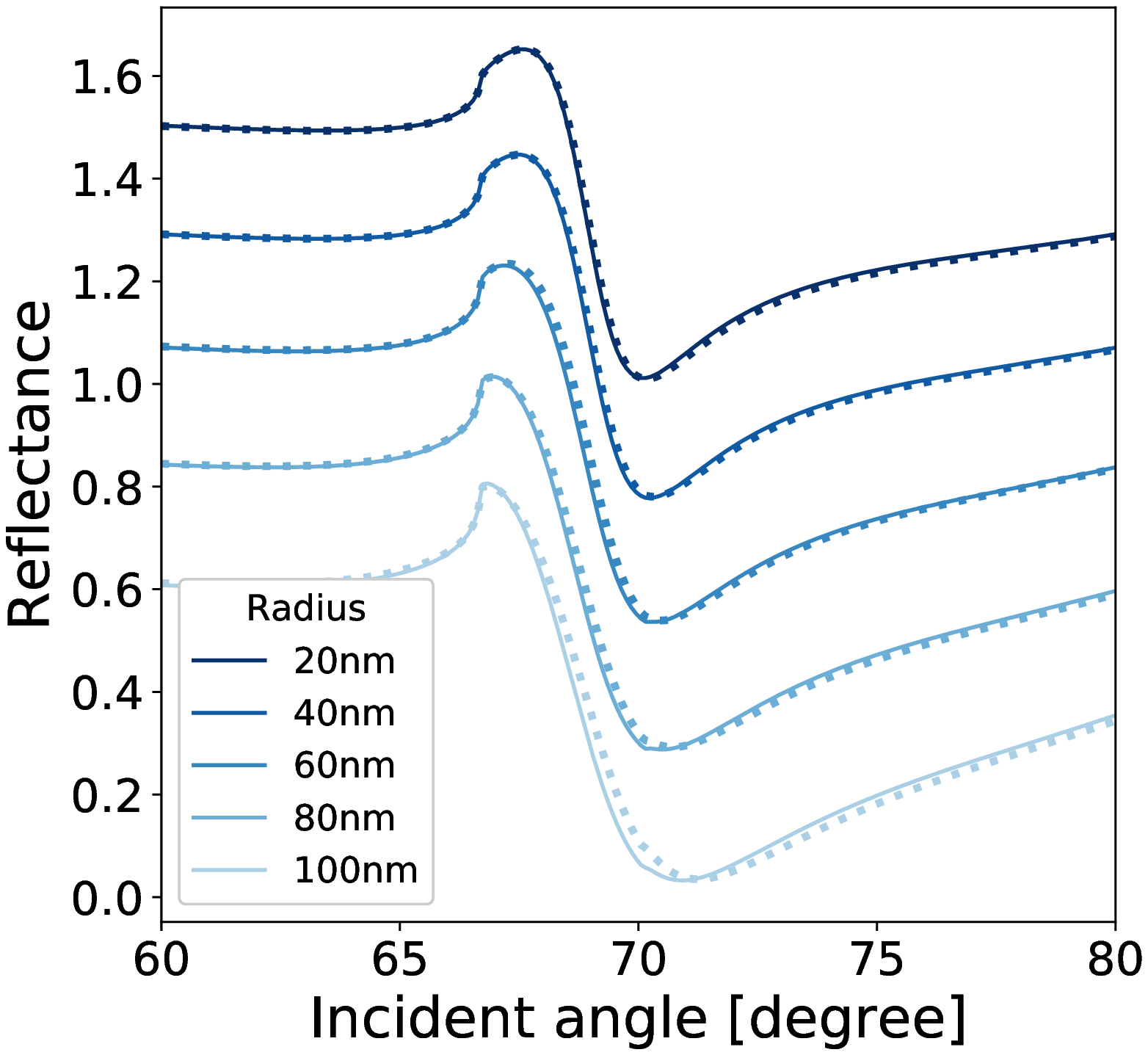} 
\par\end{flushleft}%
\end{minipage}\caption{\label{fig:RCWA_vs_CM}Comparison between the reflection spectra obtained
by the RCWA method (solid lines) and SCM method (dotted lines) for
nanodimple arrays whose period is set to 240 nm. (a) Dependence on
the depth of the dimple. The radius of the dimple is set to 50 nm.
(b) Dependence on the dimple radius. The depth of the dimple is set
to 10.2 nm.}
\end{figure}

Through the above treatments, the quantitativity of the SCM method
for a system that contains a metallic thin film is improved enough
to yield almost the same result as that obtained using the rigorous
coupled wave analysis (RCWA) method \cite{Weiss_11}. Figure \ref{fig:RCWA_vs_CM}
shows the incident angle dependence of the reflectance obtained by
the RCWA method (solid lines) and SCM method (dotted lines). Note
that the reflectances are shifted vertically by 0.2 for clarity.

\section{Results and Discussion}

\subsection{Fano resonance spectral profile in attenuated total reflection}

Consider the case where the angular spectra near the resonant angle
are obtained using a focused incident light with an angular frequency
of $\omega_{\text{i}}=\frac{2\pi}{c\lambda_{\text{i}}}$, where $c$
is the speed of light in vacuum. The wavenumber dependence of resonant
angular frequency in the narrow range near $\omega_{\text{i}}$ can
be approximated as linear, and other parameters as constants. In this
case, the concrete formula of wavenumber dependence of the reflectance
can be obtained from the formula of the TCM method (\ref{eq:r_coupled_mode}).

First, assuming that $\omega_{\text{r}}$ becomes $\omega_{\text{i}}$
when $k_{x}=k_{\text{r}}$ and the relation between $\omega_{\text{r}}$
and $k_{x}$ is linear with the gradient of the group velocity $v_{\text{SPP}}$
of SPP on the surface of the semi-infinite metal, the $k_{x}$-dependence
of $\omega_{\text{r}}$, namely the dispersion relation of the resonant
mode, is expressed as 
\begin{equation}
\omega_{\text{r}}=v_{\text{SPP}}\left(k_{x}-k_{\text{r}}\right)+\omega_{\text{i}}.\label{eq:dispersion_relation}
\end{equation}
Taking the values at $k_{x}=k_{\text{r}}$ for the parameters $\gamma_{\text{e}}$,
$\gamma$, $r_{\text{d}}$, and $\phi$ in Eq.\ (\ref{eq:r_coupled_mode}),
the reflectance $R=\left|r\right|^{2}$ becomes

\begin{align}
\frac{R\left(k_{x}\right)}{R_{\text{d}}} & =\frac{\left\{ \frac{v_{\text{SPP}}}{\gamma}\left(k_{x}-k_{\text{r}}\right)-2\chi\sin\phi\right\} ^{2}+\left(1-2\chi\cos\phi\right)^{2}}{\left\{ \left(\frac{v_{\text{SPP}}}{\gamma}\right)\left(k_{x}-k_{\text{r}}\right)\right\} ^{2}+1}.\label{eq:reflectance_for_kx}
\end{align}
Thus, the shape of the reflection spectrum as a function of the normalized
wavenumber $\frac{v_{\text{SPP}}}{\gamma}k_{x}$ is determined using
the following two parameters: ratio of external decay rate to total
decay rate modified by the magnitude of the direct reflection coefficient
$r_{\text{d}}$, 
\begin{equation}
\chi=\frac{\gamma_{\text{e}}}{r_{\text{d}}\gamma},
\end{equation}
and the phase change $\phi$ due to the absorption during direct reflection.
Here, $R_{\text{d}}=r_{\text{d}}^{2}$ is the direct reflection rate.

The wavenumbers $k_{+}$ and $k_{-}$, which provide the maximum and
minimum of Eq.\ (\ref{eq:reflectance_for_kx}), respectively, are
given by the following equations using $p=\pm1$, which is the sign
of $\sin\phi$, namely $\sin\phi=p\left|\sin\phi\right|$: 
\begin{align}
k_{\mp} & =k_{\text{r}}\pm p\frac{\gamma}{v_{\text{SPP}}}f^{\mp}\left(\phi,\chi\right),\label{eq:k_pm}\\
f^{\mp}(\phi,\chi) & =\frac{\sqrt{\left(\cos\phi-\chi\right)^{2}+\sin^{2}\phi}\mp\left(\cos\phi-\chi\right)}{\left|\sin\phi\right|}.\label{eq:SSF}
\end{align}
The local maximum $R_{+}$ and the local minimum $R_{-}$ of reflectance
are given by 
\begin{align}
R_{\text{\ensuremath{\pm}}} & =R_{\text{d}}\left\{ 1\pm\frac{2\chi\left|\sin\phi\right|}{f^{\pm}\left(\phi,\chi\right)}\right\} .\label{eq:extreme_values}
\end{align}
Thus, the total absorption ($\left|r\right|^{2}=0$) is realized when
\begin{equation}
\chi=\frac{\gamma_{\text{e}}}{r_{\text{d}}\gamma}=\frac{1}{2\cos\phi},\label{eq:critical_coupling}
\end{equation}
which is a modified critical coupling condition.

\begin{figure*}
\noindent %
\noindent\begin{minipage}[t]{0.31\textwidth}%
\begin{flushleft}
(a)\\
 \includegraphics[width=0.9\columnwidth]{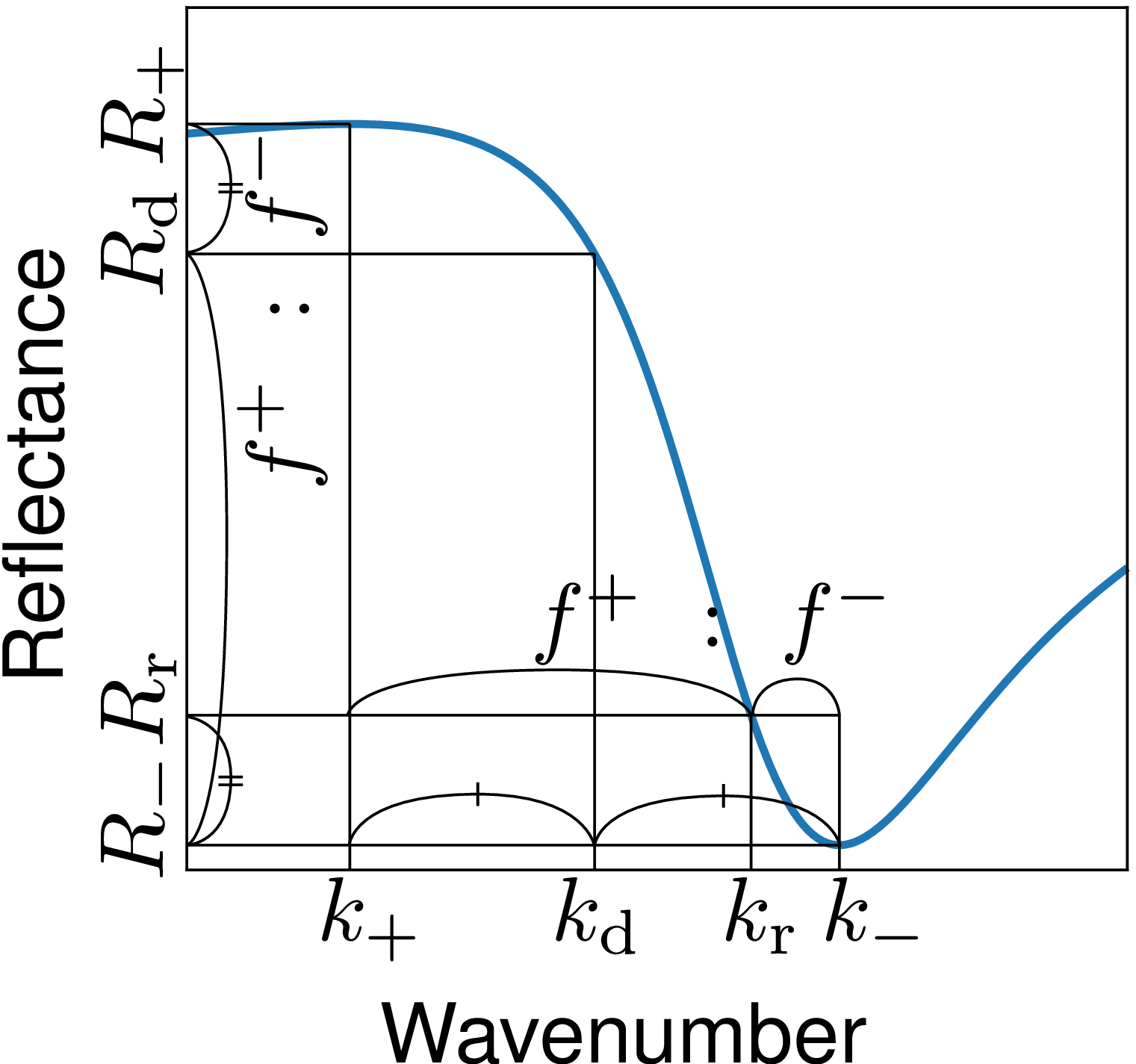}
\par\end{flushleft}%
\end{minipage}%
\begin{minipage}[t]{0.32\textwidth}%
\begin{flushleft}
(b)\\
 \includegraphics[width=0.9\columnwidth]{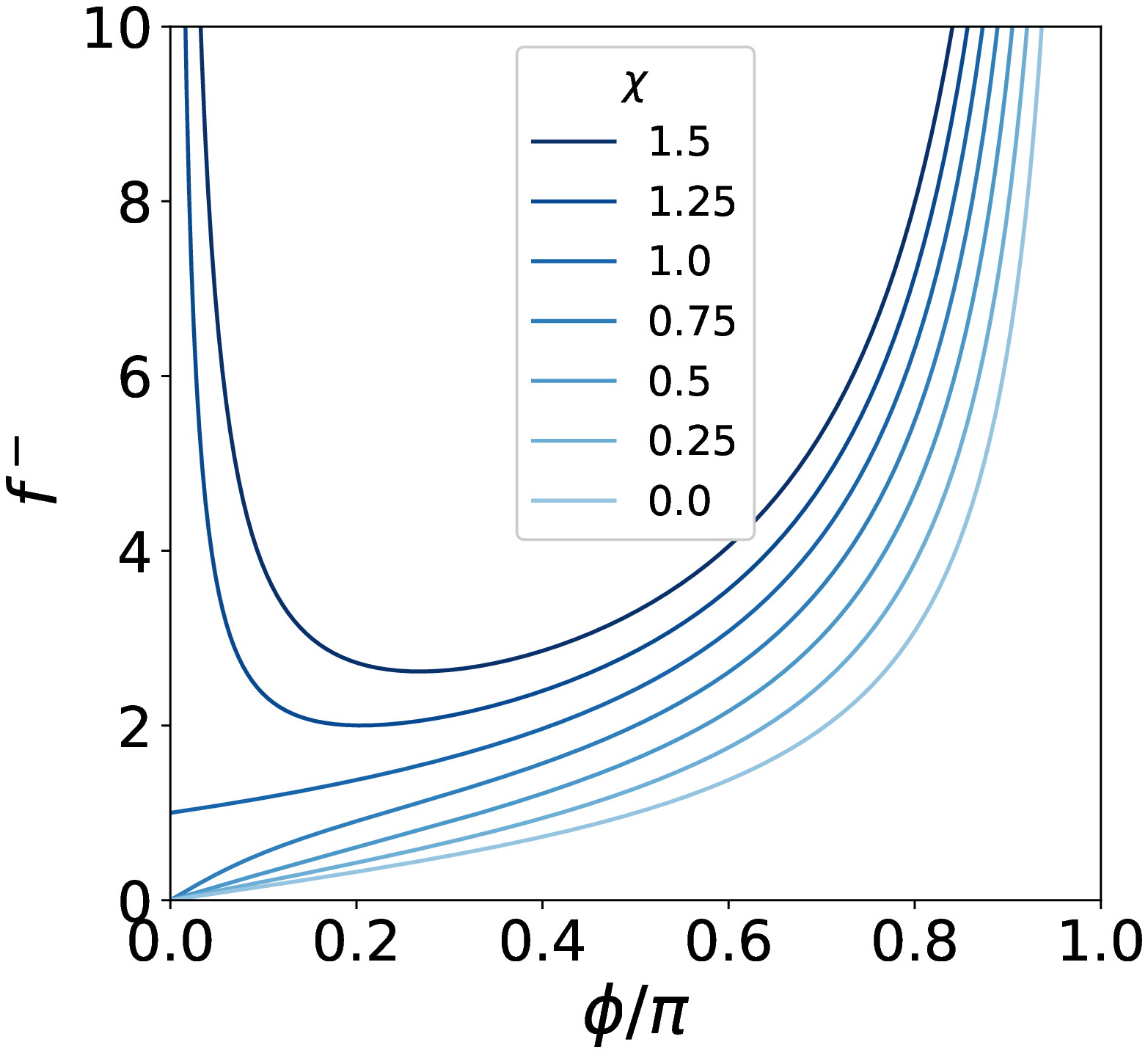}
\par\end{flushleft}%
\end{minipage}%
\begin{minipage}[t]{0.37\textwidth}%
\begin{flushleft}
(c)\\
 \includegraphics[width=0.9\columnwidth]{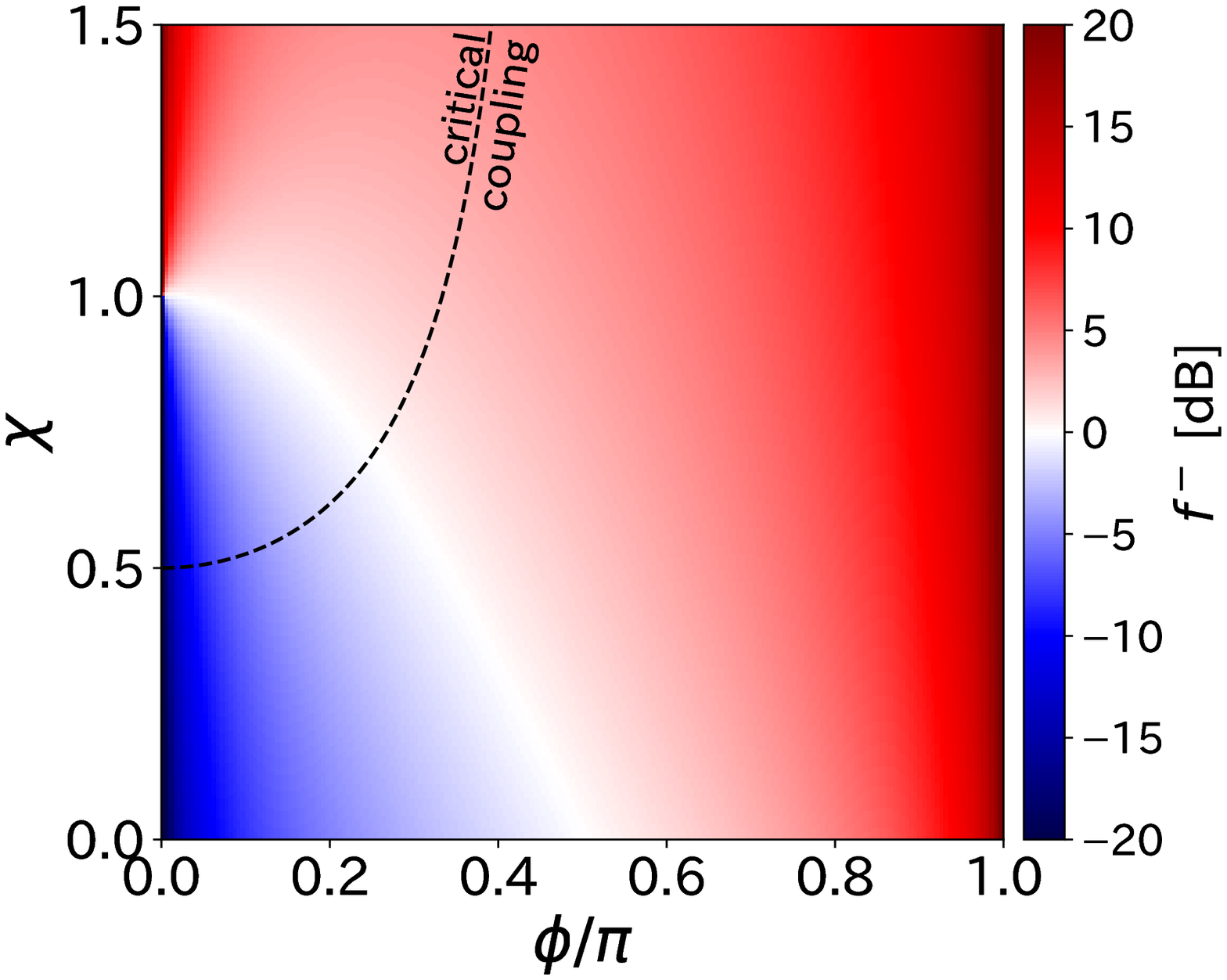}
\par\end{flushleft}%
\end{minipage}\caption{\label{fig:spedtrum_shape_factor}Characteristics of spectral profile
of Fano resonance. (a) Characteristics of the reflection spectrum
determined by SSF $f^{\pm}$. (b) and (c) Dependence of SSF on the
direct reflection phase $\phi$ and the ratio of decay rates $\chi$.
The dashed line in panel (c) denotes the critical coupling condition
Eq.\ (\ref{eq:critical_coupling}).}
\end{figure*}

At the midpoint between the wavenumbers of the reflectance maximum
and minimum, $k_{\text{d}}=\frac{k_{+}+k_{-}}{2}$, the reflectance
equals the direct reflection rate (see Supplemental Material for the
derivation): 
\begin{equation}
R\left(k_{\text{d}}\right)=R_{\text{d}}.\label{eq:Rd}
\end{equation}
In addition, 
\begin{equation}
R_{\text{r}}=R\left(k_{\text{r}}\right)=R_{-}+R_{+}-R_{\text{d}}=R_{-}+\frac{f^{-}}{f^{+}}\left(R_{\text{d}}-R_{-}\right).\label{eq:resonant_freq_estimation}
\end{equation}
Considering the relation $f^{+}f^{-}=1$, 
\begin{equation}
f^{-}=\sqrt{\frac{f^{-}}{f^{+}}}=\sqrt{\frac{R_{\text{r}}-R_{-}}{R_{\text{d}}-R_{-}}}.\label{eq:fm_estimation}
\end{equation}
These characteristics of the reflection spectrum are shown in Fig.\ \ref{fig:spedtrum_shape_factor}(a),
where $f^{\pm}$ is an important factor that determines the spectrum
shape. Therefore, we call $f^{\pm}$ the spectral shape factor (SSF)
in the following.

Eq.\ (\ref{eq:k_pm}) shows that the shift of the dip wavelength
$k_{-}$ from the resonant wavelength $k_{\text{r}}$ is determined
by the product of SSF $f^{-}$ and the decay rate $\gamma$. Though
the shift basically increases with $\gamma$, the value of $f^{-}$
can change largely depending on the parameters. Figures \ref{fig:spedtrum_shape_factor}
(b) and (c) show the behavior of the change in $f^{-}$ due to the
system parameters. We can see that $f^{-}$ can become very large
in the case where $\chi>1$ and $\phi\sim0$ or in the case of $\phi\sim\pi$.
Especially, in the region of a small phase, $f^{-}$ changes discontinuously
depending on the value of $\chi$. Therefore, it is important to investigate
how parameters $\chi$ and $\phi$ change depending on the system
configuration. Because the in-plane wavenumber $k_{x}$ is related
to the incident angle $\theta$ by the relation $k_{x}=n\frac{\omega_{\text{i}}}{c}\sin\theta$,
with $n$ being the refractive index of the prism, it is possible
to evaluate the parameters $\chi$, $\phi$, $\gamma$, and $f^{\pm}$
from the information of the angular spectrum, such as the minimum
and maximum points, as shown in the next section.

\subsection{Parameter extraction methods}

Because the reflection coefficient $r\left(\omega\right)$ can be
calculated rapidly by using the SCM method (scattering matrix method
for homogeneous multilayer films), we can find the pole $\omega_{\text{r}}-\text{i}\gamma$
and the zero $\omega_{0}-\text{i}\gamma_{0}$ easily. Then, $r_{\text{d}}$
is evaluated as the ratio between $r\left(\omega\right)$ and $\left(\omega-\omega_{\text{0}}+\text{i}\gamma_{0}\right)/\left(\omega-\omega_{\text{r}}+\text{i}\gamma\right)$.
From Eq.\ (\ref{eq:omega_0_gamma_0}), we obtain 
\begin{equation}
\phi=\arctan\left(\frac{\gamma-\gamma_{0}}{\omega_{\text{r}}-\omega_{0}}\right),
\end{equation}
\begin{equation}
\chi=\frac{\gamma_{\text{e}}}{r_{\text{d}}\gamma}=\sqrt{\left(\omega_{\text{r}}-\omega_{0}\right)^{2}+\left(\gamma-\gamma_{0}\right)^{2}}/\left(2\gamma\right).
\end{equation}
In this paper, we call this method parameter extraction by zero-point
search (ZPS).

On the other hand, using Eqs.\ (\ref{eq:k_pm}), (\ref{eq:SSF}),
(\ref{eq:extreme_values}), (\ref{eq:Rd}), and (\ref{eq:fm_estimation})
obtained by the analysis of the TCM method, the values of $k_{\text{r}}$,
$\chi$, and $\phi$ can be deduced from the angle spectrum. The value
of $R_{\text{d}}$ is obtained from the maximum point $\left(k_{+},R_{+}\right)$
and minimum point $\left(k_{-},R_{-}\right)$ using Eq.\ (\ref{eq:Rd}).
Then, $f^{-}$ is determined by reflectance $R_{\text{r}}$ at the
resonant wavelength $k_{\text{r}}$ using Eq.\ (\ref{eq:fm_estimation}).
Using Eq.\ (\ref{eq:k_pm}), the decay rate can be expressed as $\gamma=v_{\text{SPP}}\left|k_{-}-k_{\text{r}}\right|/f^{-}$
and is determined by $k_{\text{r}}$. Using Eq.\ (\ref{eq:extreme_values}),
$\chi\sin\phi$ is obtained from $R_{\text{d}}$, $f^{-}$, and $R_{-}$,
and then using Eq.\ (\ref{eq:SSF}), $\chi$ and $\phi$ can be obtained
by considering that $\gamma_{0}$ becomes 0 and $\cos\phi$ changes
its sign under the critical coupling condition (\ref{eq:critical_coupling}).
Therefore, only $k_{\text{r}}$ is left for the determination of reflectance
of the TCM method (\ref{eq:reflectance_for_kx}). In other words,
if the value of $k_{\text{r}}$ is provided, all parameters included
in the reflectance of the TCM method (\ref{eq:reflectance_for_kx})
are determined using the SCM spectral data, and the TCM spectral data,
such as $R_{\pm}$ and $R_{\text{d}}$, can be obtained. Thus, it
is possible to determine the parameters consistently with the model
of the TCM method by determining $k_{\text{r}}$ so as to minimize
the difference between the $R_{\pm}$ and $R_{\text{d}}$ values evaluated
using the SCM and TCM methods. In this paper, we call this method
parameter extraction using the TCM method.

In what follows, after checking the consistence between the two parameter
extraction methods, ZPS and TCM, we explore the possibility of characterization
of the metal surface condition by studying the variation in SSF depending
on the metal surface condition.

\begin{figure}
\begin{minipage}[t]{0.5\columnwidth}%
\begin{flushleft}
(a)\\
 \includegraphics[width=0.95\columnwidth]{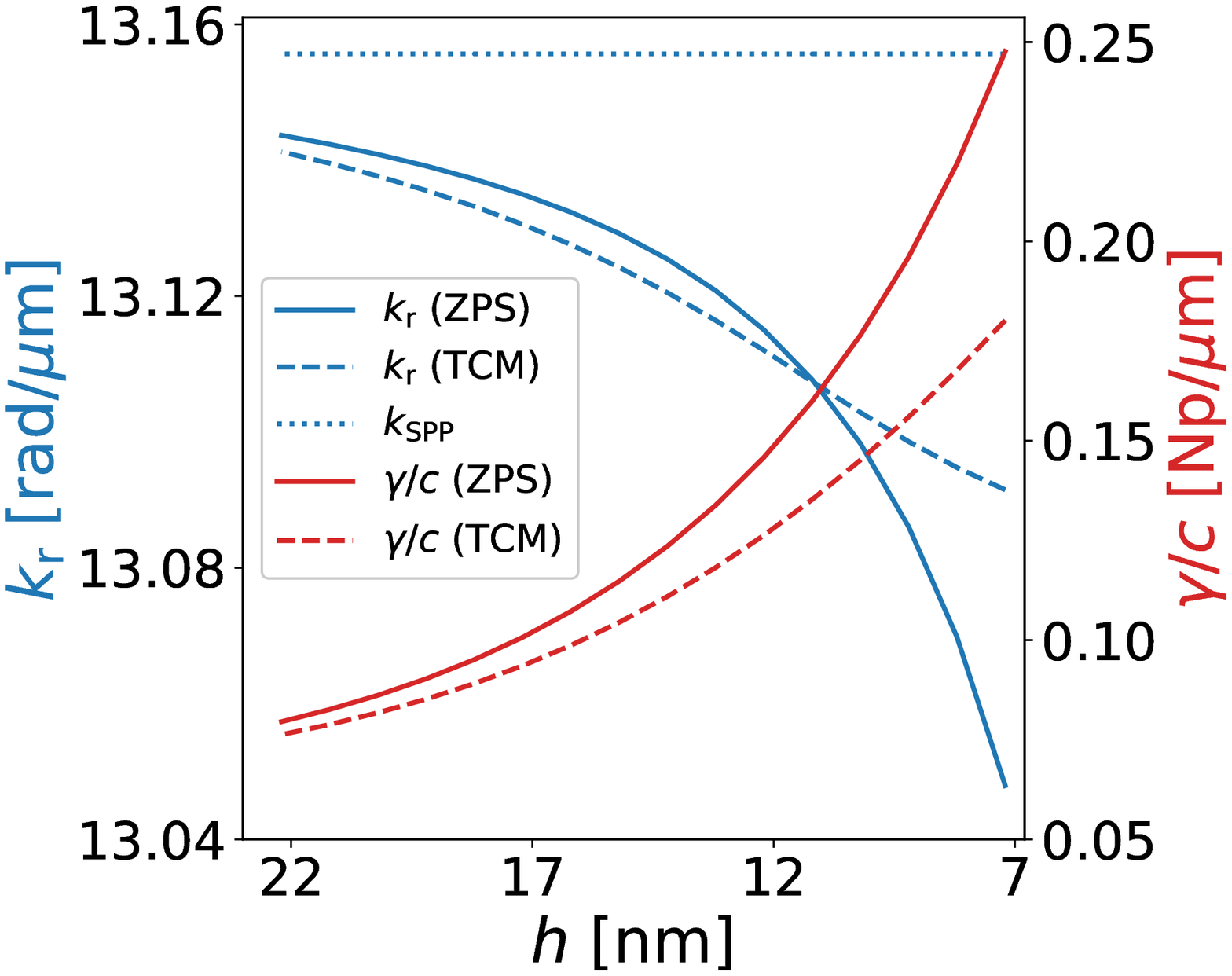}
\par\end{flushleft}%
\end{minipage}%
\begin{minipage}[t]{0.5\columnwidth}%
\begin{flushleft}
(b)\\
 \includegraphics[width=0.95\columnwidth]{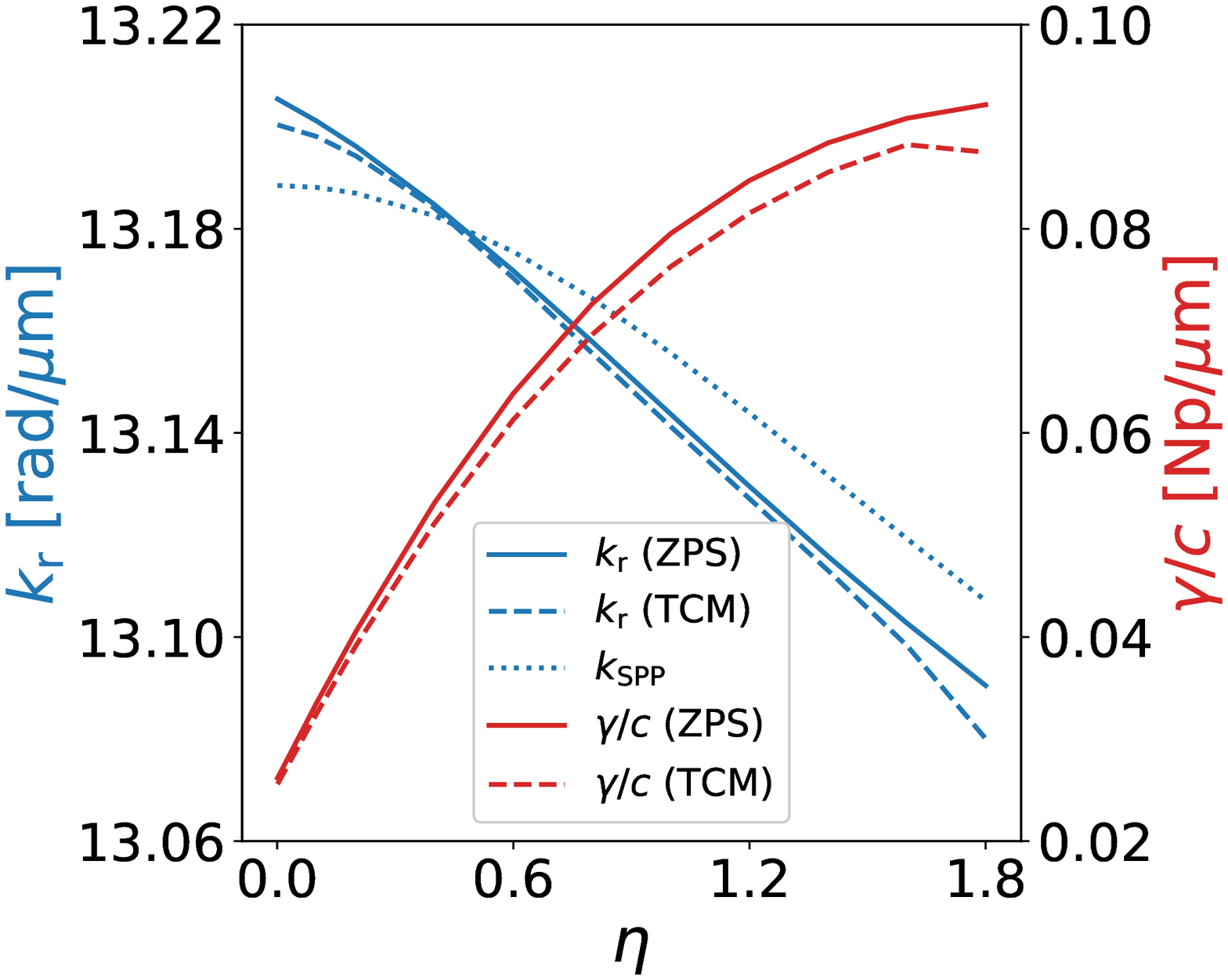}
\par\end{flushleft}%
\end{minipage}\\
\begin{minipage}[t]{0.5\columnwidth}%
\begin{flushleft}
(c)\\
 \includegraphics[width=0.95\columnwidth]{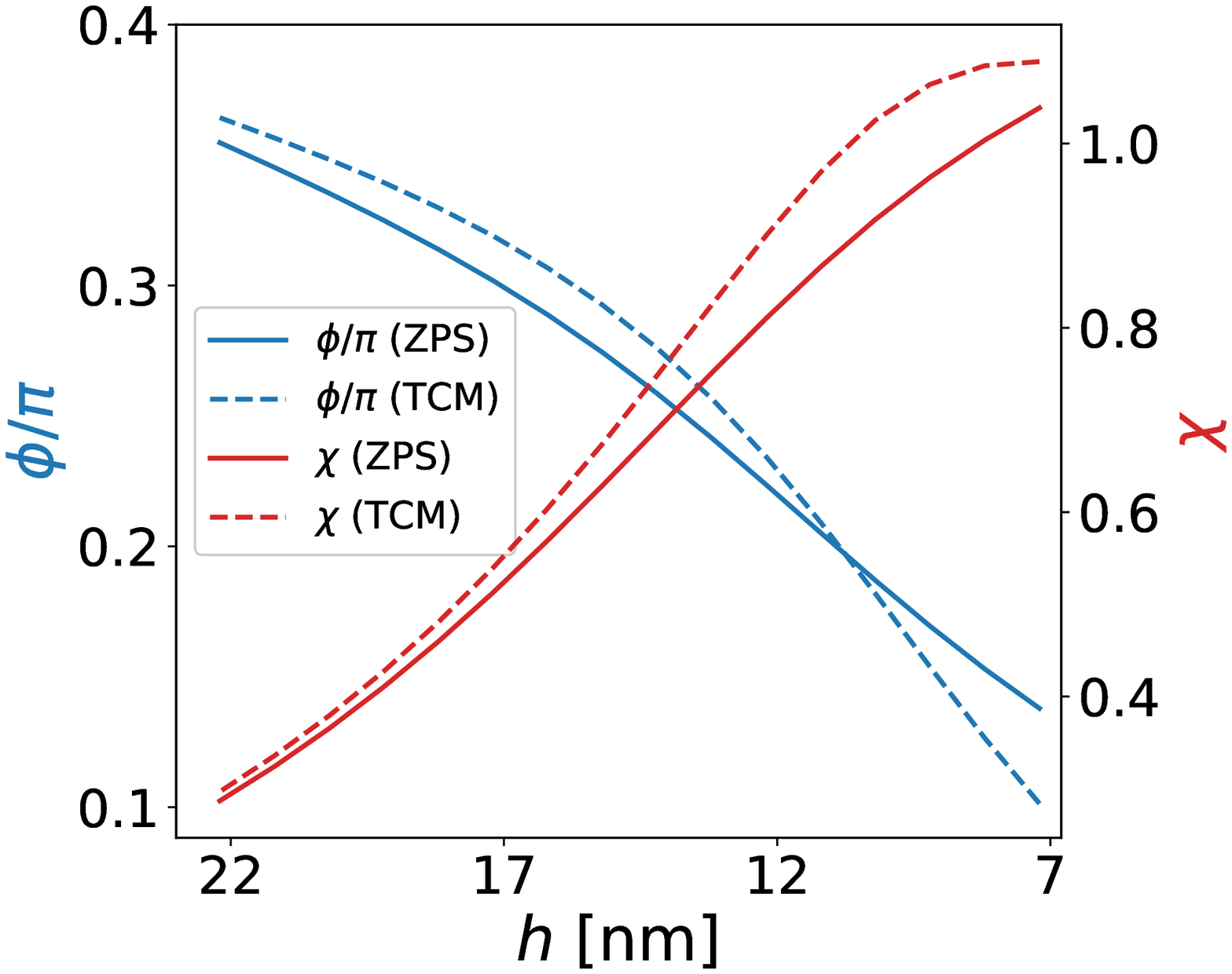}
\par\end{flushleft}%
\end{minipage}%
\begin{minipage}[t]{0.5\columnwidth}%
\begin{flushleft}
(d)\\
 \includegraphics[width=0.95\columnwidth]{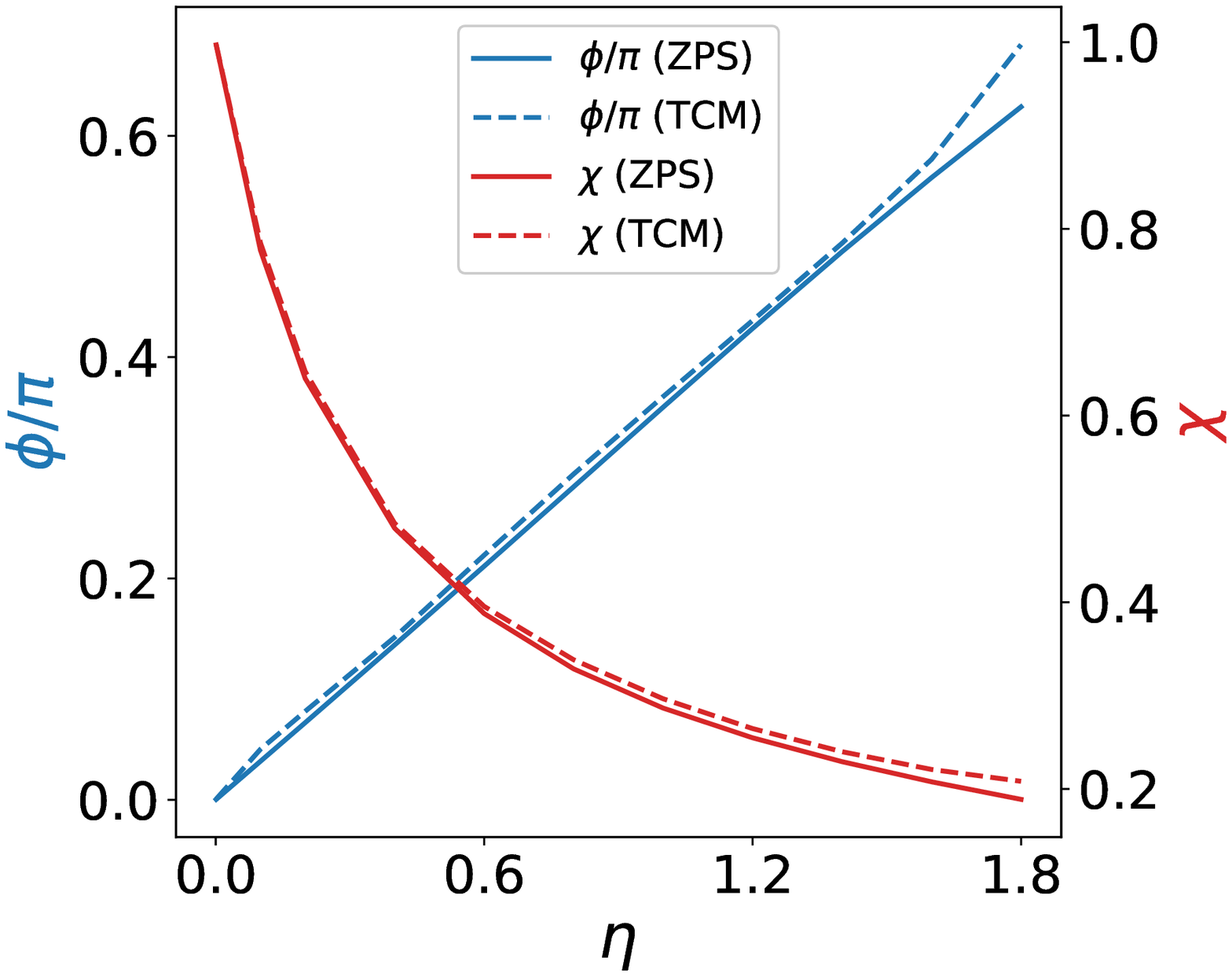}
\par\end{flushleft}%
\end{minipage}\caption{\label{fig:params_from_zero-point_vs_TCM}Extracted parameters for
a flat metal film. Solid lines denote the results obtained by parameter
extraction by ZPS\@. Dashed lines denote the results obtained by
the parameter extraction by using the TCM method. (a) and (b) Wavenumber
$k_{\text{r}}$ and decay rate $\gamma$ of the resonant mode as functions
of (a) the metal film thickness $h$ and (b) imaginary part factor
$\eta$ of metal permittivity. (c) and (d) Direct reflection phase
$\phi$ and ratio of decay rates $\chi$ as functions of (c) $h$
and (d) $\eta$.}
\end{figure}

\subsection{Flat metal film}

First, in a usual Kretschmann configuration without dimples ($d=0$),
we calculate the dependence of the reflection coefficient on the thickness
$h$ and the imaginary part of the permittivity of the metal film.
Figure \ref{fig:params_from_zero-point_vs_TCM} compares the parameters
extracted using ZPS and those extracted using the TCM method from
the reflection coefficient data.

Here, the imaginary part of the permittivity of metal is controlled
by the factor $\eta$ as $\varepsilon_{\text{m}}\left(\omega\right)=\text{Re}\left[\varepsilon_{\text{Al}}\left(\omega\right)\right]+\text{i}\eta\text{Im\ensuremath{\left[\varepsilon_{\text{Al}}\left(\omega\right)\right]}}$
with $\varepsilon_{\text{Al}}\left(\omega\right)$ being the original
permittivity of aluminum. In Fig.\ \ref{fig:params_from_zero-point_vs_TCM},
the resonant wavenumber $k_{\text{r}}$, the decay rate $\gamma$,
the change in the direct reflection phase $\phi$ due to absorption,
and the ratio of decay rates $\chi=\frac{\gamma_{\text{e}}}{r_{\text{d}}\gamma}$
are shown as functions of the metal film thickness $h$ and the imaginary
part factor $\eta$ of metal permittivity. The solid lines depict
the parameters extracted by ZPS, while the dashed lines depict the
parameters extracted by the TCM method\@. In the panels (a) and (c),
we take $\eta=1$ and change $h$ from 22.2 nm to 7.2 nm. In panels
(b) and (d), we take $h=22.2$ nm and change $\eta$ from 0.01 to
1.8. For comparison, the resonant wavenumber for the SPP on the semi-infinite
metal surface is also indicated as a dotted line. These results indicate
that appropriate values of parameters can be extracted by the TCM
method using only the shape of the angular spectrum in the cases where
the metal film thickness $h$ is not less than 10 nm or the imaginary-part
factor $\eta$ is not more than 1.5.

\begin{figure}
\begin{minipage}[t]{0.5\columnwidth}%
 
\begin{flushleft}
(a)\\
 \includegraphics[width=0.95\columnwidth]{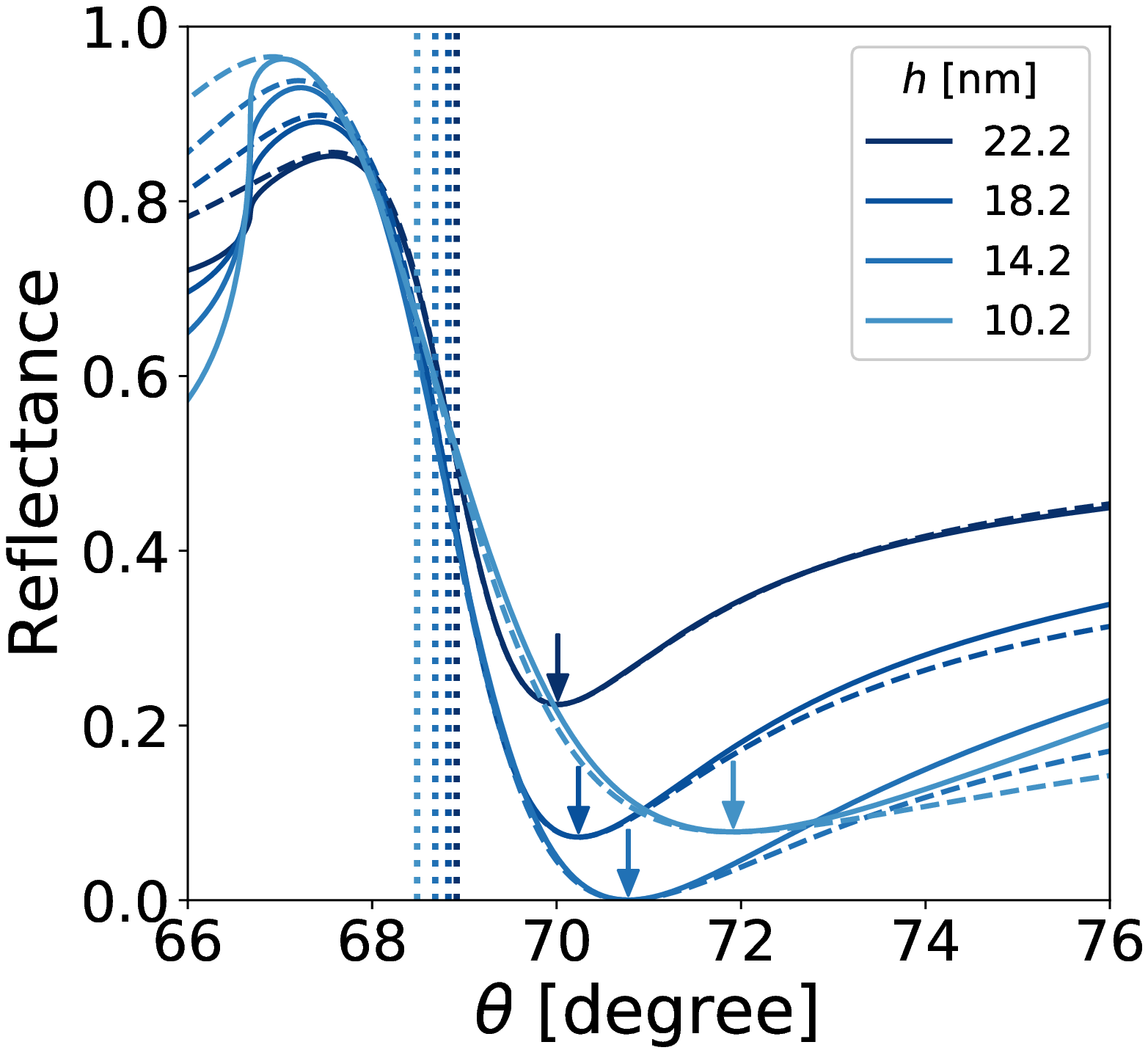} 
\par\end{flushleft}%
\end{minipage}%
\begin{minipage}[t]{0.5\columnwidth}%
 
\begin{flushleft}
(b)\\
 \includegraphics[width=0.95\columnwidth]{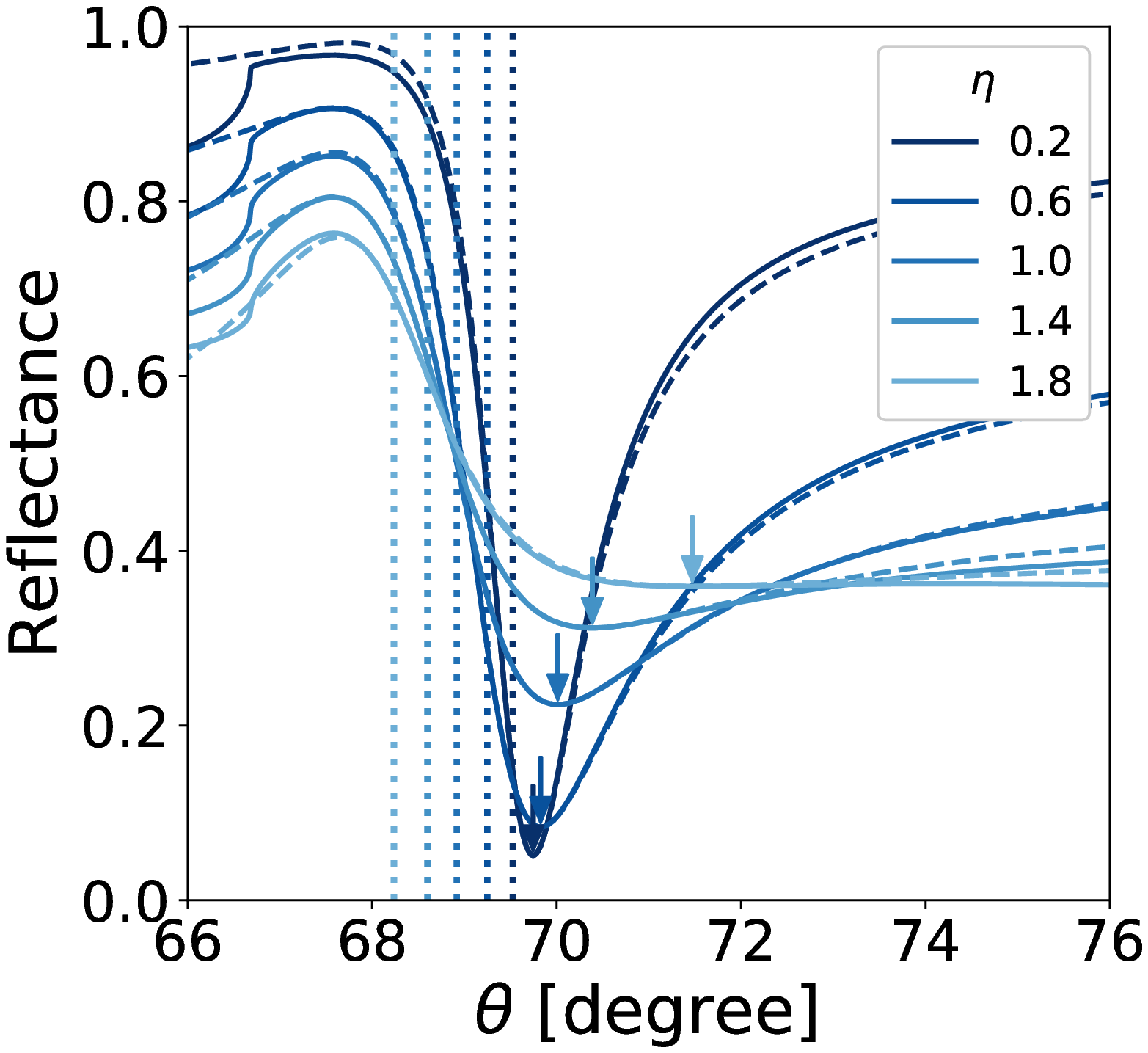} 
\par\end{flushleft}%
\end{minipage}\\
\begin{minipage}[t]{0.5\columnwidth}%
 
\begin{flushleft}
(c) \\
 \includegraphics[width=0.95\columnwidth]{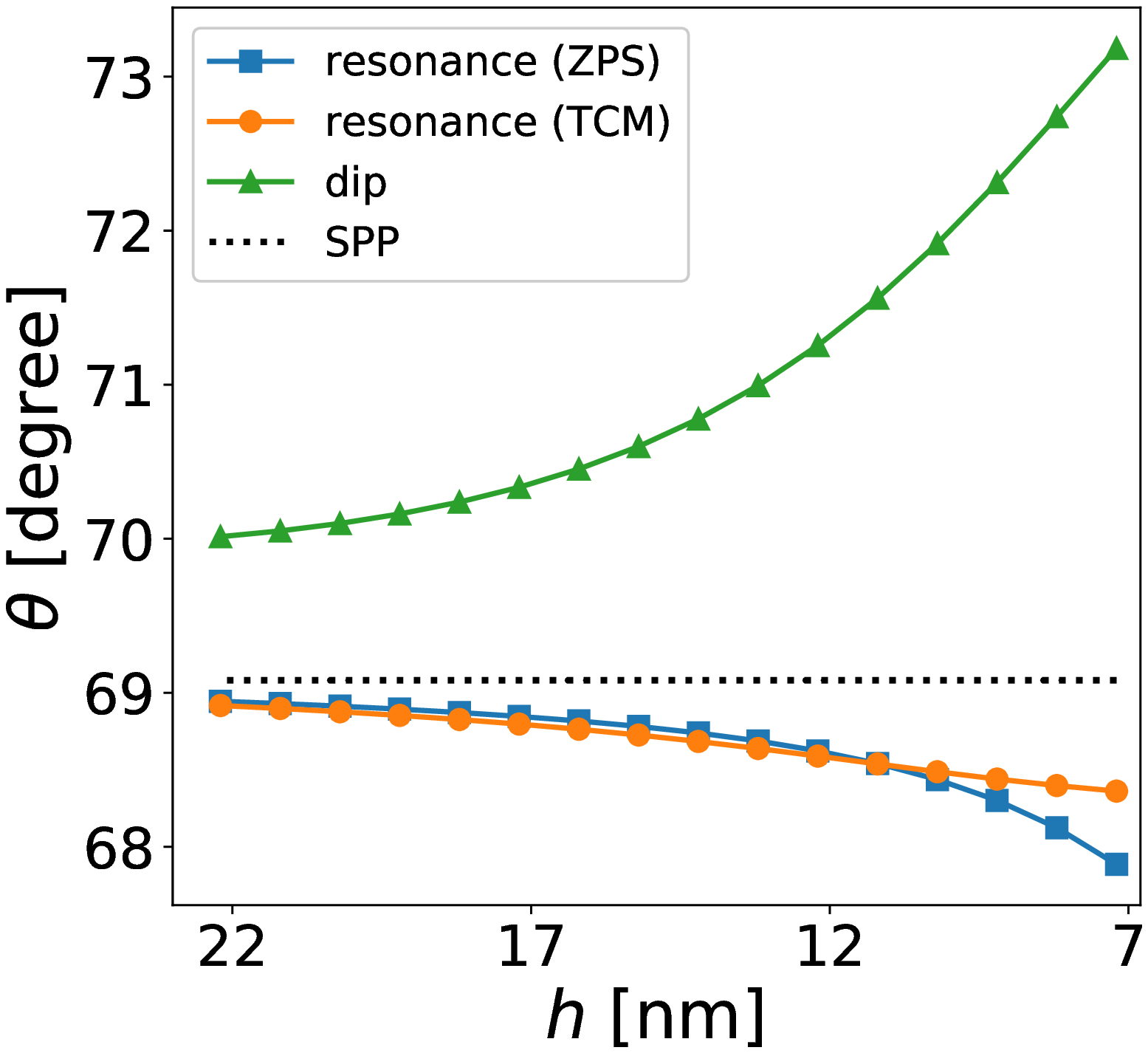} 
\par\end{flushleft}%
\end{minipage}%
\begin{minipage}[t]{0.5\columnwidth}%
 
\begin{flushleft}
(d)\\
 \includegraphics[width=0.95\columnwidth]{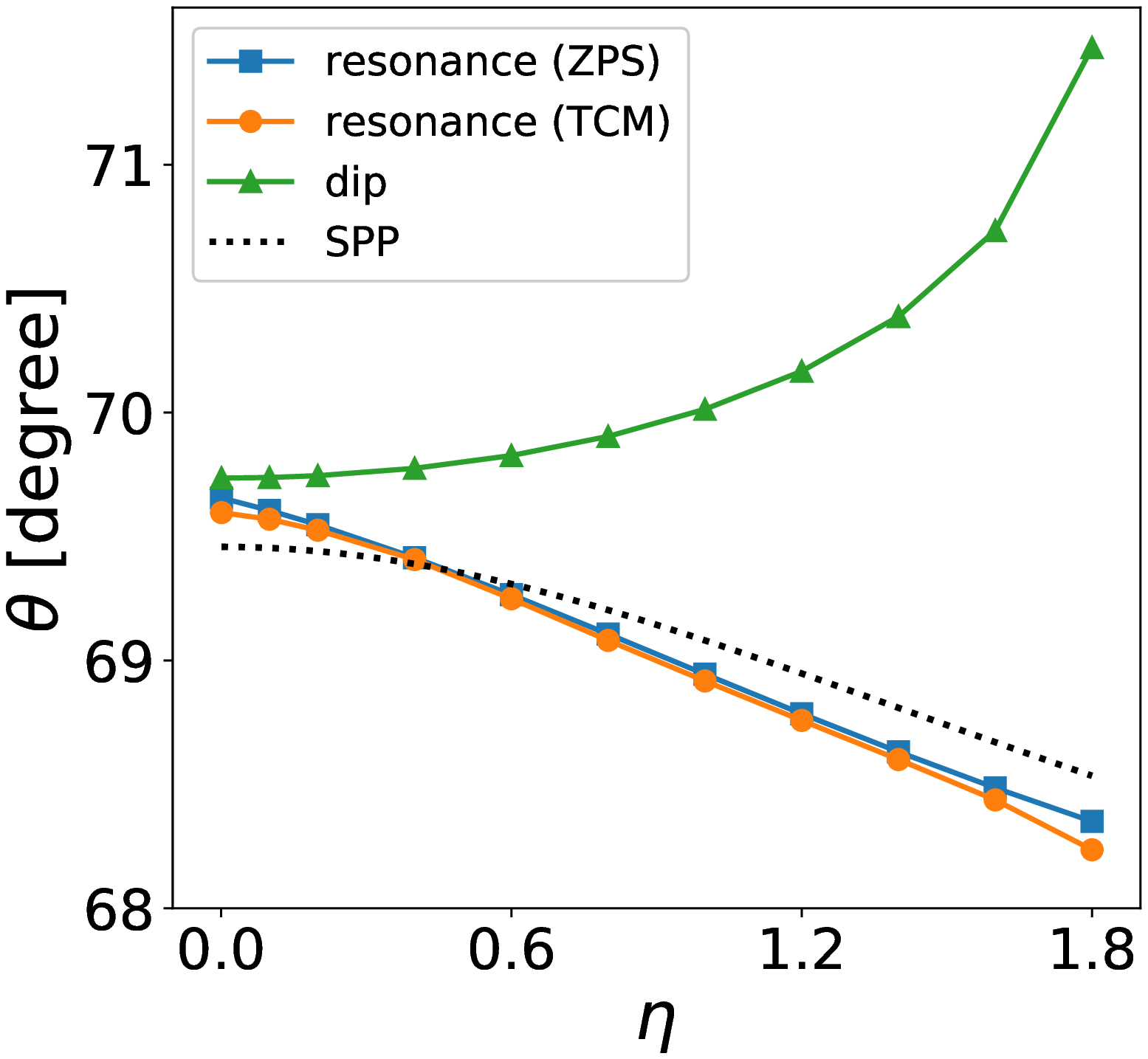} 
\par\end{flushleft}%
\end{minipage}\caption{\label{fig:Kretschmann_spectrum}Resonant and dip angles extracted
for usual a flat metal film. (a) and (b) Dependence of angular spectra
of reflection on (a) the metal film thickness $h$ and (b) imaginary
part factor $\eta$ of metal permittivity. Dotted lines depict the
resonant angles extracted by the TCM method. Arrows denote the dip
positions. (c) and (d) Resonant angle extracted by ZPS (blue squares)
and by the TCM method (orange circles) and dip angle (green triangles)
as functions of (c) $h$ and (d) $\eta$. Dotted line depicts the
resonant angle obtained using the dispersion relation of SPP.}
\end{figure}

\begin{figure*}
\begin{minipage}[t]{0.33\textwidth}%
\begin{flushleft}
(a)\\
 \includegraphics[width=0.95\columnwidth]{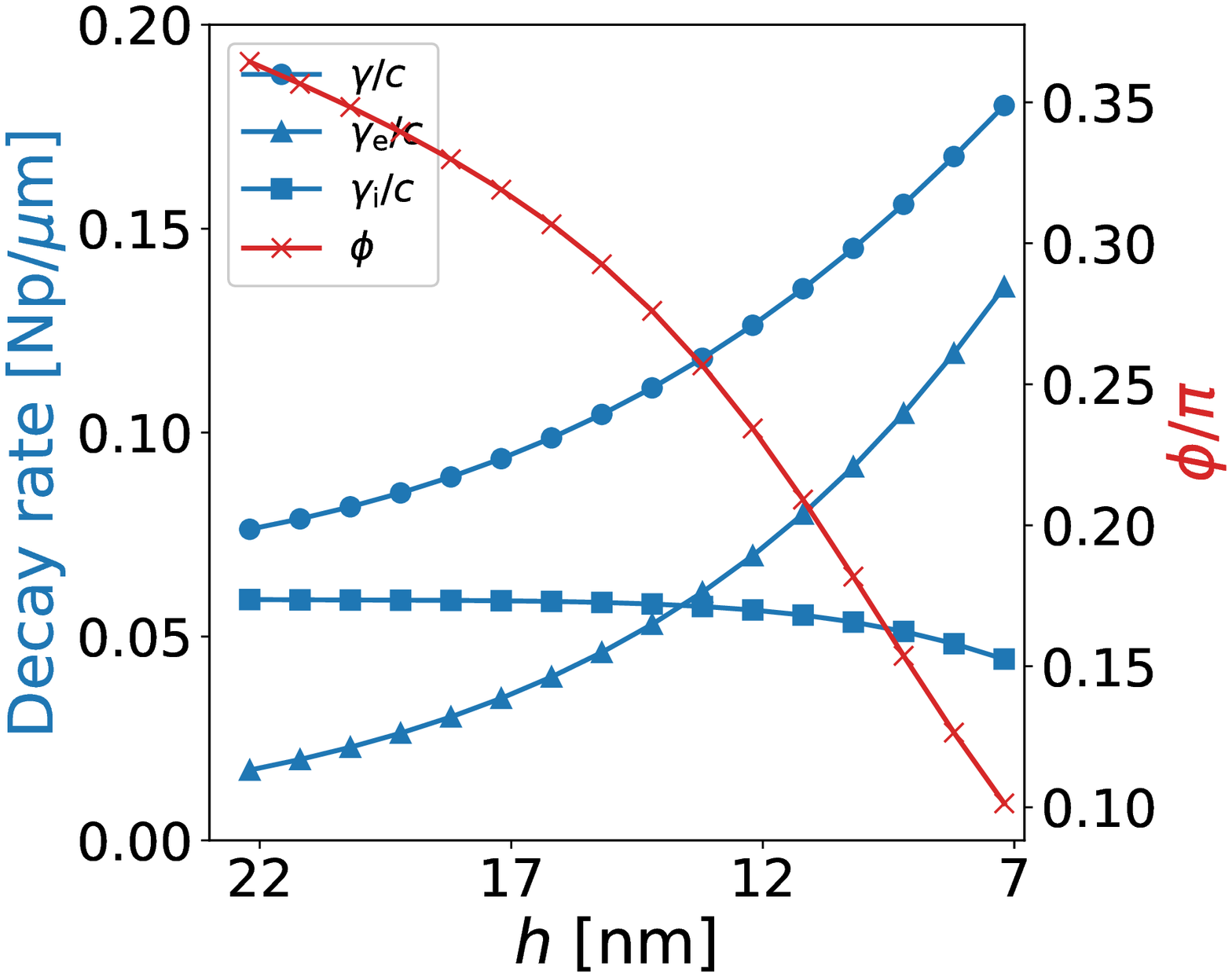}
\par\end{flushleft}%
\end{minipage}%
\begin{minipage}[t]{0.33\textwidth}%
\begin{flushleft}
(b)\\
 \includegraphics[width=0.95\columnwidth]{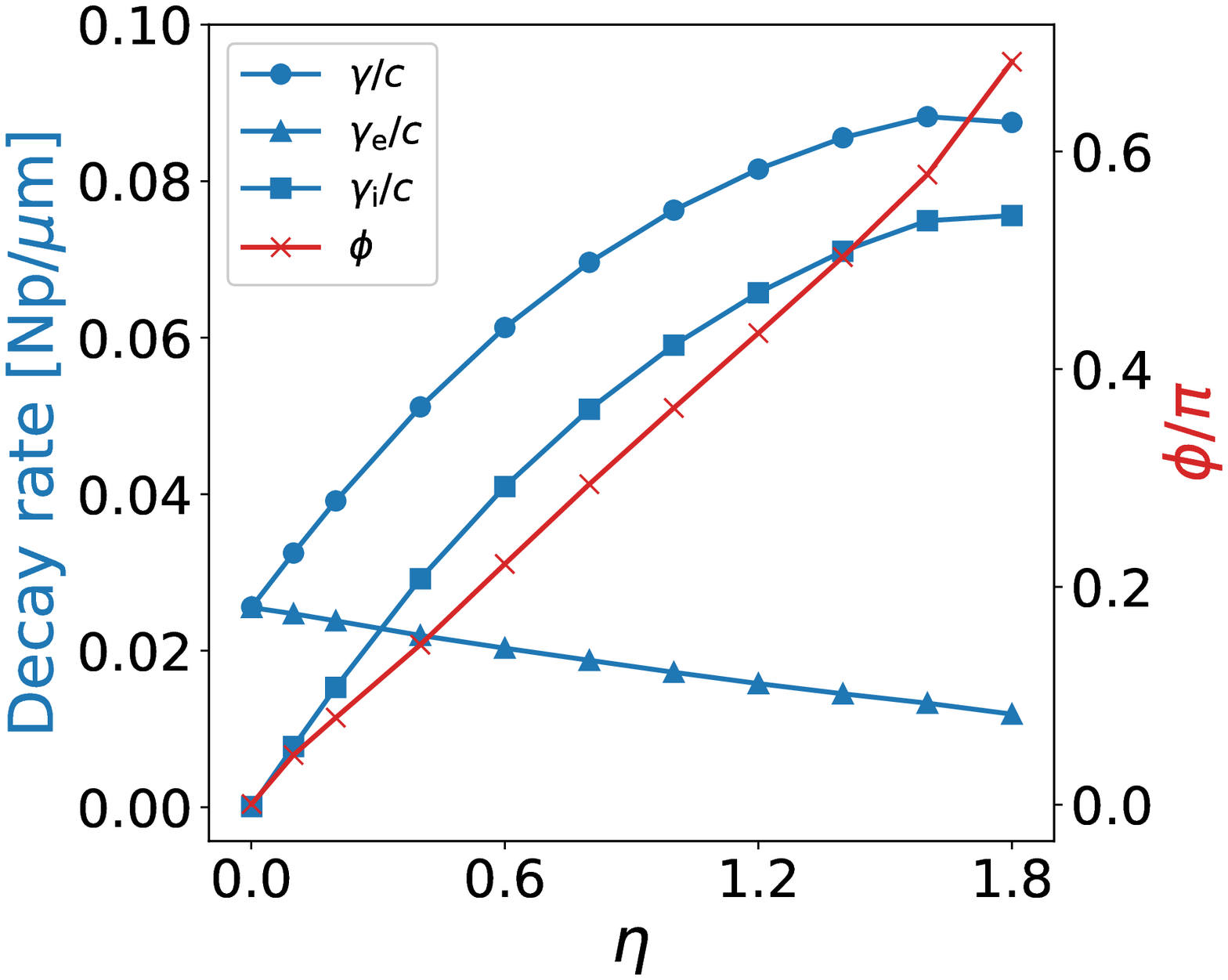}
\par\end{flushleft}%
\end{minipage}%
\begin{minipage}[t]{0.33\textwidth}%
\begin{flushleft}
(c)\\
 \includegraphics[width=0.95\columnwidth]{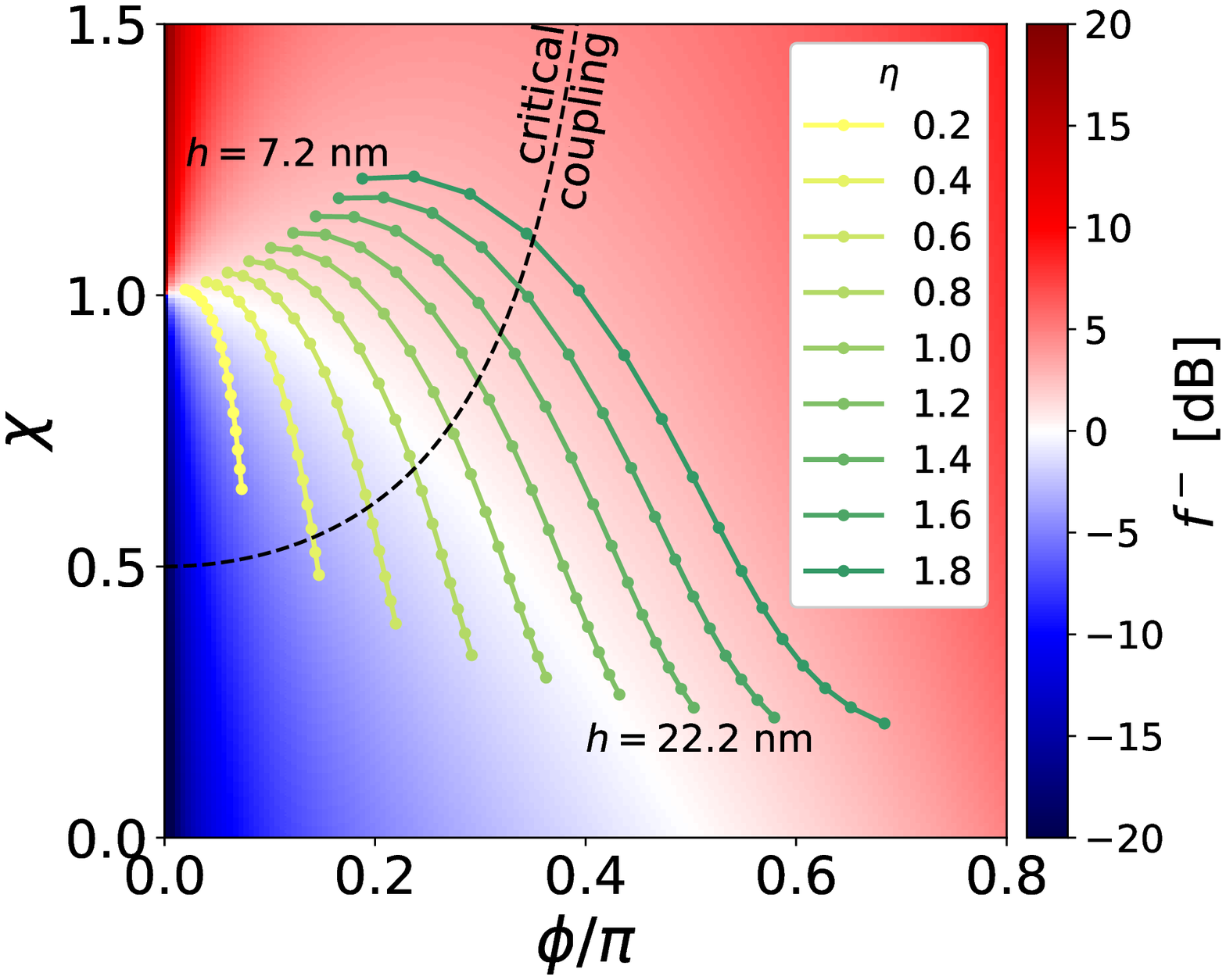}
\par\end{flushleft}%
\end{minipage}\caption{\label{fig:Kretschmann_parameters}Parameters extracted by the TCM
method from the reflection spectrum of a flat metal film. (a) and
(b) Total decay rate $\gamma$, external decay rate $\gamma_{\text{e}}$,
internal decay rate $\text{\ensuremath{\gamma_{\text{i}}}}$, and
direct reflection phase $\phi$ as functions of (a) metal film thickness
$h$ and (b) imaginary part factor $\eta$ of metal permittivity.
(c) Relation between direct reflection phase $\phi$ and the ratio
of decay rates $\chi$ depending on $\eta$ and $h$ plotted on a
color map of spectral shape factor $f^{-}$, where $h$ is varied
from 22.2 nm (lower right) to 7.2 nm (upper left) by 1 nm.}
\end{figure*}

Figure \ref{fig:Kretschmann_spectrum} shows the angular spectra of
reflection, the dip angle, and the resonant angle. Panel (a) shows
the change in the spectrum as the metal film thickness $h$ changes
from 22.2 nm to 10.2 nm. The solid lines depict the results obtained
by the SCM method, and the dashed lines depict the results obtained
from Eq.\ (\ref{eq:reflectance_for_kx}) using the parameters extracted
by the TCM method\@. We can see that the spectrum obtained by the
TCM method reproduces well the behavior around resonance. Panel (b)
shows the angular spectra of reflection calculated in the cases where
the imaginary part of the metal permittivity is multiplied by a factor
of $\eta=0.2\sim1.8$. This result also indicates the consistency
between the SCM results (solid lines) and TCM results (dashed lines).
In panels (a) and (b), the resonant angles extracted by the TCM method
are depicted by dotted lines, and the dip positions are depicted by
arrows. The changes in these values are summarized in panels (c) and
(d), which indicate that both the changes of $h$ and $\eta$ cause
the shift of the resonant angle in the opposite direction to that
of the dip angle and the shift of the resonant angle is smaller than
that of the dip angle.

\begin{figure*}
\begin{minipage}[t]{0.33\textwidth}%
\begin{flushleft}
(a)\\
 \includegraphics[width=0.9\columnwidth]{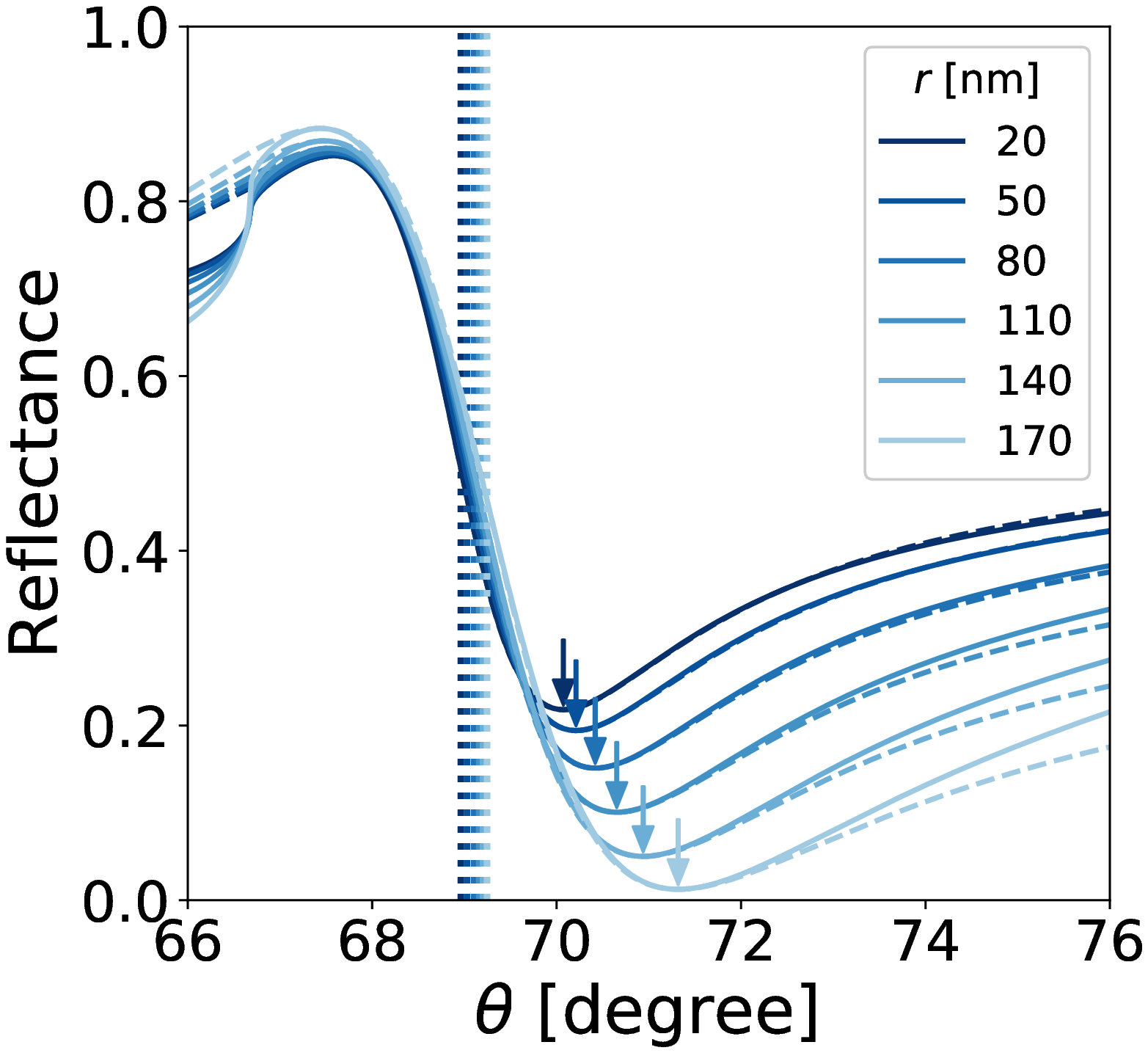}
\par\end{flushleft}%
\end{minipage}%
\begin{minipage}[t]{0.33\textwidth}%
\begin{flushleft}
(b)\\
 \includegraphics[width=0.9\columnwidth]{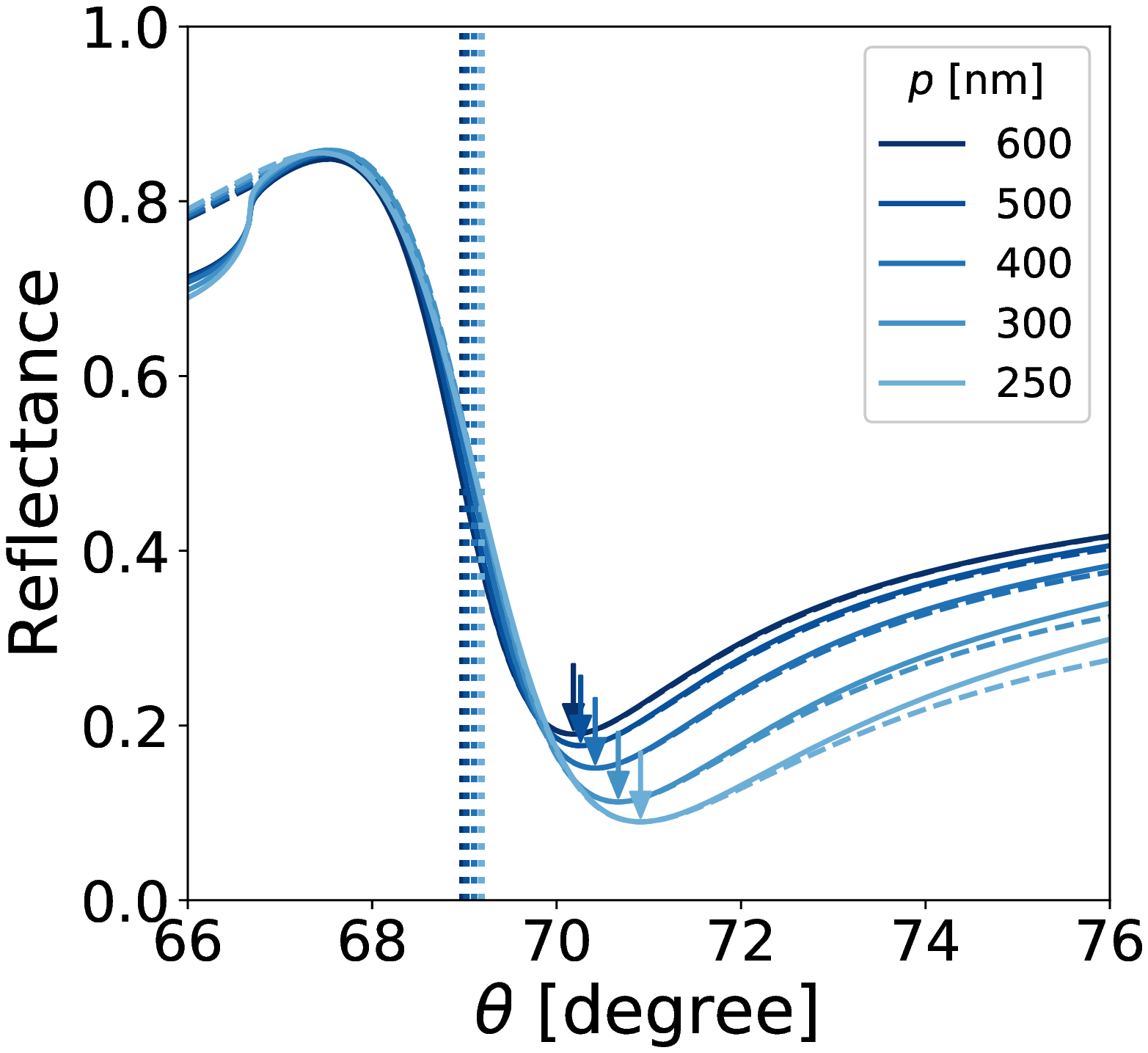}
\par\end{flushleft}%
\end{minipage}%
\begin{minipage}[t]{0.33\textwidth}%
\begin{flushleft}
(c)\\
 \includegraphics[width=0.9\columnwidth]{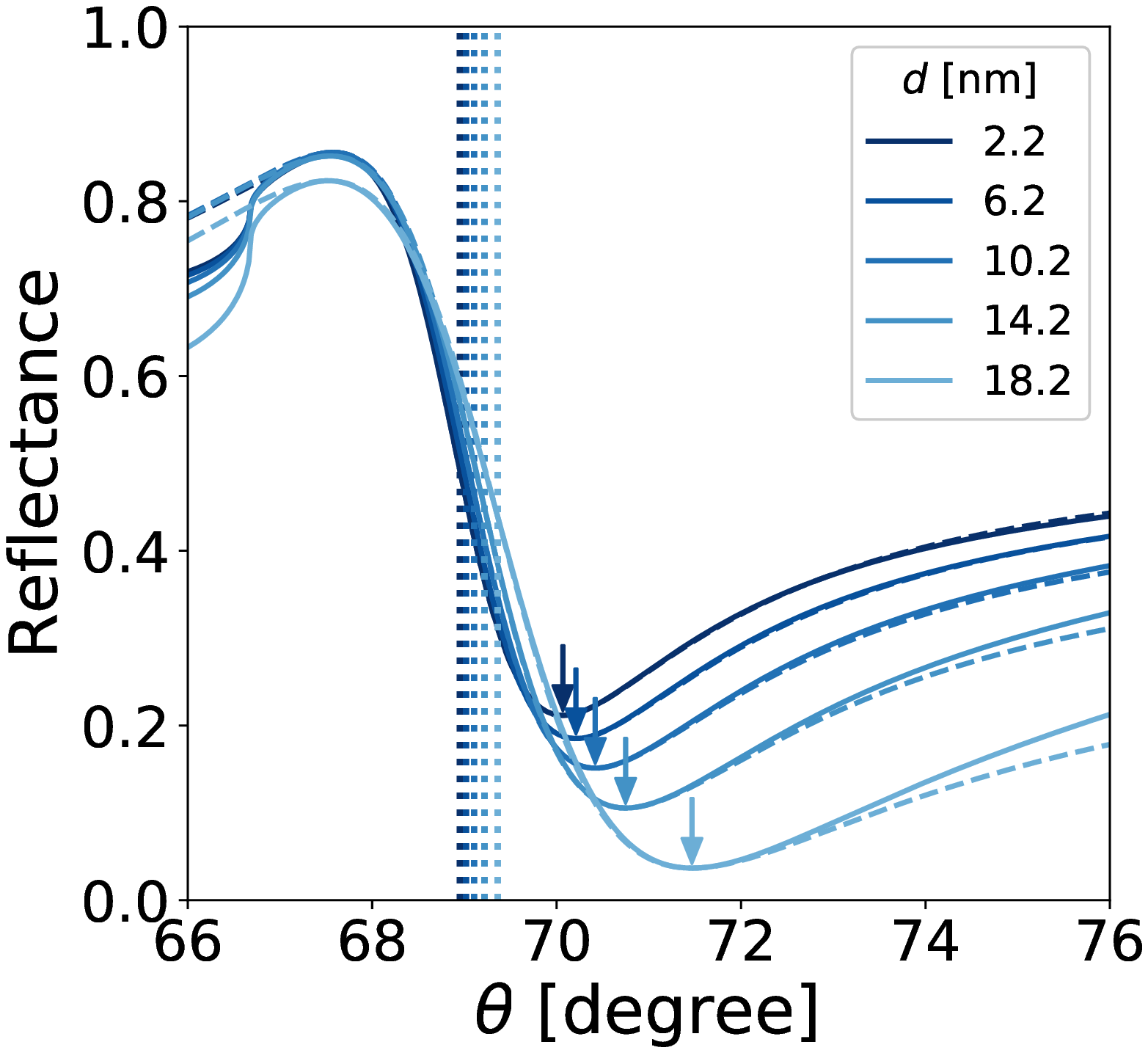}
\par\end{flushleft}%
\end{minipage}\\
\begin{minipage}[t]{0.33\textwidth}%
\begin{flushleft}
(d)\\
 \includegraphics[width=0.9\columnwidth]{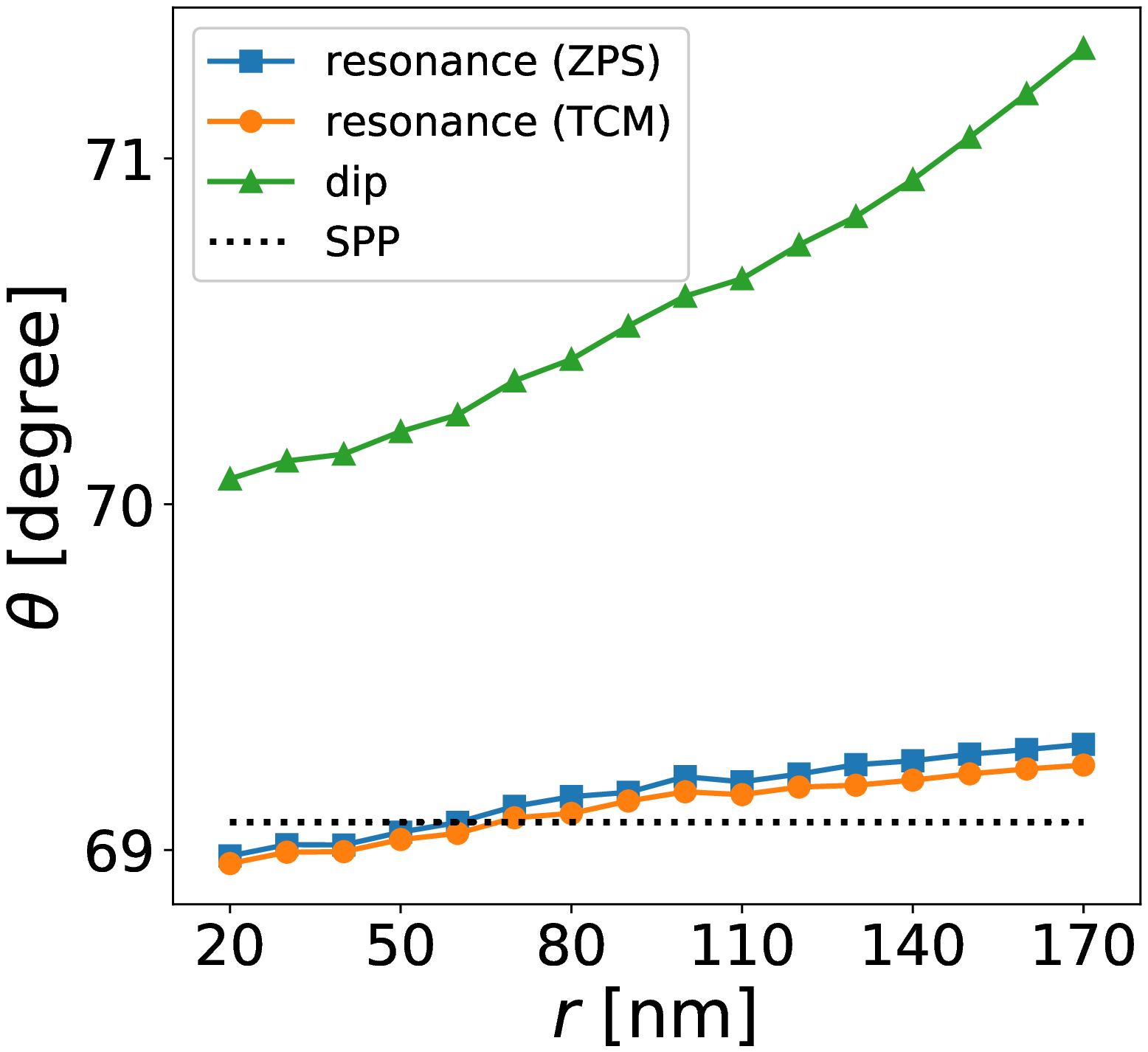}
\par\end{flushleft}%
\end{minipage}%
\begin{minipage}[t]{0.33\textwidth}%
\begin{flushleft}
(e)\\
 \includegraphics[width=0.9\columnwidth]{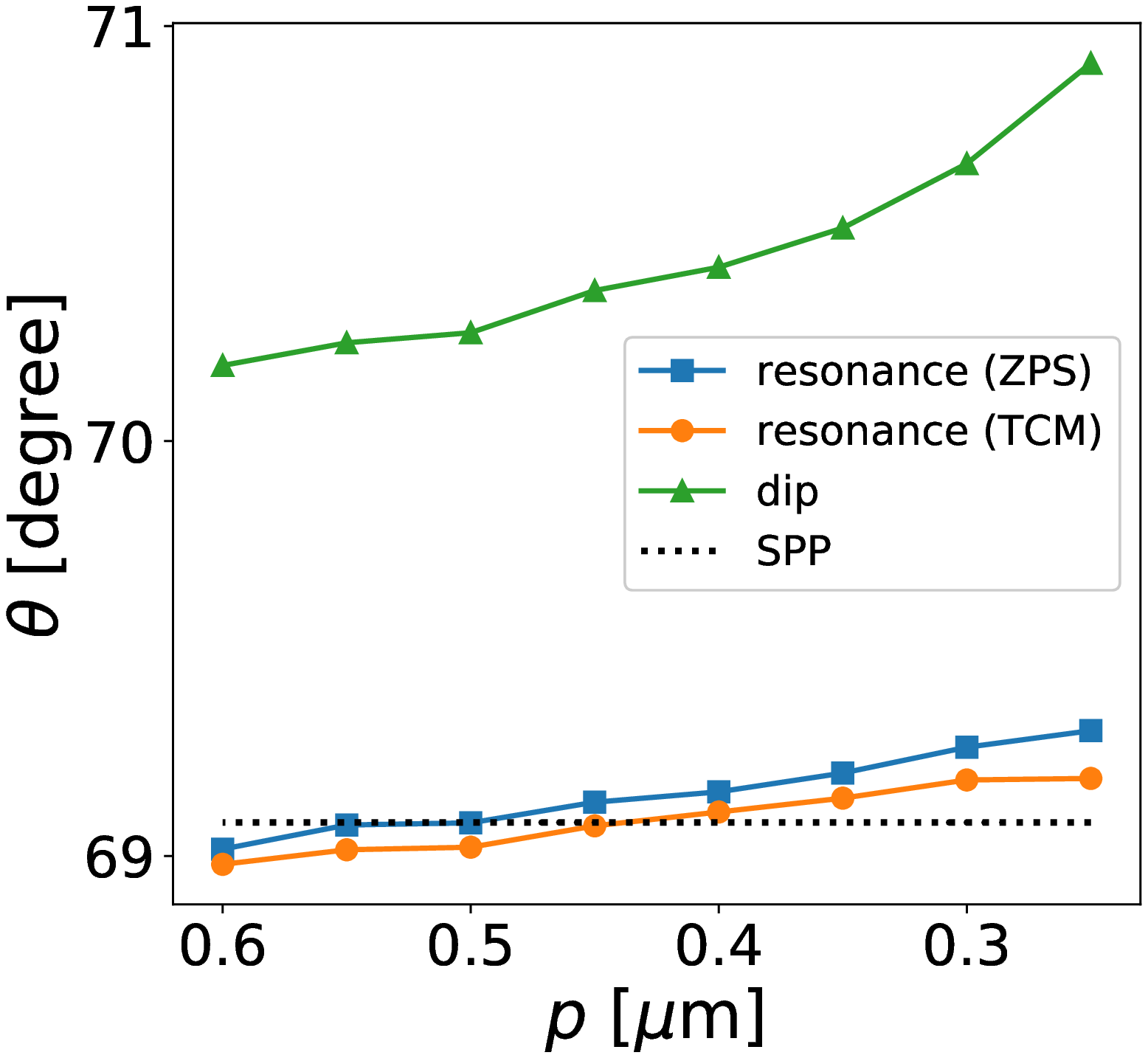}
\par\end{flushleft}%
\end{minipage}%
\begin{minipage}[t]{0.33\textwidth}%
\begin{flushleft}
(f)\\
 \includegraphics[width=0.9\columnwidth]{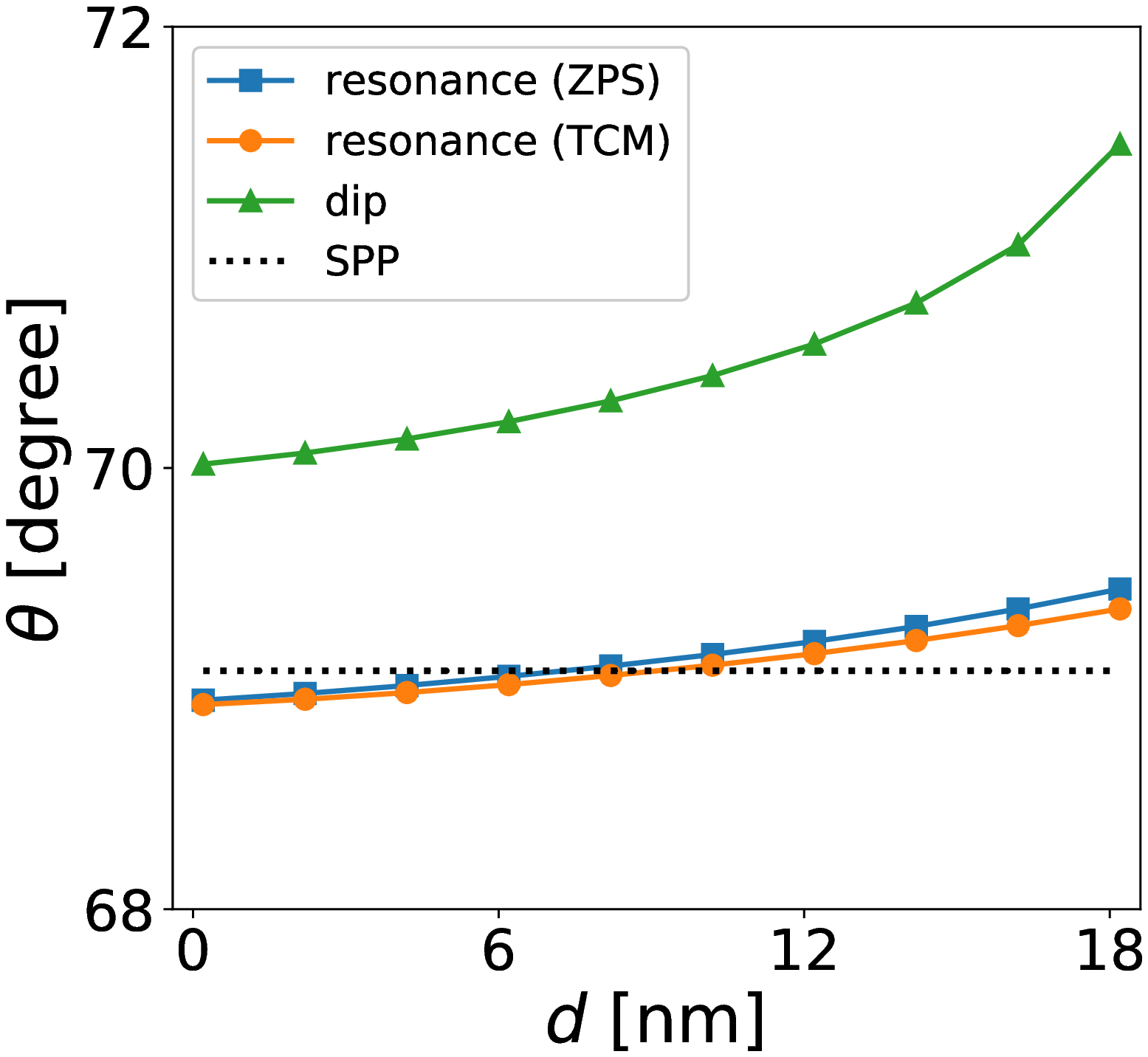}
\par\end{flushleft}%
\end{minipage}\caption{\label{fig:square_array_spectrum}Resonant and dip angles extracted
for the nanodimple periodic array. (a), (b), and (c) Dependence of
the reflection spectrum on (a) dimple radius $r$, (b) period of dimple
array $p$, and (c) depth of dimple $d$. Dotted lines depict the
resonant angles extracted by the TCM method. Arrows denote the dip
positions. (d), (e), and (f) Resonant and dip angles as functions
of (d) $r$, (e) $p$, and (f) $d$. Dotted line depicts the resonant
angle obtained using the dispersion relation of SPP\@. Here, we assume
that the thickness of the aluminum film outside the dimple is $h_{0}=$22.2
nm, and the fixed values of $r$, $p$, and $d$ are 80, 400, and
10.2 nm, respectively.}
\end{figure*}

\begin{figure*}
\begin{minipage}[t]{0.33\textwidth}%
\begin{flushleft}
(a)\\
 \includegraphics[width=0.95\columnwidth]{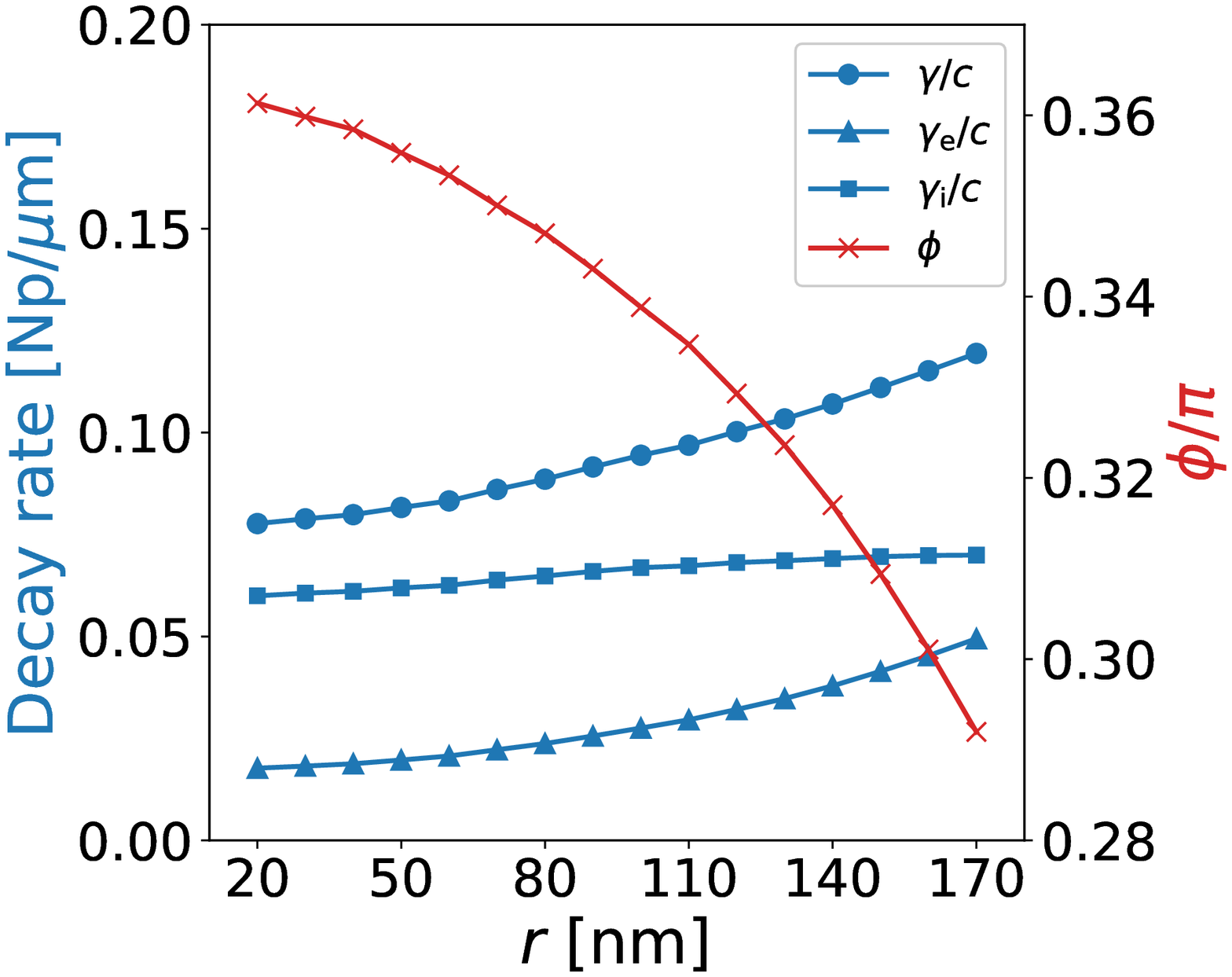}
\par\end{flushleft}%
\end{minipage}%
\begin{minipage}[t]{0.33\textwidth}%
\begin{flushleft}
(b)\\
 \includegraphics[width=0.95\columnwidth]{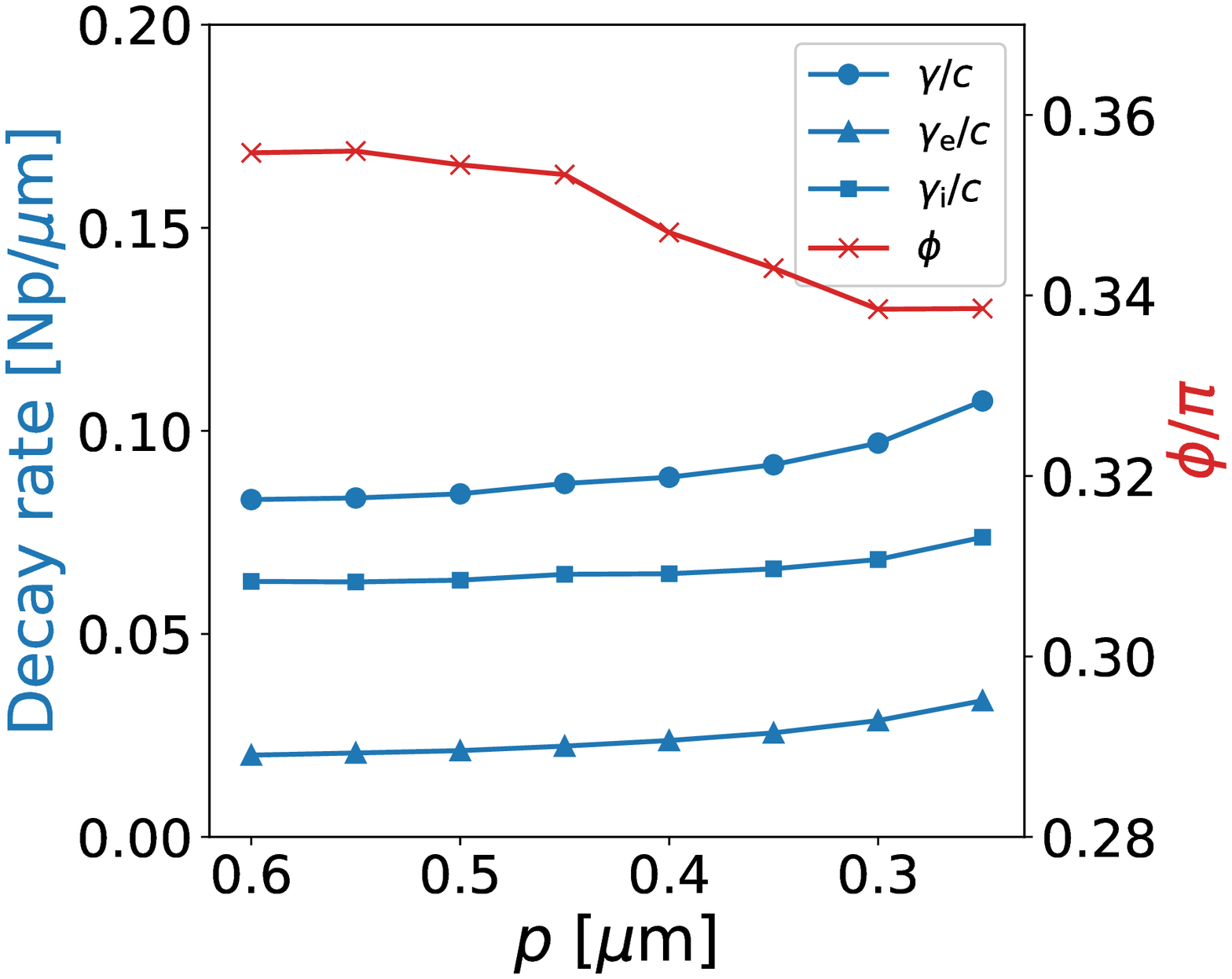}
\par\end{flushleft}%
\end{minipage}%
\begin{minipage}[t]{0.33\textwidth}%
\begin{flushleft}
(c)\\
 \includegraphics[width=0.95\columnwidth]{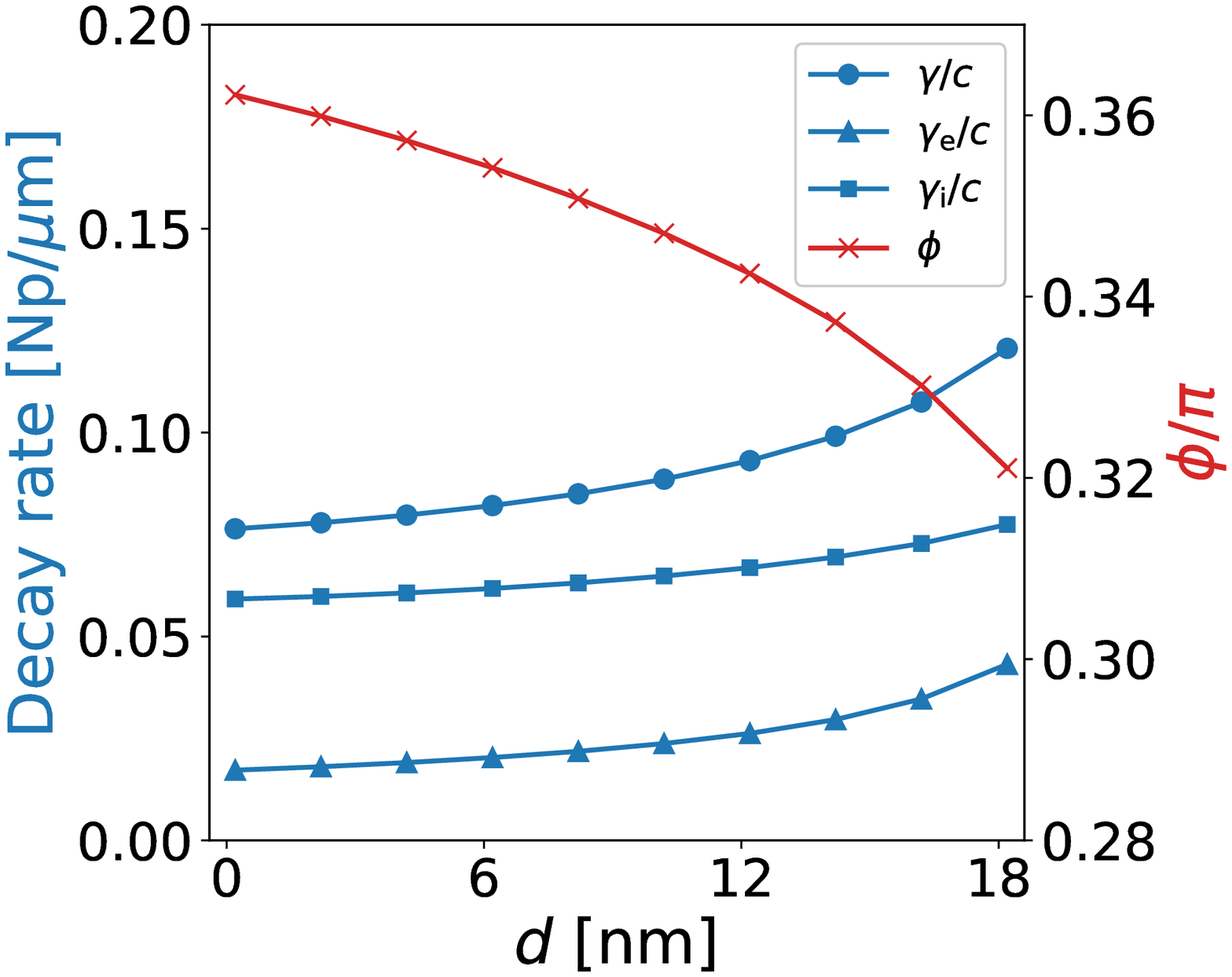}
\par\end{flushleft}%
\end{minipage}\caption{\label{fig:square_array_parameters}Parameters extracted by the TCM
method from the reflection spectrum of the nanodimple periodic array.
Total decay rate $\gamma$, external decay rate $\gamma_{\text{e}}$,
internal decay rate $\text{\ensuremath{\gamma_{\text{i}}}}$, and
direct reflection phase $\phi$ as functions of (a) dimple radius
$r$, (b) period of dimple array $p$, and (c) depth of dimple $d$.
The system configuration condition is the same as that in Fig.\ \ref{fig:square_array_spectrum}.}
\end{figure*}

Extracting parameters from the angular reflection spectrum by the
TCM method, we can obtain the total decay rate $\gamma$ of the resonant
mode, the external decay rate $\gamma_{\text{e}}$ by radiation, the
internal decay rate $\gamma_{\text{i}}=\gamma-\gamma_{\text{e}}$
caused by the metal loss, and the phase change $\phi$ caused by absorption
in the direct reflection. The changes in these parameters are shown
in Fig.\ \ref{fig:Kretschmann_parameters}(a) and (b). Panel (a)
shows that the external decay rate $\gamma_{\text{e}}$ and the total
decay rate $\gamma$ increase as the metal film thickness $h$ decreases,
while the internal decay rate $\gamma_{\text{i}}$ decrease weakly.
On the other hand, panel (b) shows that the internal decay rate $\gamma_{\text{i}}$
and the total decay rate $\gamma$ increase as the metal loss factor
$\eta$ increases, while the external decay rate $\gamma_{\text{e}}$
decreases weakly. In both panels, the direct reflection phase $\phi$
changes in the same direction as that of the change in the internal
decay rate $\gamma_{\text{i}}$.

The coincidence between the changes in $\phi$ and $\gamma_{\text{i}}$
is reasonable because both result from the absorption by the metal
loss, whose effect decreases as the metal film thickness decreases.
On the other hand, because the radiation from the resonance caused
by SPP at the NaCl solution side becomes stronger as the metal film
thickness decreases, the external decay rate $\gamma_{\text{e}}$
increases accordingly.

Figure \ref{fig:Kretschmann_parameters}(c) shows the changes in $\phi$
and $\chi$ due to those in $h$ and $\eta$ on the color map of SSF
$f^{-}$. The lines with colors ranging from yellow to green denote
the results for $\eta$ from 0.2 to 1.8, as shown in the figure legend.
The circles on the lines denote the parameter values for different
values of $h$ from 22.2 nm (lower right) to 7.2 nm (upper left).
This figure indicates that the thickness of the metal film and the
internal loss are reflected in the values of parameters $\chi$ and
$\phi$. In other words, $h$ and $\eta$ of a flat metal film can
be deduced by determining the parameters $\chi$ and $\phi$ from
the measured spectral shape of normalized reflectance.

\subsection{Nano-dimple periodic array}

The inhomogeneity on the metal surface is expected to change the internal
decay rate of SPP or the amplitude and phase of direct reflection.
To estimate this change, we study the change in the reflection spectrum
and the parameters due to the formation of a periodic array of cylindrical
dimples on the surface of the aluminum film. In what follows, we assume
that the thickness of the aluminum film outside the dimple is $h_{0}=$22.2
nm ($h+d=h_{0}$).

Figure \ref{fig:square_array_spectrum} shows the dependences of the
reflection spectrum, dip angle, and resonant angle on the radius of
the dimple cylinder, $r$; period of the dimple array, $p$; and depth
of the dimple, $d$. Panels (a) and (d) show the results in the case
where the value of $r$ is changed from 20 nm to 170 nm while fixing
$d=10.2$ nm and $p=400$ nm. It is observed that as the dimple radius
$r$ increases, the dip angle (arrows in panel (a)) and resonant angle
(dotted lines in panel (a)) shift to the wider-angle side. Note that
the change in the dip angle is much larger than that in the resonant
angle. Panels (b) and (e) show the results in the case where the value
of $p$ is changed from 600 nm to 250 nm while fixing $d=10.2$ nm
and $r=$80 nm. It is observed that as the period $p$ decreases,
the dip angle and resonant angle shift to the wider-angle side. Moreover,
the change in the dip angle is larger than that in the resonant angle.
Panels (c) and (f) show the results in the case where the value of
$d$ is changed from 2.2 nm to 18.2 nm while fixing $r=80$ nm and
$p=400$ nm. It is observed that as the depth $d$ increases, the
dip angle and resonant angle shift to the wider-angle side. Moreover,
the change in the dip angle is larger than that in the resonant angle.

\begin{figure*}
\noindent %
\noindent\begin{minipage}[t]{0.31\textwidth}%
\begin{flushleft}
(a)\\
 \includegraphics[width=0.9\columnwidth]{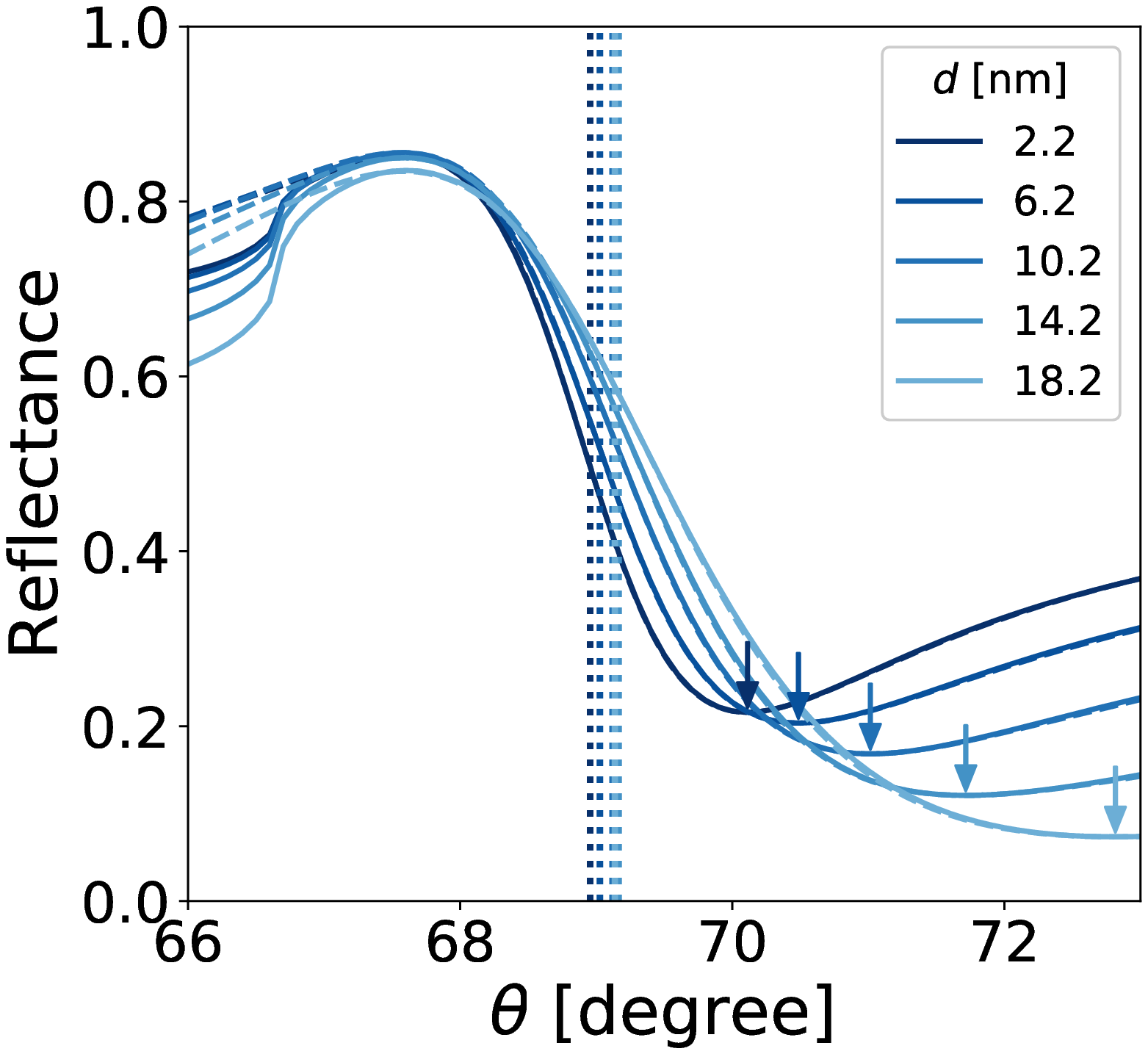}
\par\end{flushleft}%
\end{minipage}%
\noindent\begin{minipage}[t]{0.31\textwidth}%
\begin{flushleft}
(b)\\
 \includegraphics[width=0.9\columnwidth]{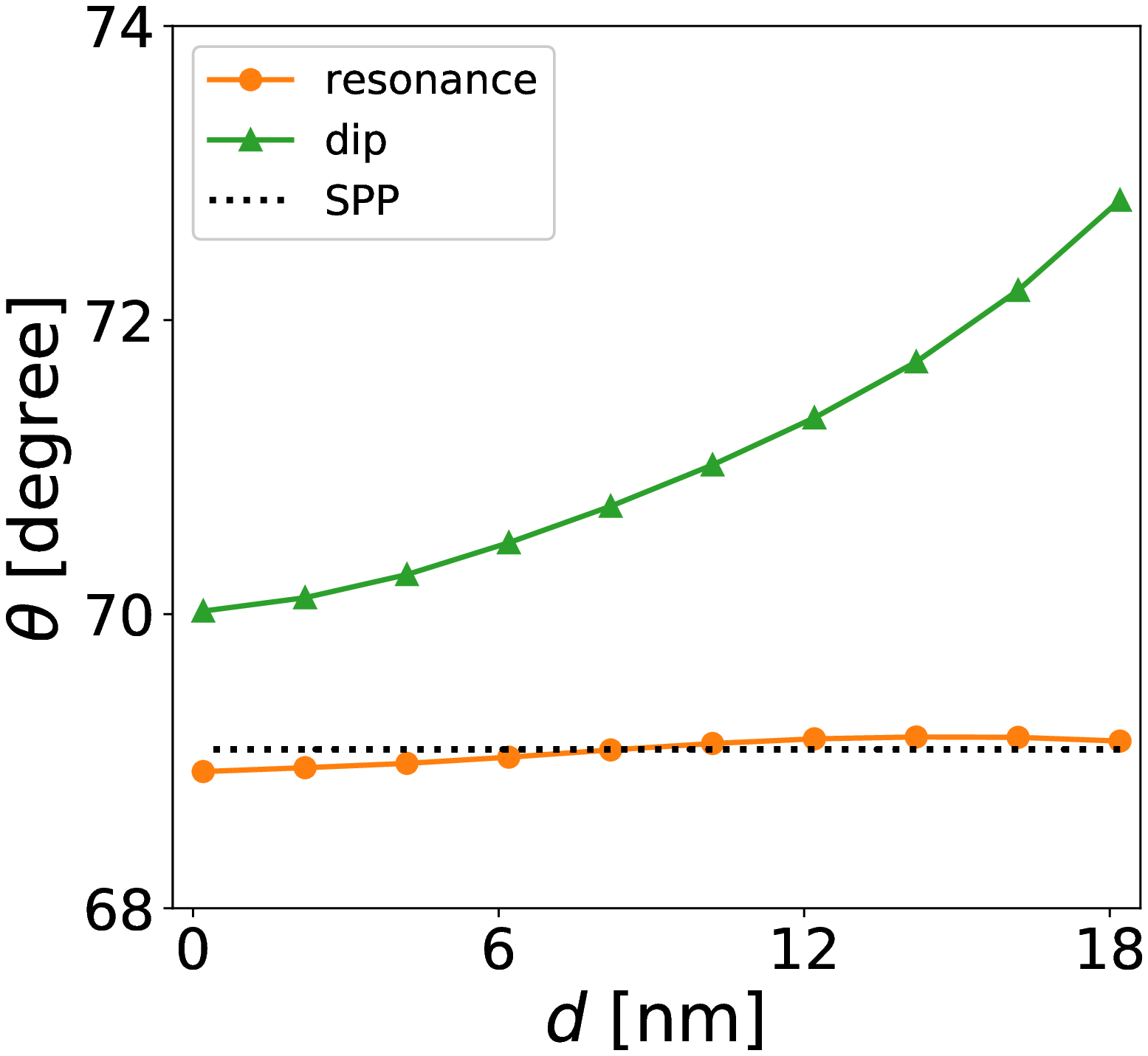}
\par\end{flushleft}%
\end{minipage}%
\begin{minipage}[t]{0.37\textwidth}%
\begin{flushleft}
(c)\\
 \includegraphics[width=0.9\columnwidth]{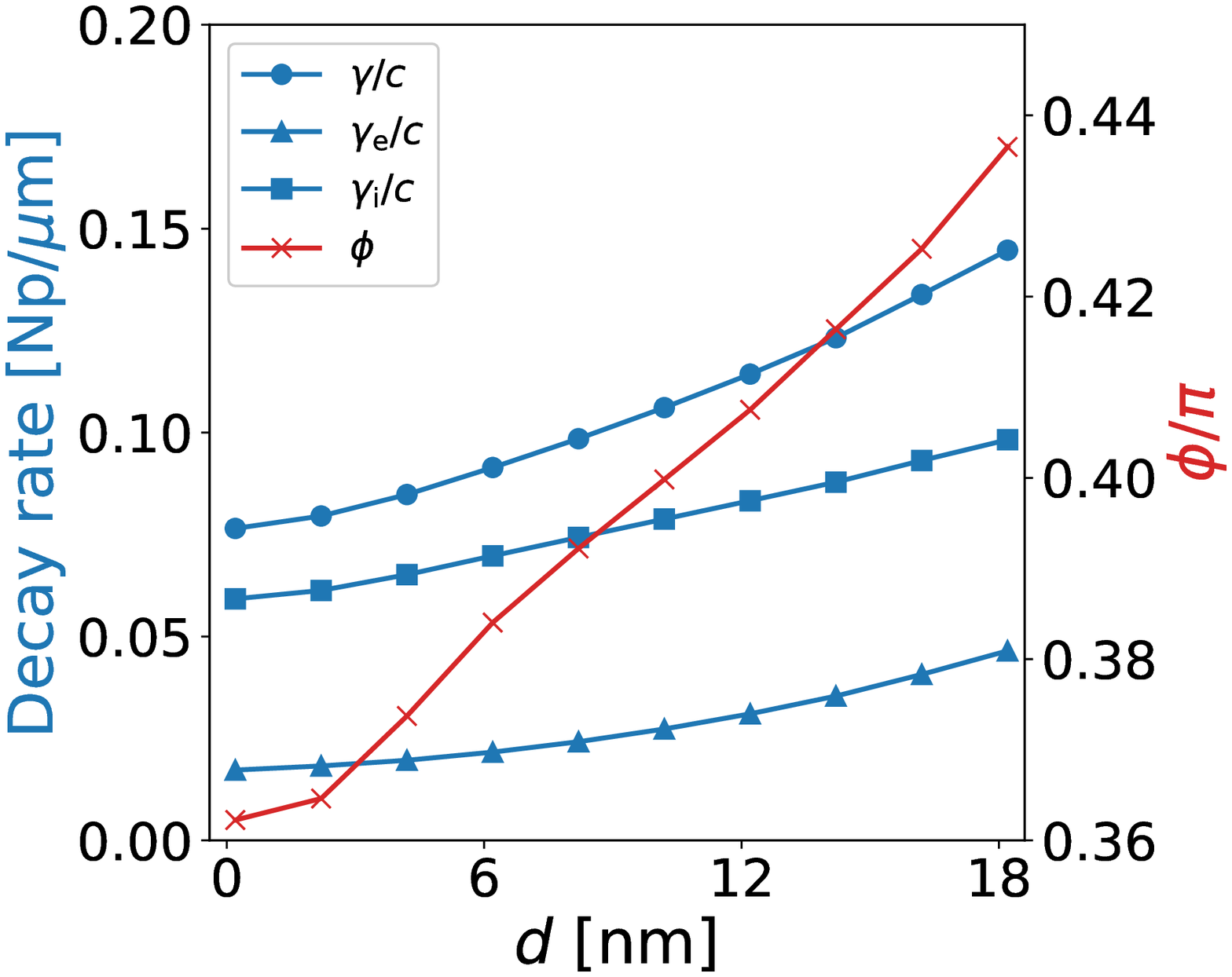}
\par\end{flushleft}%
\end{minipage}\caption{\label{fig:random_array_spectrum}Parameters extracted by the TCM
method from the reflection spectrum of the nanodimple random array.
(a) Dependence of the reflection spectrum on the depth of dimple $d$.
Dotted lines depict the resonant angles. Arrows denote the dip positions.
(b) Resonant angle and dip angle as functions of $d$. Dotted line
depicts the resonant angle obtained using the dispersion relation
of SPP\@. (c) Total decay rate $\gamma$, external decay rate $\gamma_{\text{e}}$,
internal decay rate $\text{\ensuremath{\gamma_{\text{i}}}}$, and
direct reflection phase $\phi$ as functions of $d$. Here, we assume
that the thickness of the aluminum film outside the dimple is $h_{0}=$22.2
nm, and the radius of the dimple is set to 80 nm.}
\end{figure*}

Figure \ref{fig:square_array_parameters} shows the decay rates, $\gamma$,
$\gamma_{\text{e}}$ and $\gamma_{\text{i}}$, and the direct reflection
phase $\phi$ extracted by the TCM method\@. Panels (a), (b), and
(c) show the results for the same condition as that in panels (a),
(b), and (c) of Fig.\ \ref{fig:square_array_spectrum}, respectively.
These results indicate that an increase in $r$, decrease in $p$,
or increase in $d$, namely increase in the ratio of the region of
dimple in the metal film region, results in an increase in the decay
rate and decrease in $\phi$. However, the increase in the internal
decay rate $\gamma_{\text{i}}$ is rather small, and that in the decay
rate mainly originates from the increase in the external decay rate
$\gamma_{\text{e}}$. This tendency is similar to that observed when
the thickness of a flat metal film is decreased. It can be interpreted
that the formation of dimples reduces the effective thickness of the
metal film and enhances the radiation from SPP at the NaCl solution
side to the substrate side, to increase the external decay rate. The
change in $\phi$ can be explained by the decrease in the effective
thickness of the metal film, as in a flat metal film.

The trend of $\gamma_{\text{i}}$ is opposite to that in the case
of change in the thickness of a flat metal film. This behavior indicates
that the diffraction of SPP by the dimple array enhances the internal
decay. However, the change in $\phi$ is mainly determined by the
material loss (volume of metal region) because the diffraction produced
by the periodic array with the period comparable to the wavelength
is restricted to a few diffraction orders and the influence on the
direct reflection process is weak.

\subsection{Random array}

For the periodic array discussed in the previous section, the effect
of diffusive scattering is weak because the radiative diffraction
order is restricted. In this section, we discuss a system with randomly
distributed dimples in order to include the effect of diffusive scattering.
Figure \ref{fig:random_array_spectrum} shows the dependence of (a)
the angular spectrum, (b) dip and resonant angles, and (c) decay rates
and the direct reflection phase on the depth of dimple $d$ for a
sample whose extracted values are near the averaged values of randomly
produced 100 samples. Here, $r$ is set to 80 nm, and $d$ is changed
from 0.2 nm to 18.2 nm for the same distribution of dimples in the
$xy$ plane. In addition, in this section, the thickness of the aluminum
film outside the dimple is assumed to be $h_{0}=22.2$ nm.

Comparing the results for the periodic array in Fig.\ \ref{fig:square_array_spectrum}
(c) and (f) and those for the random array in Fig.\ \ref{fig:random_array_spectrum}
(a) and (b), it is found that the extension of the width of the dip
and the shift of the dip angle due to an increase in $d$ are larger
in the random array. In addition, Fig.\ \ref{fig:square_array_parameters}
(c) and \ref{fig:random_array_spectrum} (c) show that $\phi$ increases
with $d$ in a random array, contrary to the case of periodic array.
These results are similar to those obtained when $\eta$ is increased
in a flat metal film. Thus, the introduction of a random dimple array
produces a similar effect as that of an increase in metal loss. The
diffusive scattering produced by the random array suppresses the direct
specular reflection, like absorption.

\subsection{Effect of dimple array on parameters}

From the above consideration, it is deduced that one of the effects
of dimple formation is the reduction in the effective thickness of
the aluminum film. This results in an increase in external decay,
namely the enhancement of radiation from the resonant mode. The radiation
from the resonant mode is limited by the tunneling process within
the metal region, which depends exponentially on the product of the
decay rate of the evanescent field in the $z$ direction, $\gamma_{z}=\sqrt{k_{\text{SPP}}^{2}-\text{Re}\left[\varepsilon_{\text{Al}}\right]\omega_{\text{i}}^{2}/c^{2}}$,
and the thickness of the metal region. Here, $k_{\text{SPP}}$ is
the wavenumber of SPP at an angular frequency of $\omega_{\text{i}}$.
Therefore, we define the effective thickness of the metal film as
follows: 
\begin{align}
\exp\left(-2\gamma_{z}h_{\text{eff}}\right) & =\frac{\pi r^{2}}{p^{2}}\exp\left(-2\gamma_{z}h\right)\nonumber \\
 & +\left(1-\frac{\pi r^{2}}{p^{2}}\right)\exp\left(-2\gamma_{z}h_{0}\right),\label{eq:effective_thickness}
\end{align}
where the period of array $p$ is taken to be 400 nm in the random
array case, considering the average density of the dimples.

\begin{figure}
\begin{centering}
\includegraphics[width=0.8\columnwidth]{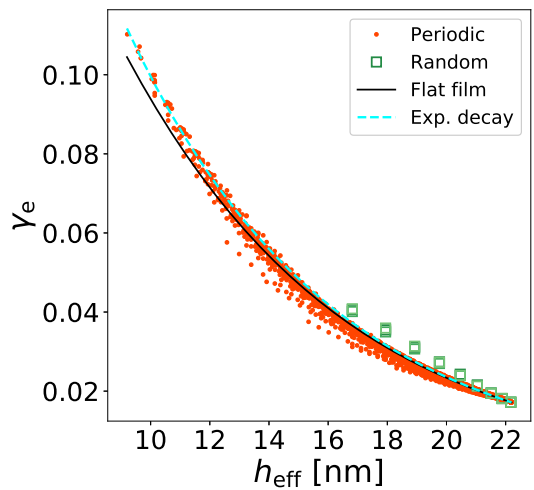} 
\par\end{centering}
\caption{\label{fig:ge_heff}Relation between the external decay rate $\gamma_{\text{e}}$
and the effective aluminum thickness $h_{\text{eff}}$ for nanodimple
periodic array (red dots), nanodimple random array (green squares),
and flat aluminum film (black solid line). The light-blue dashed line
depicts the exponential dependence on $h_{\text{eff}}$ based on the
value at $h_{\text{eff}}=22.2$ nm. }
\end{figure}

Figure \ref{fig:ge_heff} shows the relation between the value of
effective film thickness $h_{\text{eff}}$ defined by Eq.\ (\ref{eq:effective_thickness})
and the external decay rate. The black solid line depicts the results
for a flat aluminum film with thickness $h_{\text{eff}}$. The light-blue
dashed line denotes exponential dependence on $h_{\text{eff}}$, expressed
by the left- hand side of Eq.\ (\ref{eq:effective_thickness}), based
on the value at $h_{\text{eff}}=22.2$ nm. All data coincide well,
which shows the validity of the estimation of effective thickness
using Eq.\ (\ref{eq:effective_thickness}). Thus, we can evaluate
$h_{\text{eff}}$ from the value of $\gamma_{\text{e}}$ assuming
the exponential dependence between them, and use it as an index for
corrosion.

\begin{figure*}
\begin{minipage}[t]{0.33\textwidth}%
 
\begin{flushleft}
(a)\\
 \includegraphics[width=0.95\columnwidth]{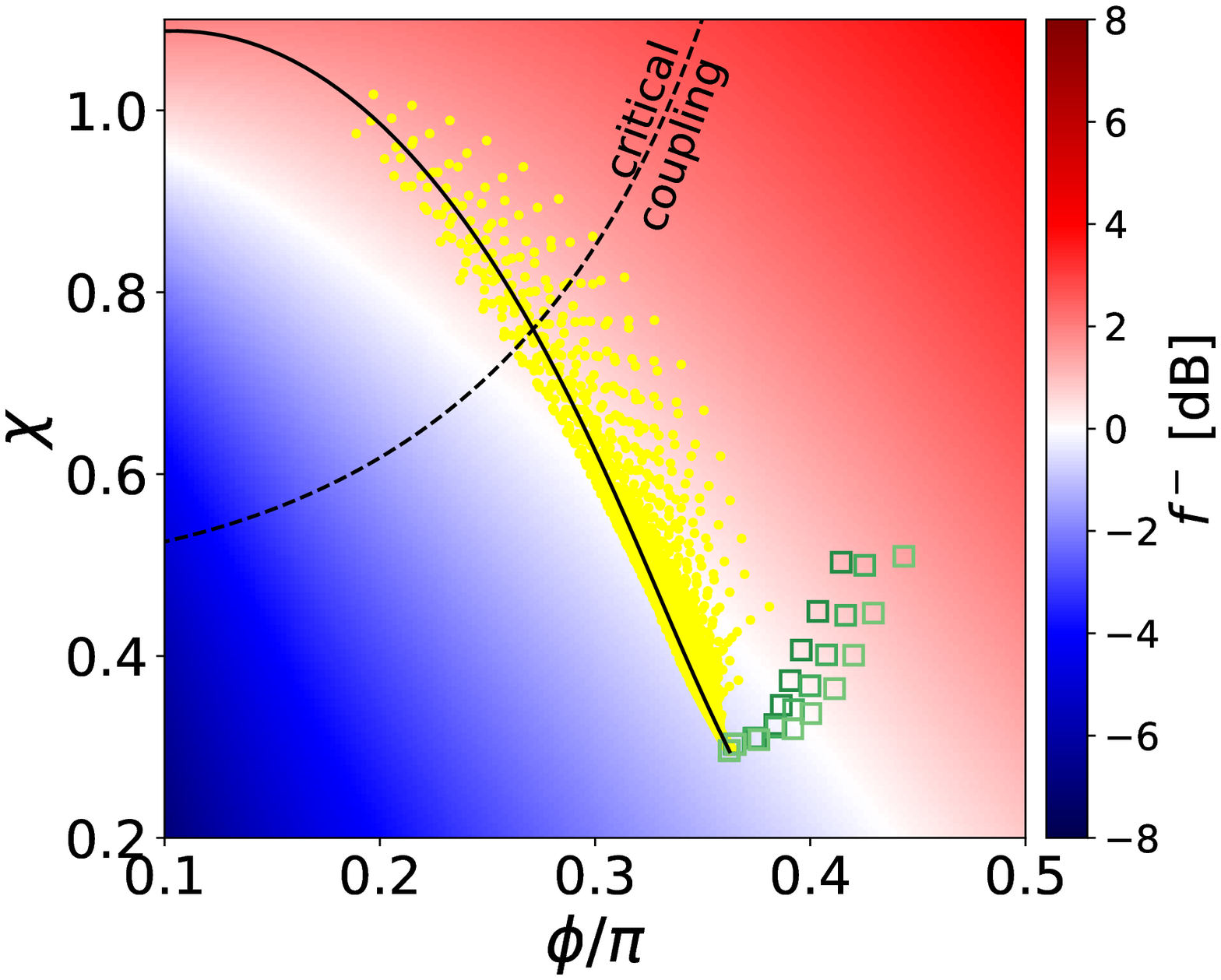} 
\par\end{flushleft}%
\end{minipage}%
\begin{minipage}[t]{0.33\textwidth}%
 
\begin{flushleft}
(b)\\
 \includegraphics[width=0.9\columnwidth]{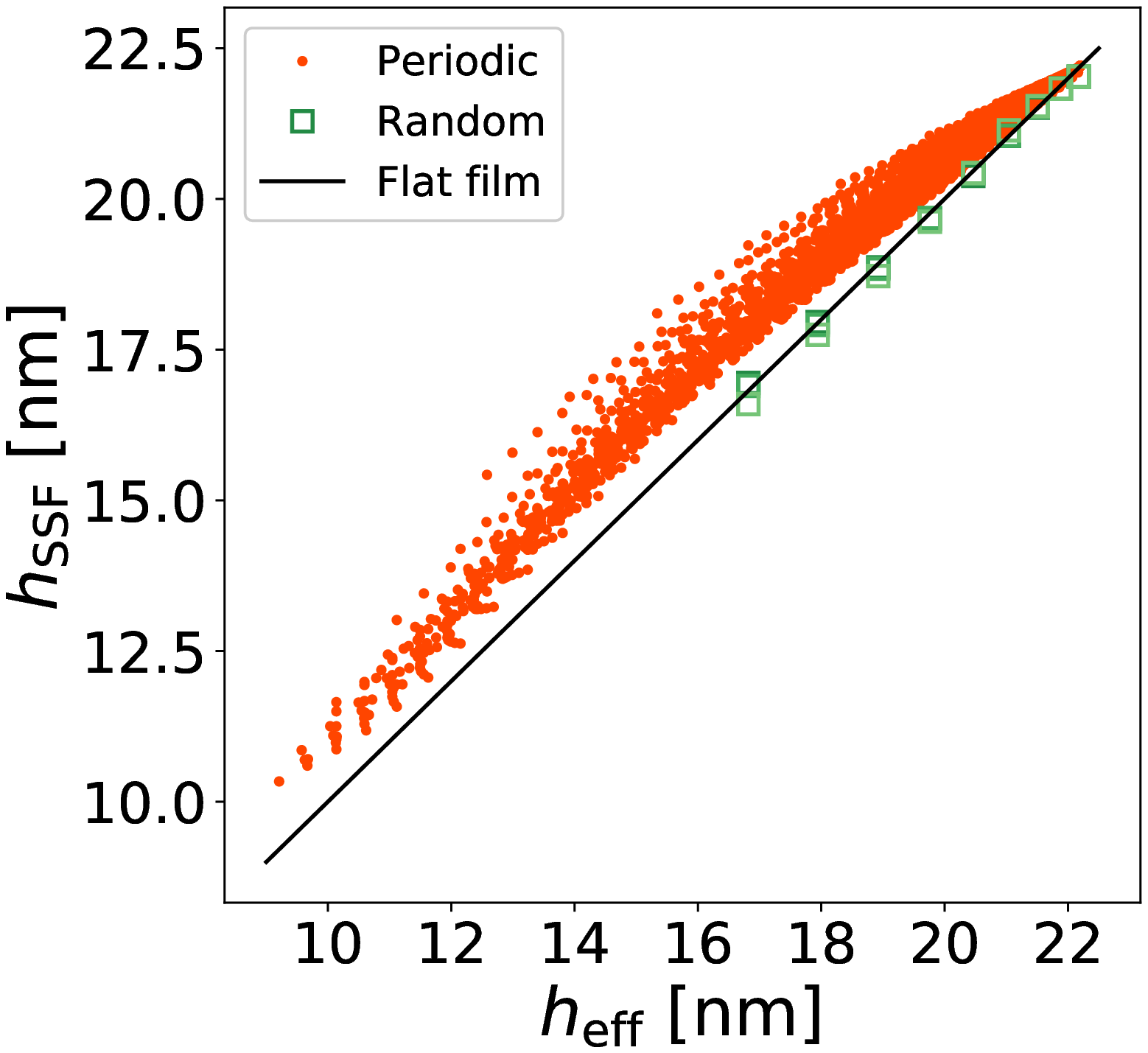} 
\par\end{flushleft}%
\end{minipage}%
\begin{minipage}[t]{0.33\textwidth}%
 
\begin{flushleft}
(c)\\
 \includegraphics[width=0.9\columnwidth]{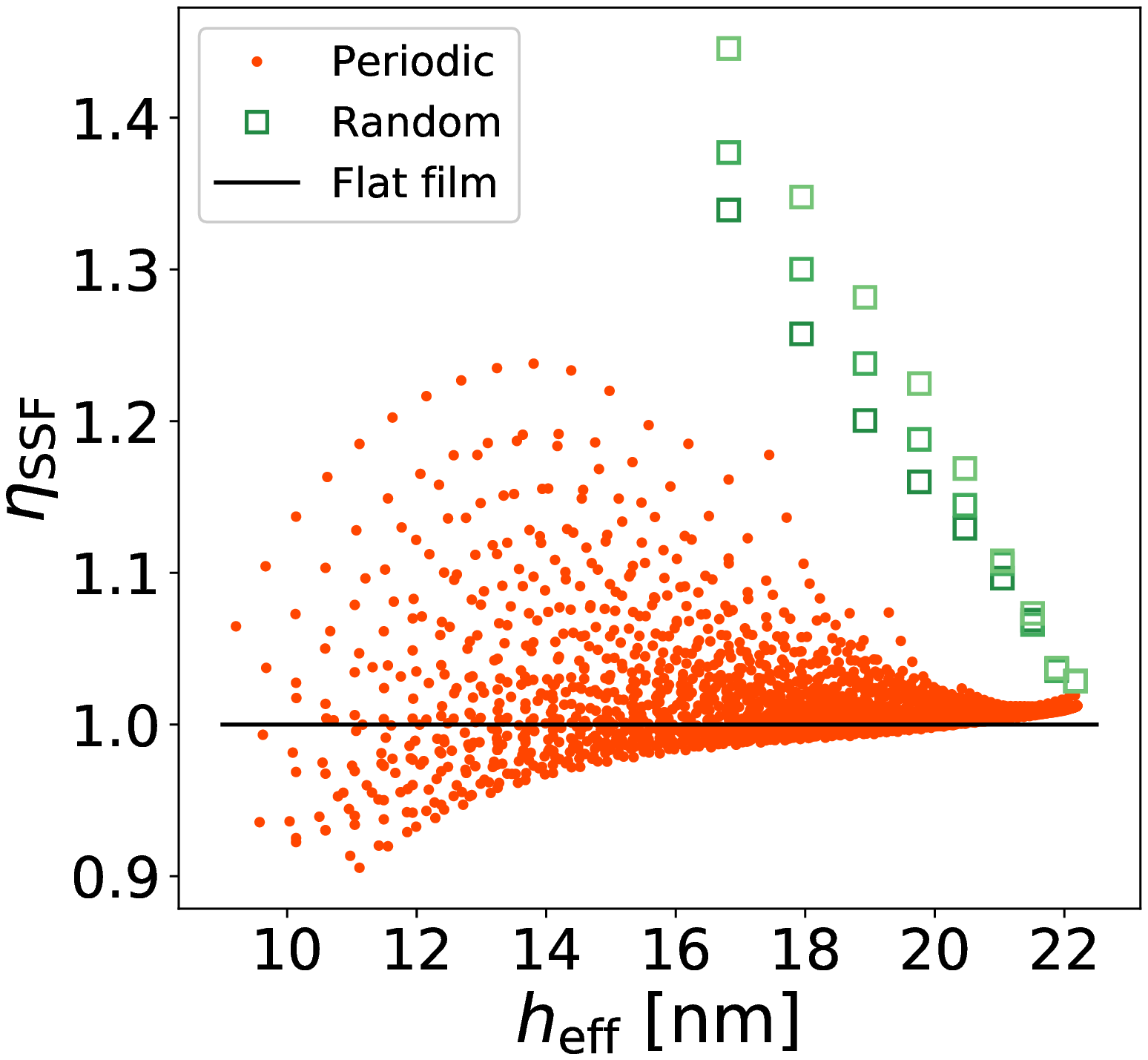} 
\par\end{flushleft}%
\end{minipage}\\
\begin{minipage}[t]{0.33\textwidth}%
 
\begin{flushleft}
(d)\\
 \includegraphics[width=0.9\columnwidth]{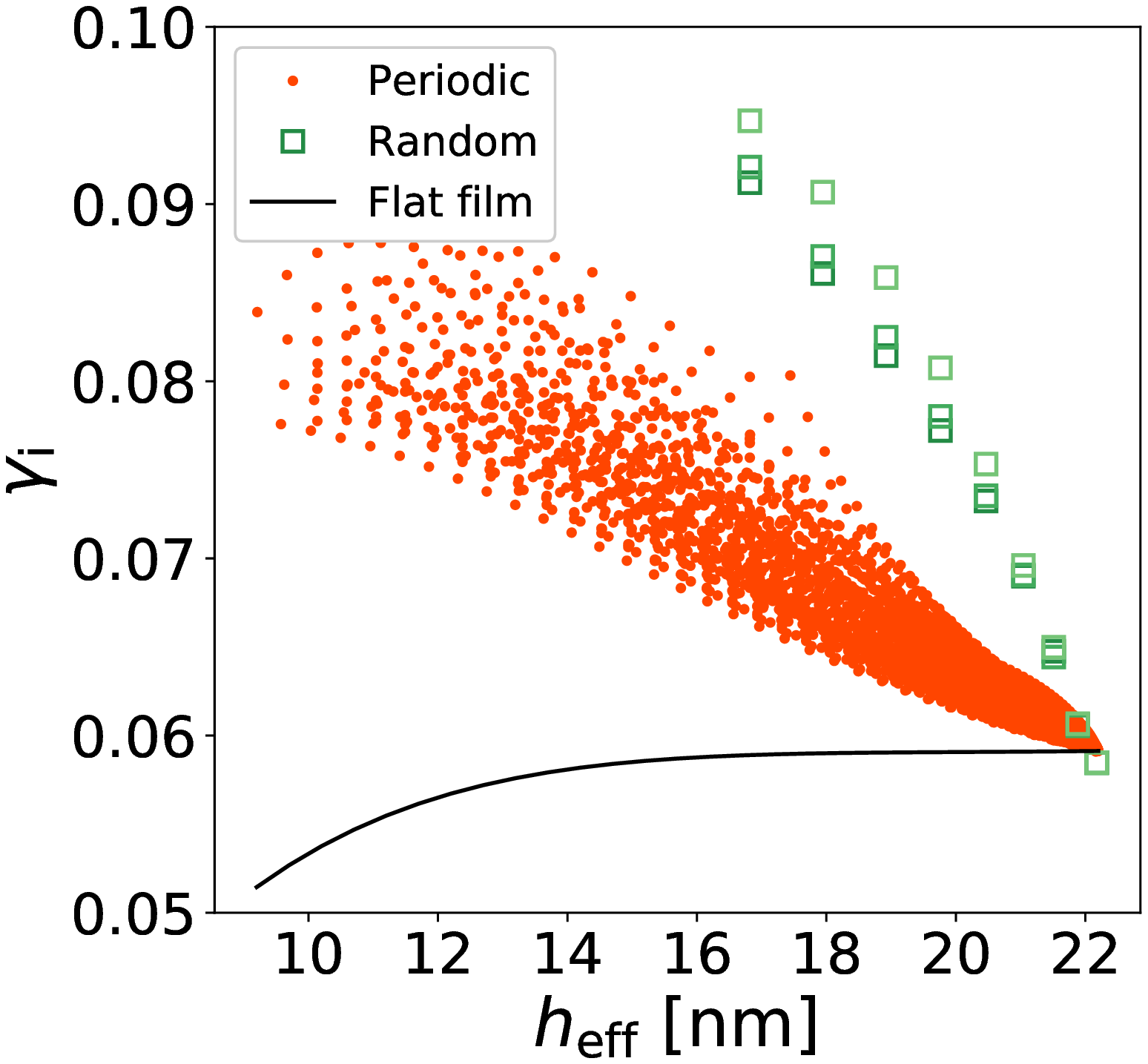} 
\par\end{flushleft}%
\end{minipage}%
\begin{minipage}[t]{0.33\textwidth}%
 
\begin{flushleft}
(e)\\
 \includegraphics[width=0.9\columnwidth]{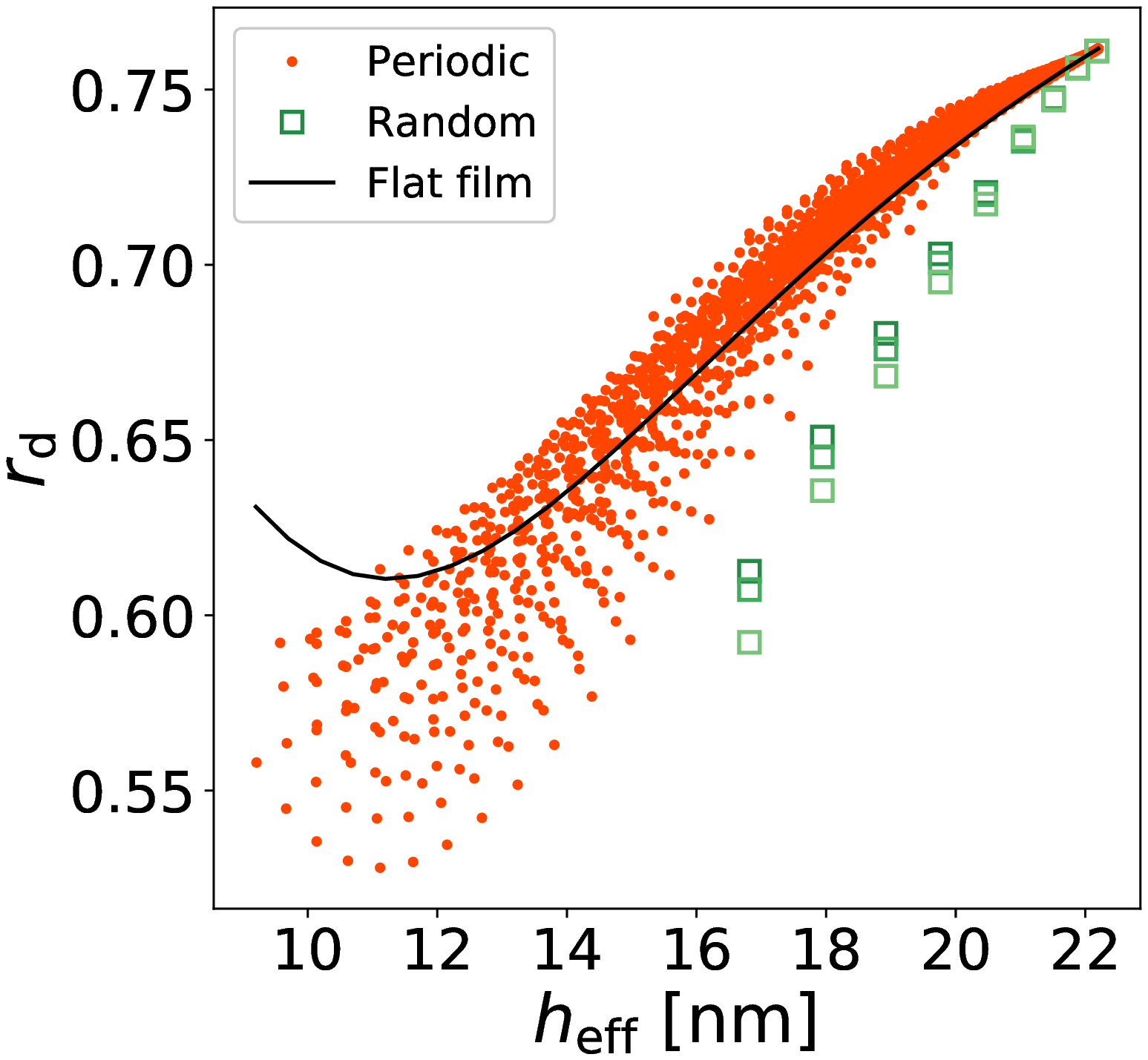} 
\par\end{flushleft}%
\end{minipage}\caption{\label{fig:all_parameters}Parameters extracted by the TCM method
from the reflection spectra for various configurations of nanodimples.
(a) Direct reflection phase $\phi$ and ratio of decay rates $\chi$
for periodic array (yellow circles) and random array (green squares)
plotted on the color map of the spectral shape factor $f^{-}$. (b)
and (c) Thickness $h_{\text{SSF}}$ and the imaginary part factor
$\eta_{\text{SSF}}$ for the flat metal film that gives the same shape
of the normalized reflection spectrum as the dimple array system with
the effective thickness $h_{\text{eff}}$. (d) and (e) Internal decay
rate $\gamma_{\text{i}}$ and direct reflection coefficient $r_{\text{d}}$
as functions of $h_{\text{eff}}$. Red dots, green squares, and the
black solid line denote the results for nanodimple periodic array,
nanodimple random array, and flat aluminum film, respectively.}
\end{figure*}

In Fig.\ \ref{fig:all_parameters} (a), the values of $\chi$ and
$\phi$ extracted from the reflection spectra for various conditions
of the periodic array (yellow circles) and random array (green squares)
of dimples are plotted on the color map of the spectral shape factor
$f^{-}$, as in Fig.\ \ref{fig:Kretschmann_parameters} (c). The
black solid line depicts the result of a flat aluminum film with various
thicknesses decreasing from the lower right to upper left, where the
imaginary part factor $\eta$ is fixed to 1. Using the results shown
in Fig.\ \ref{fig:Kretschmann_parameters} (c), we can deduce the
thickness of the aluminum film, $h_{\text{SSF}}$, and the imaginary
part factor, $\eta_{\text{SSF}}$, for the metal film that provides
the values of $\chi$, $\phi$, and $f^{-}$ at the points in Fig.\ \ref{fig:all_parameters}
(a). Thus, we obtain a flat metal film model with thickness $h_{\text{SSF}}$
and imaginary part factor $\eta_{\text{SSF}}$ for each dimple array
system. This model gives the same shape of the normalized reflection
spectrum as the dimple array system. Figure \ref{fig:all_parameters}
(b) and (c) show $h_{\text{SSF}}$ and $\eta_{\text{SSF}}$ as functions
of effective thickness $h_{\text{eff}}$ for periodic arrays (red
dots), random arrays (green squares), and a flat aluminum film (black
solid line). Panel (b) shows that $h_{\text{SSF}}$ corresponds rather
well with $h_{\text{eff}}$, which is consistent with the notion that
one of the effects of the dimple array is to reduce the effective
thickness of the aluminum film and result in the exponential enhancement
of radiative decay. While from panel (c), another effect of the dimple
array can be considered to alter (enhance in most cases) the imaginary
part of the permittivity of aluminum effectively. This effect seems
to be produced by diffusive scattering, which depends on the configuration
of the dimples. Indeed, we can see that random arrays provide higher
enhancement than the periodic array.

These effects of the dimple array appear in each parameter value,
such as the internal decay rate $\gamma_{\text{i}}$ and the direct
reflection coefficient $r_{\text{d}}$, as shown in Fig.\ \ref{fig:all_parameters}
(d) and (e). These parameters strongly depend on the structure and
distribution of dimples, even though they provide the same effective
thickness. Especially, the plots of $\gamma_{\text{i}}$ for the flat
films, periodic arrays, and random arrays are distributed separately.
These results indicate that a more detailed characterization of the
metal surface condition can be achieved by analyzing these parameters
systematically, e.g., by using machine learning. However, this will
be taken up in the future.

\section{Conclusions}

In this paper, we have shown the possibility of evaluating the surface
roughness of a metal film by analyzing the ATR spectrum from the viewpoint
of Fano resonance. Using the temporal coupled-mode method, it was
found that the geometrical feature of the ATR spectrum is characterized
by a single shape factor, which is determined by the following two
parameters: direct reflection phase and ratio of decay rates. Based
on this information, we have developed a method to extract key parameters,
such as internal and external decay rates, from spectral data, and
shown that these parameters provide corrosion information such as
the effective thickness of the metal film and the distribution of
nanosized dimples. These results form the basis for developing a novel
method for characterizing the initial stage of pitting corrosion on
a metal surface using plasmonic ATR\@.

\section*{Acknowledgments}

This study was supported by JSPS KAKENHI Grant Numbers JP18K04980
and JP18K04979.

%


\clearpage

\onecolumngrid
\begin{center}
  \textbf{\large Supplemental Material: Evaluation of surface roughness
 of metal films using plasmonic Fano resonance in attenuated total reflection}\\[.2cm]
  Munehiro Nishida,$^{1,*}$ Taisei Matsumoto,$^{1}$ Hiroya Koga,$^1$
 Terukazu Kosako,$^2$ and Yutaka Kadoya$^1$\\[.1cm]
  {\itshape ${}^1$Graduate School of Advanced Science of Matter,
Hiroshima University, Higashi-Hiroshima, 739-8530, Japan\\
  ${}^2$YAZAKI Research and Technology Center, 1500 Mishuku, Susono-city,
Shizuoka, 410-1194, Japan\\}
  ${}^*$Electronic address: mnishida@hiroshima-u.ac.jp\\
(Dated: \today)\\[1cm]
\end{center}

\setcounter{section}{0}
\setcounter{equation}{0}
\setcounter{figure}{0}
\setcounter{table}{0}
\setcounter{page}{1}
\renewcommand{\thesection}{S\arabic{section}}
\renewcommand{\theequation}{S\arabic{equation}}
\renewcommand{\thefigure}{S\arabic{figure}}
\renewcommand{\bibnumfmt}[1]{[S#1]}
\renewcommand{\citenumfont}[1]{S#1}

\section{Temporal Coupled Method Considering Absorption in Direct Reflection
Process}

The temporal coupled mode (TCM) method describes resonant scattering
phenomenon by considering the dynamics of cavities attached to ports
\cite{S_Haus_83,S_Fan_03,S_Nishida_Kadoya_18}. The Kretschmann configuration
in plasmonic ATR can be modeled by a system such that a cavity with
a single resonant mode corresponding to SPP is attached to a single
input/output port representing a connection with the incident and
reflected waves.

In the case where the cavity couples with the port weakly, the amplitude
of the resonant mode with the resonant angular frequency of $\omega_{\text{r}}$
is described by the following coupled mode equation \cite{S_Haus_83,S_Fan_03,S_Joannopoulos_08}:

\begin{align}
\frac{\text{d}a}{\text{d}t} & =-\text{i}\omega_{\text{r}}a-\gamma a+\kappa s_{+},\label{S_eq:a_motion}\\
s_{-} & =r_{\text{d}}\text{e}^{\text{i}\phi_{\text{d}}}s_{+}+da,\nonumber \\
\gamma & =\gamma_{\text{i}}+\gamma_{\text{e}}.\nonumber 
\end{align}
Here, $\gamma_{\text{i}}$ denotes the internal decay rate due to
the loss of the materials composing the cavity and $\gamma_{\text{e}}$
denotes the external decay rate due to the loss by the radiation to
the port. Variables $s_{+}$ and $s_{-}$ denote the amplitudes incoming
and outgoing radiative modes through the port, respectively, where
the mode fields are normalized so that $\left|s_{\pm}\right|^{2}$
equal the powers of the modes. The parameters $r_{\text{d}}$ and
$\phi_{\text{d}}$ denote the magnitude and phase of the direct reflection
coefficient, which determines the reflection process in which the
incoming wave from the port is reflected directly to the port without
resonance excitation. The parameter $\kappa$ ($d$) denotes the coupling
constants between the incoming (outgoing) mode and the cavity mode
through the port.

If a continuous wave $s_{+}$ with angular frequency $\omega$ is
incident, the amplitude of the resonant mode is given by 
\begin{equation}
a=\frac{\kappa s_{+}}{\text{i}\left(\omega_{\text{r}}-\omega\right)+\gamma},
\end{equation}
and the outgoing wave is given by 
\begin{equation}
s_{-}\equiv rs_{+}=\left[r_{\text{d}}\text{e}^{\text{i}\phi_{\text{d}}}+\frac{d\kappa}{\text{i}\left(\omega_{\text{r}}-\omega\right)+\gamma}\right]s_{+}.
\end{equation}

If the internal decay, energy absorption in the direct reflection
process, and coupling between the cavity and the port are all weak,
the parameters $\gamma_{\text{e}}$, $d$, and $\kappa$ are approximately
independent of the internal decay rate $\gamma_{\text{i}}$ and the
absorption of the direct reflection process. Therefore, the relations
among $\gamma_{\text{e}}$, $d$, and $\kappa$ can be derived without
considering material loss. Suppose for a while that there is no material
loss. In this case, $\gamma_{\text{i}}=0$ and the direct reflection
coefficient can be expressed as $r_{\text{c}}=\text{e}^{\text{i}\phi_{\text{c}}}$.
The same discussion as in \cite{S_Fan_03} applies to this case. From
energy-conservation and time-reversal symmetry, we obtain

\begin{align}
|d|^{2} & =2\gamma_{\text{e}},\label{S_eq:d_d_relation}\\
d & =\kappa,\\
r_{\text{c}}d^{*} & =-d.\label{S_eq:rc_d_relation}
\end{align}
Therefore, 
\begin{align}
r & =r_{\text{d}}\text{e}^{\text{i}\phi_{\text{d}}}+\frac{d^{2}}{\text{i}\left(\omega_{\text{r}}-\omega\right)+\gamma}=\left[r_{\text{d}}\text{e}^{\text{i}\phi_{\text{d}}}-\frac{2\gamma_{\text{e}}\text{e}^{\text{i}\phi_{\text{c}}}}{\text{i}\left(\omega_{\text{r}}-\omega\right)+\gamma}\right]\nonumber \\
 & =r_{\text{d}}\text{e}^{\text{i}\phi_{\text{d}}}\left(\frac{\omega-\omega_{\text{0}}+\text{i}\gamma_{0}}{\omega-\omega_{\text{r}}+\text{i}\gamma}\right),\label{S_eq:r_coupled_mode}\\
\omega_{0} & =\omega_{\text{r}}-2\sin\phi\gamma_{\text{e}}/r_{\text{d}},\\
\gamma_{0} & =\gamma-2\cos\phi\gamma_{\text{e}}/r_{\text{d}},\\
\phi & =\phi_{\text{c}}-\phi_{\text{d}}.
\end{align}

Consider the case in which the angular spectra near the resonant angle
are obtained using a focused incident light with the angular frequency
of $\omega_{\text{i}}=\frac{2\pi}{c\lambda_{\text{i}}}$, where $c$
is the speed of light in vacuum. Assuming that $\omega_{\text{r}}$
becomes $\omega_{\text{i}}$ when $k_{x}=k_{\text{r}}$ and the relation
between $\omega_{\text{r}}$ and $k_{x}$ is linear with the gradient
of the group velocity $v_{\text{SPP}}$ of SPP on the surface of a
semi-infinite metal, the $k_{x}$-dependence of $\omega_{\text{r}}$,
namely the dispersion relation of the resonant mode, is expressed
as 
\begin{equation}
\omega_{\text{r}}=v_{\text{SPP}}\left(k_{x}-k_{\text{r}}\right)+\omega_{\text{i}}.\label{S_eq:dispersion_relation-1}
\end{equation}
The parameters in Eq.\ (\ref{S_eq:r_coupled_mode}), $\gamma_{\text{e}}$,
$\gamma$, $r_{\text{d}}$, and $\phi$, are set to the values at
$k_{x}=k_{\text{r}}$, and the reflection coefficient for $\omega=\omega_{\text{i}}$
reads

\begin{align}
r\left(k_{x}\right) & =r_{\text{d}}\text{e}^{\text{i}\phi_{\text{d}}}-\frac{2\text{i}\gamma_{\text{e}}\text{e}^{\text{i}\phi_{\text{c}}}}{\omega_{\text{i}}-\left\{ v_{\text{SPP}}\left(k_{x}-k_{\text{r}}\right)+\omega_{\text{i}}\right\} +\text{i}\gamma}\nonumber \\
 & =r_{\text{d}}\text{e}^{\text{i}\phi}\frac{k_{x}-k_{0}-\text{i}\frac{\gamma_{0}}{v_{\text{SPP}}}}{k_{x}-k_{\text{r}}-\text{i}\frac{\gamma}{v_{\text{SPP}}}},\\
k_{0} & =k_{\text{r}}+2\frac{\gamma_{\text{e}}}{r_{\text{d}}v_{\text{SPP}}}\sin\phi.
\end{align}
Then, the normalized reflectance is given by the following equation:
\begin{align}
F\left(k_{x}\right) & \equiv\frac{R\left(k_{x}\right)}{R_{\text{d}}}=\frac{\left|r\left(k_{x}\right)\right|^{2}}{R_{\text{d}}}=\frac{\left(k_{x}-k_{0}\right)^{2}+\left(\frac{\gamma_{0}}{v_{\text{SPP}}}\right)^{2}}{\left(k_{x}-k_{\text{r}}\right)^{2}+\left(\frac{\gamma}{v_{\text{SPP}}}\right)^{2}}\nonumber \\
 & =\frac{\left\{ \frac{v_{\text{SPP}}}{\gamma}\left(k_{x}-k_{\text{r}}\right)-2\chi\sin\phi\right\} ^{2}+\left(1-2\chi\cos\phi\right)^{2}}{\left\{ \frac{v_{\text{SPP}}}{\gamma}\left(k_{x}-k_{\text{r}}\right)\right\} ^{2}+1},\\
\chi & \equiv\frac{\gamma_{\text{e}}}{r_{\text{d}}\gamma}.
\end{align}

When $\sin\phi\neq0$, solving the equation $F'\left(k_{x}\right)=0$,
the in-plane wavenumbers $k_{-}$ and $k_{+}$ at which $F$ is minimum
and maximum, respectively, are given as

\begin{align}
k_{\mp} & =k_{\text{r}}\pm p\frac{\gamma}{v_{\text{SPP}}}f^{\mp}\left(\phi,\chi\right),
\end{align}
and the minimum value $F_{-}$ and maximum value $F_{+}$ are given
as

\begin{equation}
F_{\mp}\equiv\frac{R_{\mp}}{R_{\text{d}}}=1\mp\frac{2\chi\left|\sin\phi\right|}{f^{\mp}\left(\phi,\chi\right)},\label{S_eq:F_mp}
\end{equation}
where $p=\pm1$ is the sign of $\sin\phi$, namely $\sin\phi=p\left|\sin\phi\right|$,
and 
\begin{equation}
f^{\mp}\left(\phi,\chi\right)=\frac{\sqrt{\left(\cos\phi-\chi\right)^{2}+\sin^{2}\phi}\mp\left(\cos\phi-\chi\right)}{\left|\sin\phi\right|}.
\end{equation}
Thus, total absorption ($F_{-}=0$) occurs when 
\begin{align}
\chi=\chi_{\text{c}} & \equiv\frac{1}{2\cos\phi}.
\end{align}

When $\sin\phi=0$, there is no maximum of $F$, and for the minimum,
\begin{align}
k_{-} & =k_{\text{r}},\\
F_{-} & =\left(1-2\chi\right)^{2}.
\end{align}
Thus, total absorption ($F_{-}=0$) occurs when 
\begin{equation}
\chi_{\text{c}}=\frac{1}{2}.
\end{equation}

Noting that 
\begin{align}
f^{+}f^{-} & =\frac{\left(\cos\phi-x\right)^{2}+\sin^{2}\phi-\left(\cos\phi-x\right)^{2}}{\sin^{2}\phi}=1,\\
F\left(k_{\text{r}}\right) & =\frac{v_{\text{SPP}}^{2}\left(k_{\text{r}}-k_{0}\right)^{2}+\gamma_{0}^{2}}{\gamma^{2}}=1-4\chi\cos\phi+4\chi^{2},
\end{align}
we obtain 
\begin{align}
F_{-}+F_{+} & =2+2\chi\left|\sin\phi\right|\left\{ \frac{f^{-}-f^{+}}{f^{+}f^{-}}\right\} =2-4\chi\cos\phi+4\chi^{2}\nonumber \\
 & =1+F\left(k_{\text{r}}\right),\nonumber \\
\therefore R_{\text{r}}=R\left(k_{\text{r}}\right) & =R_{-}+R_{+}-R_{\text{d}}.\label{S_eq:Rr_Rmp_Rd}
\end{align}
Then, from Eq.\ (\ref{S_eq:F_mp}) and Eq.\ (\ref{S_eq:Rr_Rmp_Rd}),
\begin{align}
\frac{R_{\text{+}}-R_{\text{d}}}{R_{\text{d}}-R_{-}} & =\frac{R_{\text{r}}-R_{\text{-}}}{R_{\text{d}}-R_{-}}=\frac{f^{-}}{f^{+}}.
\end{align}
In addition, we obtain

\begin{align}
F\left(\frac{k_{-}+k_{+}}{2}\right) & =F\left(k_{\text{r}}+\frac{p\gamma}{2v_{\text{SPP}}}\left(f^{+}-f^{-}\right)\right)=1,\nonumber \\
\therefore R\left(k_{\text{d}}\right) & =R_{\text{d}},\\
k_{\text{d}} & =\frac{k_{-}+k_{+}}{2}.
\end{align}

\section{Improvement in Spatial Coupled Mode Method}

The spatial coupled mode (SCM) method derives a set of coupled equations
for waveguide modes of nanoholes perforated in a metal film \cite{S_Garcia-Vidal_Martin-Moreno_05,S_Leon-Perez_Martin-Moreno_08,S_Nishida_Kadoya_15,S_Nishida_Kadoya_18}.
The electromagnetic (EM) fields outside the metal film are expressed
by a linear combination of plane-wave modes specified by the parallel
wave vector $\vec{k}=\vec{k}_{0}+\vec{K}$ and the polarization $\sigma=\text{p or s}$,
where $\vec{k}_{0}$ and $\vec{K}$ are the incident parallel wave
vector and reciprocal lattice vector of the nanohole array, respectively.
The EM fields in the metal film region are expressed by the superposition
of the waveguide modes of a nanohole and evanescent plane waves in
order to account for the penetration of the EM field into the metal
region. Here, we assume that the dielectrics inside the nanohole and
outside the metal film are the same with the relative permittivity
of $\varepsilon$.

We use Dirac's notation to describe the electric field components
parallel to the $xy$-plane for mode $\alpha$, such that 
\begin{equation}
\vec{E}_{\alpha}(\vec{r})=\left(E_{\alpha x},E_{\alpha y}\right)=\left\langle \left.\vec{r}\right|\alpha\right\rangle ,
\end{equation}
Here, the mode index $\alpha$ represents the full information of
the modes of a nanohole, such as the ``HE$_{11}$ horizontal mode''
\cite{S_Roberts_87}. This may also represent the parallel wave vector
$\vec{k}$ and the polarization $\sigma$ $(=\text{p or s)}$ for
plane-wave modes. Because the magnetic field components parallel to
the $xy$-plane are determined by the position-dependent admittance
\cite{S_Nishida_Kadoya_15,S_Nishida_Kadoya_18}, we use the admittance
operator $\hat{Y}$ to express them, such that

\begin{equation}
\vec{H}_{\alpha}(\vec{r})\equiv-\bm{e}_{z}\times\bm{H}_{\alpha}(\vec{r})=\left(H_{\alpha y},-H_{\alpha x}\right)=\left\langle \vec{r}\Bigr|\hat{Y}\alpha\right\rangle .
\end{equation}
For the plane-wave mode in the dielectric, this relation is reduced
to 
\begin{align}
\left\langle \vec{r}\Bigr|\hat{Y}\vec{k}\sigma\right\rangle  & =Y_{\overrightarrow{k}\sigma}\left\langle \vec{r}\Bigr|\vec{k}\sigma\right\rangle ,\\
Y_{\overrightarrow{k}\text{p}} & =\frac{1}{Z_{0}}\frac{k_{z}}{k_{\omega}},\quad Y_{\overrightarrow{k}\text{s}}=\frac{\varepsilon}{Z_{0}}\frac{k_{\omega}}{k_{z}},
\end{align}
where $Z_{0}$ and $k_{\omega}$ are the impedance and wavenumber
in the vacuum, respectively, and $k_{z}$ is the $z$-component of
the wave vector.

We define the internal product of the two fields as 
\begin{align}
\left\langle \alpha\left|\beta\right.\right\rangle  & \equiv\iint\text{d}x\text{d}y\vec{E}_{\alpha}^{*}\cdot\vec{E}_{\beta},\\
\left\langle \alpha\Bigr|\hat{Y}\beta\right\rangle  & \equiv\iint\text{d}x\text{d}y\vec{E}_{\alpha}^{*}\cdot\vec{H}_{\beta}=\iint\text{d}x\text{d}y\left[\bm{E}_{\alpha}^{*}\times\bm{H}_{\beta}\right]_{z},
\end{align}
where $*$ denotes the complex conjugate. Here, the mode fields are
normalized by $\left\langle \alpha\left|\alpha\right.\right\rangle =1$.
Then, the orthogonality condition for the plane-wave modes is expressed
as

\begin{equation}
\left\langle \left.\vec{k}\sigma\right|\hat{Y}\vec{k}'\sigma'\right\rangle =Y_{\vec{k}\sigma}\delta_{\vec{k}\vec{k}'}\delta_{\sigma\sigma'}.
\end{equation}
However, due to the metal loss, the orthogonality condition for the
waveguide modes should be modified as 
\begin{equation}
\left\langle \alpha^{*}\Bigr|\hat{Y}\beta\right\rangle \equiv\int\vec{E}_{\alpha}\cdot\vec{H}_{\beta}\text{d}x\text{d}y=Y_{\alpha}\delta_{\alpha\beta},\quad Y_{\alpha}\equiv\left\langle \alpha^{*}\Bigr|\hat{Y}\alpha\right\rangle ,
\end{equation}
based on the Lorentz reciprocity theorem \cite{S_Lorentz_1895,S_deHoop1960,S_Lalanne_Rodier_05}.

Consider a metal film with nanohole array located at the region $0\leq z\leq d$.
At the interface between the metal and the dielectric at $z=0$, the
EM fields on the dielectric side, $|0^{-}\rangle$, $|\hat{Y}0^{-}\rangle$,
and on the metal side, $|0^{+}\rangle$,$|\hat{Y}0^{+}\rangle$, are
given by 
\begin{align}
|0^{-}\rangle & =|\vec{k_{0}}\sigma_{0}\rangle+\sum_{\vec{k}\sigma}r_{\vec{k}\sigma}|\vec{k}\sigma\rangle,\\
|\hat{Y}0^{-}\rangle & =|\hat{Y}\vec{k_{\text{0}}}\sigma_{0}\rangle-\sum_{\vec{k}\sigma}r_{\vec{k}\sigma}|\hat{Y}\vec{k}\sigma\rangle,\\
|0^{+}\rangle & =\sum_{\alpha}\left(A_{\alpha\vec{k}_{0}\sigma_{0}}+B_{\alpha\vec{k}_{0}\sigma_{0}}\text{e}^{\text{i}q_{\alpha}d}\right)|\alpha\rangle+\sum_{\vec{k}\sigma}\left(A_{\vec{k}\sigma}^{\text{m}}+B_{\vec{k}\sigma}^{\text{m}}\text{e}^{\text{i}k_{z}^{\text{m}}d}\right)|\vec{k}\sigma\rangle,\\
|\hat{Y}0^{+}\rangle & =\sum_{\alpha}\left(A_{\alpha\vec{k}_{0}\sigma_{0}}-B_{\alpha\vec{k}_{0}\sigma_{0}}\text{e}^{\text{i}q_{\alpha}d}\right)|\hat{Y}\alpha\rangle+\sum_{\vec{k}\sigma}\left(A_{\vec{k}\sigma}^{\text{m}}-B_{\vec{k}\sigma}^{\text{m}}\text{e}^{\text{i}k_{z}^{\text{m}}d}\right)|\hat{Y}\vec{k}\sigma\rangle.
\end{align}
Similarly, the EM fields at $z=d$ are given by

\begin{align}
|d^{+}\rangle & =\sum_{\vec{k}\sigma}t_{\vec{k}\sigma}|\vec{k}\sigma\rangle,\\
|\hat{Y}d^{+}\rangle & =\sum_{\vec{k}\sigma}t_{\vec{k}\sigma}|\hat{Y}\vec{k}\sigma\rangle,\\
|d^{-}\rangle & =\sum_{\alpha}\left(A_{\alpha\vec{k}_{0}\sigma_{0}}\text{e}^{\text{i}q_{\alpha z}d}+B_{\alpha\vec{k}_{0}\sigma_{0}}\right)|\alpha\rangle+\sum_{\vec{k}\sigma}\left(A_{\vec{k}\sigma}^{\text{m}}\text{e}^{\text{i}k_{z}^{\text{m}}d}\delta_{\vec{k}\vec{k_{0}}}\delta_{\sigma\sigma_{0}}+B_{\vec{k}\sigma}^{\text{m}}\right)|\vec{k}\sigma\rangle,\\
|\hat{Y}d^{-}\rangle & =\sum_{\alpha}\left(A_{\alpha\vec{k}_{0}\sigma_{0}}\text{e}^{\text{i}q_{\alpha z}d}-B_{\alpha\vec{k}_{0}\sigma_{0}}\right)|\hat{Y}\alpha\rangle+\sum_{\vec{k}\sigma}\left(A_{\vec{k}\sigma}^{\text{m}}\text{e}^{\text{i}k_{z}^{\text{m}}d}\delta_{\vec{k}\vec{k_{0}}}\delta_{\sigma\sigma_{0}}-B_{\vec{k}\sigma}^{\text{m}}\right)|\hat{Y}\vec{k}\sigma\rangle.
\end{align}
Here, $q_{\alpha z}$ is the propagation constant of the waveguide
mode $\alpha$ and $k_{z}^{\text{m}}=\sqrt{\varepsilon_{\text{m}}\left(\omega\right)k_{\omega}^{2}-\left|\vec{k}\right|^{2}}$,
with $\varepsilon_{\text{m}}\text{\ensuremath{\left(\omega\right)}}$
being the dielectric function for the metal.

Using these definitions, the coupled-mode equations can be derived
in a similar manner as the original derivation \cite{S_Leon-Perez_Martin-Moreno_08}.
Under the condition that the projection of the EM field onto the electric
field of the plane wave $|\vec{k}\sigma\rangle$ and that of the magnetic
field onto the electric field of the waveguide mode $|\alpha\rangle$
are continuous at the two interfaces at $z=0$ and $z=h$, we can
derive a coupled system of equations for the coefficients of waveguide
modes as follows:

\begin{align}
 & \left\{ \begin{array}{c}
\sum_{\beta}\left(G_{\alpha\beta}^{-}A_{\beta\vec{k}_{0}\sigma_{0}}+G_{\alpha\beta}^{+}B_{\beta\vec{k}_{0}\sigma_{0}}\right)=I_{\alpha\vec{k}_{0}\sigma_{0}},\\
\sum_{\beta}\left(G_{\alpha\beta}^{+}A_{\beta\vec{k}_{0}\sigma_{0}}+G_{\alpha\beta}^{-}B_{\beta\vec{k}_{0}\sigma_{0}}\right)=I_{\alpha\vec{k}_{0}\sigma_{0}}^{\prime},
\end{array}\right.\\
 & \quad I_{\alpha}=2Y_{\vec{k}_{0}\sigma_{0}}\frac{f_{\vec{k}_{0}\sigma_{0}}^{+}-\text{e}^{2\text{i}k_{0z}^{\text{m}}d}f_{\vec{k}_{0}\sigma_{0}}^{-}}{\left(f_{\vec{k}_{0}\sigma_{0}}^{+}\right)^{2}-\text{e}^{2\text{i}k_{0z}^{\text{m}}d}\left(f_{\vec{k}_{0}\sigma_{0}}^{-}\right)^{2}}\langle\alpha^{*}|\vec{k}_{0}\sigma_{0}\rangle,\\
 & \quad I_{\alpha}^{\prime}=Y_{\vec{k}_{0}\sigma_{0}}\frac{\text{e}^{\text{i}k_{0z}^{\text{m}}d}\left\{ \left(f_{\vec{k}_{0}\sigma_{0}}^{-}\right)^{2}-\left(f_{\vec{k}_{0}\sigma_{0}}^{+}\right)^{2}\right\} }{\left(f_{\vec{k}_{0}\sigma_{0}}^{+}\right)^{2}-\text{e}^{2\text{i}k_{0z}^{\text{m}}d}\left(f_{\vec{k}_{0}\sigma_{0}}^{-}\right)^{2}}\langle\alpha^{*}|\vec{k}_{0}\sigma_{0}\rangle,\\
 & \quad G_{\alpha\beta}^{+}=\sum_{\vec{k}\sigma}\langle\alpha^{*}|\vec{k}\sigma\rangle Y_{\vec{k}\sigma}\left\{ \frac{\left(\text{e}^{\text{i}q_{\beta z}d}-\text{e}^{\text{i}k_{z}^{\text{m}}d}\right)\langle\vec{k}\sigma|\beta+\rangle f_{\vec{k}\sigma}^{+}+\text{e}^{\text{i}k_{z}^{\text{m}}d}\left(1-\text{e}^{\text{i}q_{\beta z}d+\text{i}k_{z}^{\text{m}}d}\right)\langle\vec{k}\sigma|\beta-\rangle f_{\vec{k}\sigma}^{-}}{\left(f_{\vec{k}_{0}\sigma_{0}}^{+}\right)^{2}-\text{e}^{2\text{i}k_{0z}^{\text{m}}d}\left(f_{\vec{k}_{0}\sigma_{0}}^{-}\right)^{2}}\right\} -Y_{\alpha}\text{e}^{\text{i}q_{\alpha z}d}\delta_{\alpha\beta},\\
 & \quad G_{\alpha\beta}^{-}=\sum_{\vec{k}\sigma}\langle\alpha^{*}|\vec{k}\sigma\rangle Y_{\vec{k}\sigma}\left\{ \frac{\left(1-\text{e}^{\text{i}q_{\beta z}d+\text{i}k_{z}^{\text{m}}d}\right)\langle\vec{k}\sigma|\beta-\rangle f_{\vec{k}\sigma}^{+}+\text{e}^{\text{i}k_{z}^{\text{m}}d}\left(\text{e}^{\text{i}q_{\beta z}d}-\text{e}^{\text{i}k_{z}^{\text{m}}d}\right)\langle\vec{k}\sigma|\beta+\rangle f_{\vec{k}\sigma}^{-}}{\left(f_{\vec{k}_{0}\sigma_{0}}^{+}\right)^{2}-\text{e}^{2\text{i}k_{0z}^{\text{m}}d}\left(f_{\vec{k}_{0}\sigma_{0}}^{-}\right)^{2}}\right\} +Y_{\alpha}\delta_{\alpha\beta},
\end{align}
\begin{multline}
\left(\left(f_{\vec{k}_{0}\sigma_{0}}^{+}\right)^{2}-\text{e}^{2\text{i}k_{0z}^{\text{m}}d}\left(f_{\vec{k}_{0}\sigma_{0}}^{-}\right)^{2}\right)r_{\vec{k}\sigma\vec{k}_{0}\sigma_{0}}=-\left(f_{\vec{k}\sigma}^{+}f_{\vec{k}\sigma}^{-}-\text{e}^{2\text{i}k_{z}^{\text{m}}h}f_{\vec{k}\sigma}^{\prime-}f_{\vec{k}\sigma}^{+}\right)\delta_{\vec{k}\vec{k}_{0}}\delta_{\sigma\sigma_{0}}\\
+\sum_{\alpha}\left[\left\{ \left(1-\text{e}^{\text{i}q_{\alpha z}h+\text{i}k_{z}^{\text{m}}h}\right)\langle\vec{k}\sigma|\alpha-\rangle f_{\vec{k}\sigma}^{+}+\left(\text{e}^{\text{i}q_{\alpha z}h+\text{i}k_{z}^{\text{m}}h}-\text{e}^{2\text{i}k_{z}^{\text{m}}h}\right)\langle\vec{k}\sigma|\alpha+\rangle f_{\vec{k}\sigma}^{-}\right\} A_{\alpha\vec{k}_{0}\sigma_{0}}\right.\\
\left.+\left\{ \left(\text{e}^{\text{i}q_{\alpha z}h}-\text{e}^{\text{i}k_{z}^{\text{m}}h}\right)\langle\vec{k}\sigma|\alpha+\rangle f_{\vec{k}\sigma}^{+}+\left(\text{e}^{\text{i}k_{z}^{\text{m}}h}-\text{e}^{\text{i}q_{\alpha z}h+2\text{i}k_{z}^{\text{m}}h}\right)\langle\vec{k}\sigma|\alpha-\rangle f_{\vec{k}\sigma}^{-}\right\} B_{\alpha\vec{k}_{0}\sigma_{0}}\right],\label{S_eq:r}
\end{multline}
\begin{multline}
\left(\left(f_{\vec{k}_{0}\sigma_{0}}^{+}\right)^{2}-\text{e}^{2\text{i}k_{0z}^{\text{m}}d}\left(f_{\vec{k}_{0}\sigma_{0}}^{-}\right)^{2}\right)t_{\vec{k}\sigma\vec{k}_{0}\sigma_{0}}=\text{e}^{\text{i}k_{z}^{\text{m}}d}\left\{ \left(f_{\vec{k}\sigma}^{+}\right)^{2}-\left(f_{\vec{k}\sigma}^{-}\right)^{2}\right\} \delta_{\vec{k}\vec{k}_{0}}\delta_{\sigma\sigma_{0}}\\
+\sum_{\alpha}\left\{ \left(\left(\text{e}^{\text{i}q_{\alpha z}d}-\text{e}^{\text{i}k_{z}^{\text{m}}d}\right)\langle\vec{k}\sigma|\alpha+\rangle f_{\vec{k}\sigma}^{+}+\left(\text{e}^{\text{i}k_{z}^{\text{m}}d}-\text{e}^{\text{i}q_{\alpha z}d+2\text{i}k_{z}^{\text{m}}d}\right)f_{\vec{k}\sigma}^{-}\langle\vec{k}\sigma|\alpha-\rangle\right)A_{\alpha\vec{k}_{0}\sigma_{0}}\right.\\
\left.+\left(\left(1-\text{e}^{\text{i}q_{\alpha z}d+\text{i}k_{z}^{\text{m}}d}\right)\langle\vec{k}\sigma|\alpha-\rangle f_{\vec{k}\sigma}^{+}+\left(\text{e}^{\text{i}q_{\alpha z}d+\text{i}k_{z}^{\text{m}}d}-\text{e}^{2\text{i}k_{z}^{\text{m}}d}\right)f_{\vec{k}\sigma}^{-}\langle\vec{k}\sigma|\alpha+\rangle\right)B_{\alpha\vec{k}_{0}\sigma_{0}}\right\} ,\label{S_eq:t}
\end{multline}
where we assume that in the metal film region, the electric field
inside the nanohole, $|\alpha\rangle$, and the magnetic field outside
the nanohole, $\sum_{\vec{k}\sigma}\left(A_{\vec{k}\sigma}^{\text{m}}-B_{\vec{k}\sigma}^{\text{m}}\right)|\hat{Y}\vec{k}\sigma\rangle$,
are orthogonal, and

\begin{align}
 & \left\langle \left.\vec{k}\sigma\right|\alpha\pm\right\rangle =\left\langle \left.\vec{k}\sigma\right|\alpha\right\rangle \pm\left\langle \left.\vec{k}\sigma\right|\hat{Y}\alpha\right\rangle /Y_{\vec{k}\sigma}^{\text{m}},\\
 & f_{\vec{k}\sigma}^{\pm}=1\pm Y_{\vec{k}\sigma}/Y_{\vec{k}\sigma}^{\text{m}},\\
 & Y_{\overrightarrow{k}\text{p}}^{\text{m}}=\frac{1}{Z_{0}}\frac{k_{z}^{\text{m}}}{k_{\omega}},\quad Y_{\overrightarrow{k}\text{s}}^{\text{m}}=\frac{\varepsilon_{\text{m}}}{Z_{0}}\frac{k_{\omega}}{k_{z}^{\text{m}}}.
\end{align}

However, the scattering coefficients $r_{\vec{k}\sigma\vec{k}_{0}\sigma_{0}}$
and $t_{\vec{k}\sigma\vec{k}_{0}\sigma_{0}}$, calculated using Eqs.\ (\ref{S_eq:r})
and (\ref{S_eq:t}), do not obey the reciprocity relation \cite{S_deHoop1960,S_Potton_04,S_Mansuripur_11}:
\begin{align}
Y_{\vec{k}_{0}\sigma_{0}}r_{\vec{k}\sigma\vec{k}_{0}\sigma_{0}} & =Y_{\vec{k}\sigma}r_{\vec{k}_{0}\sigma_{0}\vec{k}\sigma},\\
Y_{\vec{k}_{0}\sigma_{0}}t_{\vec{k}\sigma\vec{k}_{0}\sigma_{0}} & =Y_{\vec{k}\sigma}t_{\vec{k}_{0}\sigma_{0}\vec{k}\sigma}.
\end{align}
Therefore, we define the scattering coefficients that guarantee reciprocity,
as 
\begin{align}
\tilde{r}_{\vec{k}\sigma\vec{k}_{0}\sigma} & =\frac{1}{2}r_{\vec{k}\sigma\vec{k}_{0}\sigma_{0}}+\frac{Y_{\vec{k}\sigma}}{2Y_{\vec{k}_{0}\sigma_{0}}}r_{\vec{k}_{0}\sigma_{0}\vec{k}\sigma},\label{S_eq:r_tilde}\\
\tilde{t}_{\vec{k}\sigma\vec{k}_{0}\sigma} & =\frac{1}{2}t_{\vec{k}\sigma\vec{k}_{0}\sigma_{0}}+\frac{Y_{\vec{k}\sigma}}{2Y_{\vec{k}_{0}\sigma_{0}}}t_{\vec{k}_{0}\sigma_{0}\vec{k}\sigma}.\label{S_eq:t_tilde}
\end{align}
Using Eqs.\ (\ref{S_eq:r_tilde}) and (\ref{S_eq:t_tilde}), we can
calculate the reflection and transmission coefficients between various
diffraction orders to form reflection and transmission matrices $\tilde{\bm{r}}$
and $\tilde{\bm{t}}$. Then, we can define a scattering matrix $\bm{S}$
as 
\begin{equation}
\bm{S}=\begin{pmatrix}\tilde{\bm{t}} & \tilde{\bm{r}}\\
\tilde{\bm{r}} & \tilde{\bm{t}}
\end{pmatrix}.
\end{equation}
The full scattering matrix for the multilayer system, including nanohole
array layers, can be calculated using recurrence formula \cite{S_Weiss_11}.
If the system is divided into two parts whose scattering matrices
are given by $\bm{S}_{1}$ and $\bm{S}_{2}$, then the total scattering
matrix $\bm{S}_{2+1}$ is given by

\begin{equation}
\bm{S}_{2+1}=\bm{S}_{2}\star\bm{S}_{1},\label{S_eq:recurrence}
\end{equation}
where the operation $\star$ is defined by 
\begin{equation}
\bm{X}\star\bm{Y}=\begin{pmatrix}\bm{X}_{11}\left(1-\bm{Y}_{12}\bm{X}_{21}\right)^{-1}\bm{Y}_{11} & \bm{X}_{12}+\bm{X}_{11}\left(1-\bm{Y}_{12}\bm{X}_{21}\right)^{-1}\bm{Y}_{12}\bm{X}_{22}\\
\bm{Y}_{21}+\bm{Y}_{22}\left(1-\bm{X}_{21}\bm{Y}_{12}\right)^{-1}\bm{X}_{21}\bm{Y}_{11} & \bm{Y}_{22}\left(1-\bm{X}_{21}\bm{Y}_{12}\right)^{-1}\bm{X}_{22}
\end{pmatrix}.
\end{equation}

\end{document}